\documentclass[pre,twocolumn,floats,superscriptaddress]{revtex4}
\usepackage{amsmath}
\usepackage{amssymb}
\usepackage{graphicx}
\usepackage{color}

\newcommand{\abs}[1]{\left|#1\right|}
\newcommand{\fun}[2]{\mathop{{#1}{\left(#2\right)}}}

\newcommand{\D}[1]{\mathop{\mathrm{d}#1}}

\begin{document}
\title{Kinetic description of one-dimensional stochastic dynamics with small inertia}
%\title{Kinetic description of 1D motion of passive and active Brownian particles with small inertia}
%\titlerunning{Kinetic description of 1D motion with small inertia}
%
%\author{Denis~S.~Goldobin\inst{1,2,3,
%        }\thanks{e-mail: Denis.Goldobin@gmail.com}
%        \and Lyudmila~S.~Klimenko\inst{1,2}
%        \and Irina~V.~Tyulkina\inst{1,3}}
%%\authorrunning{D.~S.~Goldobin, L.~S.~Klimenko, I.~V.~Tyulkina}
%
%\institute{
%Institute of Continuous Media Mechanics, UB RAS, Academician Korolev Street 1, 614013 Perm, Russia
%\and Institute of Physics and Mathematics, Perm State University, Bukirev Street 15, 614990 Perm, Russia
%\and Department of Control Theory, Nizhny Novgorod State University, Gagarin Avenue 23, 603022 Nizhny Novgorod, Russia
%}
\author{Denis S.\ Goldobin}
\affiliation{Institute of Continuous Media Mechanics, Ural Branch of RAS, Acad.\ Korolev Street 1,
 614013 Perm, Russia}
\affiliation{Institute of Physics and Mathematics, Perm State University, Bukirev Street 15,
 614990 Perm, Russia}
\affiliation{Department of Control Theory, Lobachevsky State University of Nizhny Novgorod, Gagarin Avenue 23,
 603022 Nizhny Novgorod, Russia}
\author{Lyudmila S.\ Klimenko}
\affiliation{Institute of Continuous Media Mechanics, Ural Branch of RAS, Acad.\ Korolev Street 1,
 614013 Perm, Russia}
\affiliation{Institute of Physics and Mathematics, Perm State University, Bukirev Street 15,
 614990 Perm, Russia}
\author{Irina V.\ Tyulkina}
\affiliation{Institute of Continuous Media Mechanics, Ural Branch of RAS, Acad.\ Korolev Street 1,
 614013 Perm, Russia}
\affiliation{Department of Control Theory, Lobachevsky State University of Nizhny Novgorod, Gagarin Avenue 23,
 603022 Nizhny Novgorod, Russia}
\author{Vasily A.\ Kostin}
\affiliation{Department of Control Theory, Lobachevsky State University of Nizhny Novgorod, Gagarin Avenue 23,
 603022 Nizhny Novgorod, Russia}
\affiliation{Gaponov-Grekhov Institute of Applied Physics of the Russian Academy of Sciences, Ul'yanova Street 46,
 603950 Nizhny Novgorod, Russia}
\author{Lev A.\ Smirnov}
\affiliation{Department of Control Theory, Lobachevsky State University of Nizhny Novgorod, Gagarin Avenue 23,
 603022 Nizhny Novgorod, Russia}
\affiliation{Research and Education Mathematical Center ``Mathematics of Future Technologies,''\\
 Lobachevsky State University of Nizhny Novgorod, Gagarin Avenue 23,
 603022 Nizhny Novgorod, Russia}
\date{\today}

%
%\title[Low-dimensional reduction for a sparse balanced network of neurons]{Low-dimensional reduction for kinetics of a sparse balanced synaptic network of quadratic integrate-and-fire neurons}
%
%\author*[1]{\fnm{Mariya V.} \sur{Ageeva}}%\email{iauthor@gmail.com}
%%\equalcont{These authors contributed equally to this work.}
%
%\author[1,2,3]{\fnm{Denis S.} \sur{Goldobin}}\email{Denis.Goldobin@gmail.com}
%%\equalcont{These authors contributed equally to this work.}
%
%%\author[1,2]{\fnm{Third} \sur{Author}}\email{iiiauthor@gmail.com}
%%\equalcont{These authors contributed equally to this work.}
%
%\affil*[1]{\orgdiv{Department}, \orgname{Institute of Continuous Media Mechanics of UB RAS}, \orgaddress{\street{Academician Korolev Street 1}, \city{Perm}, \postcode{614013}, %\state{State},
%\country{Russia}}}
%
%\affil[2]{\orgdiv{Institute of Physics and Mathematics}, \orgname{Perm State University}, \orgaddress{\street{Bukirev Street 15}, \city{Perm}, \postcode{614990}, %\state{State},
%\country{Russia}}}
%
%\affil[3]{\orgdiv{Department of Control Theory}, \orgname{ Nizhny Novgorod State University}, \orgaddress{\street{Gagarin Avenue 23}, \city{Nizhny Novgorod}, \postcode{603022}, %\state{State},
%\country{Russia}}}

\begin{abstract}
%\color{blue}
We study single-variable approaches for describing stochastic dynamics with small inertia. The basic models we deal with describe passive Brownian particles and phase elements (phase oscillators, rotators, superconducting Josephson junctions) with an effective inertia in the case of a linear dissipation term and active Brownian particles in the case of a nonlinear dissipation.
Elimination of a fast variable (velocity) reduces the characterization of the system state to a single variable and is formulated in four representations: moments, cumulants, the basis of Hermite functions, and the formal cumulant variant of the last.
This elimination provides rigorous mathematical description for the overdamped limit in the case of linear dissipation and the overactive limit of active Brownian particles.
For the former, we derive a low-dimensional equation system which generalizes the Ott--Antonsen Ansatz to systems with small effective inertia. In the latter case, we derive a Fokker--Planck-type equation with a forced drift term and an effective diffusion in one dimension, where the standard two-/three-dimensional mechanism is impossible. In the four considered representations, truncated equation chains are demonstrated to be utilitary for numerical simulation for a small finite inertia.
\end{abstract}

%\keywords{Synchronization, Circular cumulants, Quadratic integrate-and-fire neurons, Balanced networks}

%%\pacs[JEL Classification]{D8, H51}

%%\pacs[MSC Classification]{35A01, 65L10, 65L12, 65L20, 65L70}


\maketitle

\section{Introduction}
Mathematical description of the dynamics of a system in the limit of high dissipation rate (overdamped systems) can often be reduced to a single variable. This variable is the coordinate of a mechanical system in a viscous medium (like for Brownian particles)~\cite{Haken-1977,Gardiner-1983-1997,Becker-1985} or the oscillation phase for periodic self-sustained oscillators~\cite{Winfree-1967,Kuramoto-1975}, where transversal deviations from the limit cycle decay fast enough to be negligible. However, in stochastic systems with $\delta$-correlated noise, such reduction becomes nontrivial as the inertia term is not small for fast fluctuations in mechanical systems~\cite{Haken-1977,Gardiner-1983-1997,Becker-1985,Wilemski-1976,Gardiner-1984,Goldobin-Klimenko-2020} and, in oscillatory systems, deviations from the limit cycle are non-negligible~\cite{Yoshimura-Arai-2008,Teramae-etal-2009,Goldobin-etal-2010}.
In the phase equations for oscillatory systems, a similar inertia-like term often appears, which makes the system dynamics much more complex~\cite{Acebron-Bonilla-Spigler-2000,Komarov-Gupta-Pikovsky-2014,Olmi-etal-2014,Olmi-2015,Laing-2019,Bountis-etal-2014,Jaros-Maistrenko-Kapitaniak-2015,Munyayev-etal-2020,Munyayev-etal-2022,Munyayev-etal-2023,Zharkov-Altudov-1978,Alexandrov-Gorsky-2024}.

The problem of transition to the limit of small (vanishing) inertia, in other words, the problem of adiabatic elimination of a fast variable (velocity) has been thoroughly studied for passive Brownian particles~\cite{Haken-1977,Gardiner-1983-1997,Becker-1985,Wilemski-1976,Gardiner-1984,Goldobin-Klimenko-2020,Schoner-Haken-1987} and for certain types of active Brownian particles~\cite{Milster-etal-2017}. The latter framework has also proved useful for understanding the behavior of ``overactive'' Brownian particles in potential force fields~\cite{Aranson-Pikovsky-2022,Pikovsky-2023}.

Recently, a regular approach to constructing low-dimensional reduction models of the collective dynamics of oscillator populations was introduced on the basis of the formalism of so-called circular cumulants~\cite{Tyulkina-etal-2018,Goldobin-etal-2018,Goldobin-Dolmatova-2019}. This approach generalizes the Ott--Antonsen Ansatz~\cite{Ott-Antonsen-2008,Ott-Antonsen-2009}, which itself builds on the Watanabe--Strogatz partial integrability~\cite{Watanabe-Strogatz-1993,Watanabe-Strogatz-1994,Pikovsky-Rosenblum-2008,Marvel-Mirollo-Strogatz-2009}. Applying the circular cumulant formalism to systems with non-negligible inertia necessitates a systematic analysis of possible approaches to the problem of fast variable elimination. Furthermore, the development of mean-field theories is of interest for ``swarmalators''~\cite{Tanaka-2007,OKeeffe-etal-2017}---active elements with intercoupled spatial dynamics and internal self-oscillations. The use of circular cumulants may prove fruitful for constructing such theories.

In this paper we provide a detailed analysis of the fast variable elimination problem, with emphasis on unconventional approaches and the potential for employing the circular cumulant formalism. Mathematically, this is more sophisticated than the plain moment or cumulant formalism for the joint distribution of two variables. First, these two variables can have different geometric nature: the fast variable is always on the infinite line, but the ``normal'' one is cyclic in the case of phase oscillators. Second, for the fast and normal variables we have completely different limiting cases that underlie possible macroscopic reduction. This added sophistication opens up more options in the technical details of possible approaches.

%{\color{blue}
The paper is organized as follows. In Sec.~\ref{sec2}, we formulate the mathematical model of stochastic dynamics with small inertia, provide synopses of the Ott--Antonsen theory and the circular cumulant formalism (Sec.~\ref{ssec:CCs}), and assess the scaling laws of the velocity moments (Sec.~\ref{ssec:scaling}), which are helpful for the analysis in subsequent sections. For the linear dissipation law, relevant to oscillators with small inertia and passive Brownian particles, we provide detailed analyses of the moment representation (Sec.~\ref{ssec31}), the cumulant representation (Sec.~\ref{ssec32}), the representation on the basis of Hermit functions (Sec.~\ref{ssec33}), and the formal cumulant variant for the Hermite basis (Sec.~\ref{ssec34}).
In Sec.~\ref{sec:ABP}, we construct the moment and cumulant representations for an active Brownian particle, address the problem of fast variable elimination, and derive the forced drift and diffusion terms for one-dimensional motion.
In Sec.~\ref{sec:litr} we place the analysis and results of this paper against the backdrop of the reference works in the field.
In Sec.~\ref{sec:applic} we derive a low-dimensional model reduction that generalizes the Ott--Antonsen Ansatz to oscillator populations with small inertia and examine its accuracy and utility.
Conclusions are summarized in Sec.~\ref{sec:concl}.
%}

\section{Kinetic description of populations of Brownian particles and phase oscillators with inertia}
\label{sec2}
The analysis we conduct in this paper is valid for both Brownian particles and phase elements with small effective inertia. We consider the Langevin equation with inertia:
\begin{equation}
\mu\ddot\varphi+\dot\varphi=F(\varphi,t)+\sigma\xi(t)\,, %\qquad\mu\ll1\,,
\label{eq001}
\end{equation}
where $\mu$ is the mass for Brownian particles~\cite{Juniper-etal-2015,Tierno-Johansen-Straube-2021,Kourov-Samoilova-Straube-2025} or a measure of dynamics inertia for such systems as superconducting Josephson junctions~\cite{Zharkov-Altudov-1978}, some models of electric power grids~\cite{Morren-etal-2006,Short-Infield-Freris-2007}, etc.; $F(\varphi,t)$ is a deterministic force, $\sigma$ is the noise amplitude, $\xi(t)$ is the normalized $\delta$-correlated Gaussian noise: $\langle\xi\rangle=0$, $\langle\xi(t)\,\xi(t')\rangle=2\delta(t-t')$. For many realistic physical systems, dimensionless parameter $\mu$ is small, but in the case of a $\delta$-correlated noise the limit $\mu\to0$ does not correspond to a simple dropping of the first term of Langevin equation~(\ref{eq001}). Indeed, for the fluctuating part of $\varphi=\langle\varphi\rangle+\widetilde{\varphi}$, where $\langle\cdots\rangle$ indicates the averaging over noise realizations, in the presence of such noise, one finds $|\ddot{\widetilde{\varphi}}|/|\dot{\widetilde{\varphi}}|\to\infty$; whence for any small but finite $\mu$ the reference values of the first term of Eq.~(\ref{eq001}) are infinitely large against the background of the reference values of the second term. The problem of taking the limit $\mu\to0$ for such problems is known in statistical physics and thermodynamics as the problem of fast variable elimination~\cite{Haken-1977,Gardiner-1983-1997,Becker-1985}.

Prior to turning to the main analysis of this paper, we would like to mention one of important motives for this work---recently introduced formalism of circular cumulants~\cite{Tyulkina-etal-2018,Goldobin-etal-2018,Goldobin-Dolmatova-2019}. Within the framework of this formalism, the generalization of the Ott--Antonsen theory~\cite{Ott-Antonsen-2008,Ott-Antonsen-2009} to nonideal situations, where the conditions of the original theory are violated, became possible. The presence of small inertia is an important peculiar case of nonideal situation. In our analysis we will bear in mind the issue of employment of the circular cumulant representation for the problems where the applicability conditions of the original Ott--Antonsen theory are violated by the presence of inertia and its smallness allows one to raise the question of construction of a perturbation theory.

\subsection{Representation of circular cumulants and Ott--Antonsen theory}
\label{ssec:CCs}
Here we provide a brief synopsis of the Ott--Antonsen (OA) theory and its parts relevant for our work. The OA theory is valid for a sinusoidal shape of $F(\varphi,t)=\omega(t)+b(t)\sin\varphi+c(t)\cos\varphi$ or, which is the same, $F(\varphi,t)=\omega(t)+\mathrm{Im}[2h(t)e^{-i\varphi}]$ with $2h(t)=-b(t)+ic(t)$. This shape is found for many classical problems of Nonlinear dynamics; for instance, for the Kuramoto ensemble~\cite{Kuramoto-1975}, chain of superconducting Josephson junctions~\cite{Watanabe-Strogatz-1993,Watanabe-Strogatz-1994}, ensemble of coupled active rotators~\cite{Klinshov-Franovic-2019}, theta-neurons and quadratic integrate-and-fire neurons~\cite{Pazo-Montbrio-2014,Laing-2014}. For Eq.~(\ref{eq001}) without the inertia term and with a sinusoidal shape of $F(\varphi,t)$,
\begin{equation}
\dot\varphi=\omega(t)+\mathrm{Im}[2h(t)e^{-i\varphi}]+\sigma\xi(t)\,,
\label{eqCC01}
\end{equation}
the evolution of the probability density function $w(\varphi,t)$ is governed by the Fokker--Planck equation:
\begin{align}
\partial_tw(\varphi,t)&+\partial_\varphi\left[\left(\omega(t)-ih(t)e^{-i\varphi}
 +ih^\ast(t)e^{i\varphi}\right)w(\varphi,t)\right]
\nonumber\\
 &\qquad\qquad
 =\sigma^2\partial_\varphi^2w(\varphi,t)\,.
\label{eqCC02}
\end{align}
In Fourier space, $w(\varphi,t)=(2\pi)^{-1}\sum_{n=-\infty}^\infty a_n(t)e^{-in\varphi}$, where $a_{-n}=a_n^\ast$ and $a_0=1$, since $w(\varphi,t)$ is real and normalized,  $\int_0^{2\pi}w(\varphi,t)\mathrm{d}\varphi=1$, and the Fokker--Planck equation acquires the form
\begin{equation}
\dot{a}_n=n\big[i\omega(t)\,a_n +h(t)\,a_{n-1} -h^\ast(t)\,a_{n+1}\big]
 -\sigma^2n^2a_n\,.
\label{eqCC03}
\end{equation}
For a large population of identical oscillators $\varphi_j$ obeying Eq.~(\ref{eqCC01}) with independent noise inputs $\xi(t)$, the quantities $a_n(t)=\langle{e^{in\varphi}}\rangle$ are also Kuramoto--Daido order parameters~\cite{Kuramoto-1975,Daido-1996} (for $n=1$ we have the standard Kuramoto order parameter~\cite{Kuramoto-1975}). From the view point of statistics of a random variable on the circumference~\cite{Ley-Verdebout-2017}, $a_n$ can be called {\em circular moments}.

For $\sigma=0$ (no individual noise), the infinite chain of equations~(\ref{eqCC03}) admits ansatz $a_n=(a_1)^n$ for $n\ge0$, which is called the ``Ott--Antonsen Ansatz.'' With this ansatz for all $n\ge1$ we obtain the same equation:
\begin{equation}
\dot{a}_1=i\omega(t)\,a_1 +h(t) -h^\ast(t)\,a_1^2\,.
\label{eqCC04}
\end{equation}
This exact low-dimensional equation for the dynamics of the Kuramoto order parameter is the main result of the OA theory and allowed obtaining important analytical results in nonlinear dynamics.

The problem of generalization of the OA theory to nonideal situations even in the cases where one has obvious small parameter (for instance, $\sigma$) was persisting for 10 years after the pioneering work~\cite{Ott-Antonsen-2008} in 2008, since, in the representation of circular moments $a_n$, even a small violation of the applicability of the OA Ansatz $a_n=a_1^n$ does not give an obvious hierarchy of small corrections to the solution. In~\cite{Tyulkina-etal-2018} the representation of so-called circular cumulants $\kappa_n$ was introduced; $\kappa_n$ are related to circular moments by the recursive formula (see Appendix~\ref{sec:app2})
\begin{equation}
\kappa_n=\frac{a_n}{(n-1)!}-\sum_{l=1}^{n-1}\frac{\kappa_l a_{n-l}}{(n-l)!}\,;
\label{eqCC05}
\end{equation}
in particular,
%\[
%\kappa_1=a_1,\qquad
%\kappa_2,\qquad
%\kappa_3=(a_3-3a_2a_1+2a_1^3)/2.
%\]
$\kappa_1=a_1$ and $\kappa_2=a_2-a_1^2$.
Recursive formula~(\ref{eqCC05}) differs from its analog for the conventional moments and cumulants, since for the circular cumulants a different normalization is adopted. The conventional normalization would give $\kappa_n^\prime=(n-1)!\kappa_n$. The choice of unconven\-tion\-al normalization is admissible because $\kappa_n$ are not genuine analogs of cumulants and only possess formal similarities to them on the one hand, and, on the other hand, the equations of dynamics of $\kappa_n$ acquire the simplest form for this normalization.

In terms of circular cumulants the OA Ansatz corresponds to a very simple form of solutions: $\kappa_1=a_1$, $\kappa_{n\ge2}=0$; and weak violations of the applicability of the original theory generate hierarchies of smallness of $\kappa_n$, which allows one to construct a perturbation theory. The specific form of hierarchy depends on the specific form of a weak applicability violation~\cite{Tyulkina-etal-2018,Goldobin-Dolmatova-2020,diVolo-etal-2022}, but always allows one to obtain expansions with respect to a small parameter. For instance, in the presence of noise ($\sigma\ne0$) for Eq.~(\ref{eqCC03}) one finds~\cite{Tyulkina-etal-2018}:
\begin{align}
\dot\kappa_n=in\omega\kappa_n &+h\delta_{1n}
%\\
%\qquad
 -h^\ast\Big(n^2\kappa_{n+1}+n\sum\limits_{m=0}^{n-1}\kappa_{n-m}\kappa_{m+1}\Big)
\nonumber
\\
& -\sigma^2\Big(n^2\kappa_n+n\sum\limits_{m=0}^{n-2}\kappa_{n-1-m}\kappa_{m+1}\Big)\,.
\label{eqCC06}
\end{align}
The infinite equation chain~(\ref{eqCC06}) cannot be truncated as the dynamics of $\kappa_n$ is subject to forcing by $-n^2h^\ast\kappa_{n+1}$.
However, for small $\sigma$ the chain~(\ref{eqCC06}) generates the smallness hierarchy $\kappa_n\propto\sigma^{2(n-1)}$, which allows one to construct a perturbation theory of prescribed accuracy.
The leading order corrections are practically important; these corrections are fully provided by the first two equations of the chain~(\ref{eqCC06}): $n=1,2$. The chain can be formally truncated by setting higher order cumulant $\kappa_3=0$. This delivers a two circular cumulant generalization of the Ott--Antonsen theory:
\begin{align}
\dot{\kappa}_1&=i\omega\kappa_1+h-h^\ast(\kappa_1^2+\kappa_2) -\sigma^2\kappa_1\,,
\label{eqCC07}
\\[5pt]
\dot{\kappa}_2&=2i\omega\kappa_2-4h^\ast\kappa_1\kappa_2 -\sigma^2(4\kappa_2+2\kappa_1^2)\,,
\label{eqCC08}
\end{align}
the accuracy of which was thoroughly examined in~\cite{Goldobin-etal-2018}.

The presence of small inertia is a peculiar and important nontrivial case of violation of the applicability conditions of the original Ott--Antonsen theory.

\subsection{Asymptotic scaling law for velocity moments for $\mu\to0$}
\label{ssec:scaling}
The moments of the microscopic velocity of Brownian particles or
$\dot\varphi$ for oscillators diverge as $\mu\to0$. Understanding of the asymptotic laws of this divergence assists in constructing expansions with respect to $\mu$ in the subsequent sections. For the derivation of the scaling laws we decompose the velocity into the mean and fluctuating parts $\varphi=\langle\varphi\rangle+\widetilde{\varphi}$ (where $\langle\widetilde{\varphi}\rangle=0$) and substitute to Langevin equation~(\ref{eq001}). One finds
\[
\langle\dot\varphi\rangle=\langle F(\varphi,t)\rangle
\]
and, keeping only the leading terms
(in particular, notice $|\dot{\widetilde{\varphi}}|\gg|F(\varphi,t)-\langle F(\varphi,t)\rangle|$\,),
\[
\ddot{\widetilde{\varphi}}+\frac{1}{\mu}\dot{\widetilde{\varphi}}\approx\frac{\sigma}{\mu}\xi(t)\,.
\]
The solution of this equation is
\[
\dot{\widetilde{\varphi}}(t) =\frac{\sigma}{\mu}\int\limits_0^{+\infty}\mathrm{d}\tau\,\xi(t-\tau)e^{-\frac{\tau}{\mu}}\,;
\]
therefore, $\dot{\widetilde{\varphi}}$ is a Gaussian random variable. One can calculate its variance:
\[
\langle[\dot{\widetilde{\varphi}}(t)]^2\rangle=\frac{\sigma^2}{\mu^2} \int\limits_0^{+\infty}\mathrm{d}\tau_1 \int\limits_0^{+\infty}\mathrm{d}\tau_2\,
2\delta(\tau_1-\tau_2)e^{-\frac{\tau_1+\tau_2}{\mu}}=\frac{\sigma^2}{\mu}\,.
\]
Hence, one can write
\[
\dot{\widetilde{\varphi}}=\frac{\sigma}{\sqrt{\mu}}R\,,
\]
where $R$ is a normalized Gaussian random number $\mathcal{N}(0,1)$.
Finally,
%\begin{widetext}
\begin{align}
\langle v^n\rangle&=
\langle\big[\langle\dot\varphi(t)\rangle+\dot{\widetilde{\varphi}}(t)\big]^n\rangle
\nonumber\\[7pt]
&%\quad
\approx\left\{
\begin{array}{cc}
\displaystyle
\langle\big[\dot{\widetilde{\varphi}}(t)\big]^n\rangle\,%,
& \mbox{ for even }n\,,\\
\displaystyle
\langle\big[\dot{\widetilde{\varphi}}(t)\big]^n\rangle +n\langle\dot\varphi(t)\rangle\,\langle\big[\dot{\widetilde{\varphi}}(t)\big]^{n-1}\rangle\,%,
& \mbox{ for odd }n\\
\end{array}
\right.
\nonumber\\[5pt]
&%\qquad
\propto\left\{
\begin{array}{cr}
\displaystyle
\frac{\sigma^n}{\mu^{n/2}}\,%,
& \mbox{ for even }n\,,\\
\displaystyle
n\langle\dot\varphi\rangle\,\frac{\sigma^{n-1}}{\mu^{(n-1)/2}}\,%,
& \mbox{ for odd }n\,.
\end{array}
\right.
%\nonumber
\label{eq0Sc}
\end{align}
%\end{widetext}
The asymptotic scaling laws for even and odd moments are different; in particular, the magnitude of the odd moments is defined by the average dynamics.

%Hence,
%\[
%w_n
%\sim\left\{
%\begin{array}{cl}
%\displaystyle
%\frac{\sigma^n}{\mu^{n/2}}\,,
%& \mbox{ for even }n\,;\\
%\displaystyle
%n\langle\dot\varphi\rangle\,\frac{\sigma^{n-1}}{\mu^{(n-1)/2}}\,,
%& \mbox{ for odd }n\,.
%\end{array}
%\right.
%\]

\section{Passive Brownian particles and phase oscillators with inertia}
\label{sec3}
For the Langevin equation with inertia (\ref{eq001}) the evolution of the probability density $\rho(v,\varphi)$, where $v\equiv\dot\varphi$, is governed by the Fokker--Planck equation (FPE)
\begin{equation}
\partial_t\rho=-v\partial_\varphi\rho+\partial_v\left\{\frac{1}{\mu}\big[v-F(\varphi,t)\big]\rho\right\}
+\frac{\sigma^2}{\mu^2}\partial_v^2\rho\,,
\label{eq002}
\end{equation}
where $\varphi$ can be defined in a rotating reference frame if needed~\cite{Pikovsky-Rosenblum-Kurths-2003}.
Our goal is to exclude the velocity $v$ and describe the effective dynamics of a single variable $\varphi$. We examine four possible approaches to accomplishing this task. The diversity of approaches is motivated by the difference between the representations in terms of circular moments and cumulants (Sec.~\ref{ssec:CCs}).

\subsection{Moment representation for Fokker--Planck equation}
\label{ssec31}
We deal with the moments of velocity $v$
\[
w_n(\varphi,t)=\int\limits_{-\infty}^{+\infty}v^n\rho(v,\varphi,t)\,\mathrm{d}v\,.
\]
Multiplying FPE~(\ref{eq002}) by $v^n$ and integrating over $v$, one finds
\begin{align}
&\partial_tw_0+\partial_\varphi w_1=0\,,
\label{eq003}\\
&w_1+\mu\partial_tw_1=F w_0-\mu\partial_\varphi w_2\,,
\label{eq004}\\
&w_n+\frac{\mu}{n}\partial_t w_n=F w_{n-1}-\frac{\mu}{n}\partial_\varphi w_{n+1}
\nonumber\\
&\hspace{2cm} {}
 +(n-1)\frac{\sigma^2}{\mu}w_{n-2}\quad\mbox{ for } n\ge2\,.
\label{eq005}
\end{align}

For constructing a perturbation theory with small parameter $\mu$, convenient is to account for the scaling  $\langle{v^n}\rangle$~(\ref{eq0Sc}) and rescale moments
\begin{equation}
w_n
=\left\{
\begin{array}{cr}
\displaystyle
\frac{1}{\mu^{n/2}}W_n
& \mbox{ for even }n\,,\\
\displaystyle
\frac{1}{\mu^{(n-1)/2}}W_n
& \mbox{ for odd }n\,.
\end{array}
\right.
\label{eq:wW}
\end{equation}
Now one can rewrite Eqs.~(\ref{eq003})--(\ref{eq005}) in a form which is free of diverging coefficients $\propto1/\mu$:
\begin{align}
&\partial_t W_0+\partial_\varphi W_1=0\,,
\label{eq006}\\
&W_1+\mu\partial_t W_1=F W_0-\partial_\varphi W_2\,,
\label{eq007}\\
&W_n+\frac{\mu}{n}\partial_t W_n=\mu F W_{n-1}-\frac{\mu}{n}\partial_\varphi W_{n+1}
\nonumber\\
&\hspace{1.5cm} {}
+(n-1)\sigma^2W_{n-2}\qquad\mbox{ for } n=2m\,,
\label{eq008}\\
&W_n+\frac{\mu}{n}\partial_t W_n=F W_{n-1}-\frac{1}{n}\partial_\varphi W_{n+1}
\nonumber\\
&\hspace{1.5cm} {}
+(n-1)\sigma^2W_{n-2}\quad\mbox{ for } n=2m+1\,.
\label{eq009}
\end{align}
By regrouping terms, one can obtain:
\begin{align}
&\partial_t W_0+\partial_\varphi W_1=0\,,
\label{eq010}\\
&W_1=F W_0-\partial_\varphi W_2-\mu\partial_t W_1\,,
\label{eq011}\\
&W_n=(n-1)\sigma^2W_{n-2}+\mu\Big[F W_{n-1}-\frac{1}{n}\partial_\varphi W_{n+1}
\nonumber\\
&\hspace{1.5cm} {}
-\frac{1}{n}\partial_t W_n\Big]\qquad\mbox{ for } n=2m\,,
\label{eq012}\\
&W_n=(n-1)\sigma^2W_{n-2}+F W_{n-1}-\frac{1}{n}\partial_\varphi W_{n+1}
\nonumber\\
&\hspace{1.5cm} {}
-\frac{\mu}{n}\partial_t W_n\qquad\mbox{ for } n=2m+1\,.
\label{eq013}
\end{align}
The derived equation system contains only $\mu^0$- and $\mu^1$-terms, which makes taking the limit $\mu\to0$ trivial.

\subsubsection{Adiabatic elimination of fast variable}
System~(\ref{eq010})--(\ref{eq013}) for $\mu=0$ acquires the form
\begin{align}
&%\textstyle
\partial_t W_0+\partial_\varphi W_1=0\,,
\label{eq014}\\
&%\textstyle
W_1=F W_0-\partial_\varphi W_2\,,
\label{eq015}\\
&%\textstyle
W_{2m}=(2m-1)\sigma^2W_{2(m-1)}\,,
\label{eq016}\\
&%\textstyle
W_{2m+1}=2m\sigma^2W_{2m-1}+F W_{2m}-\frac{\partial_\varphi W_{2(m+1)}}{2m+1}\,.
\label{eq017}
\end{align}
Eq.~(\ref{eq016}) yields
\[
W_{2m}=(2m-1)!!\,\sigma^{2m}W_0\,,
\]
where we use notation $(2m-1)!!\equiv 1\times 3\times 5\times 7\times\dots\times(2m-1)$.
From Eq.~(\ref{eq017}),
\[
W_{2m+1}=2m\sigma^2W_{2m-1}+(2m-1)!!\,\sigma^{2m}(F-\sigma^2\partial_\varphi)W_0\,.
\]
With $W_2=\sigma^2W_0$, Eqs.~(\ref{eq014}) and (\ref{eq015}) give
\[
W_1=(F-\sigma^2\partial_\varphi)W_0\,,
\]
\begin{equation}
\partial_t W_0+\partial_\varphi(F W_0)=\sigma^2\partial_\varphi^2W_0\,.
\label{eq018}
\end{equation}
Thus, we obtain a usual Fokker--Planck-type equation for $W_0$, and all higher $W_{n\ge1}$ can be calculated from $W_0$ in a trivial way. Notice, the derivation of Eq.~(\ref{eq018}) required employment of Eqs.~(\ref{eq014})--(\ref{eq016}). Thus, if one deals with truncations of infinite chain~(\ref{eq010})--(\ref{eq013}) for a finite small $\mu$, then the {\em adiabatic elimination of a fast variable}~\cite{Becker-1985,Haken-1977,Gardiner-1983-1997} corresponds to truncation after the first three equations.

\subsubsection{Corrected Smoluchowski equation ($\mu^1$-correction)}
Here we derive the $\mu^1$-correction to Eq.~(\ref{eq018}) --- so-called cor\-rect\-ed Smoluchowski equation~\cite{Gardiner-1983-1997,Wilemski-1976}.
Keeping the $\mu^1$-cor\-rec\-tions to $W_0$, one can obtain from the infinite equation chain~(\ref{eq010})--(\ref{eq013})
\begin{align}
&%\textstyle
\partial_t W_0+\partial_\varphi W_1=0\,,
\label{eq019}\\
&%\textstyle
W_1=F W_0-\partial_\varphi W_2-\mu\partial_t W_1\,,
\label{eq020}\\
&%\textstyle
W_2=\sigma^2W_0+\mu\left[-\frac12\partial_t W_2 +F W_1-\frac12\partial_\varphi W_3\right]\,,
\label{eq021}\\
&%\textstyle
W_3=2\sigma^2W_1+F W_2-\frac13\partial_\varphi W_4+\mathcal{O}(\mu)\,,
\label{eq022}\\
&%\textstyle
W_4=3\sigma^2W_2+\mathcal{O}(\mu)\,.
\label{eq023}
\end{align}

Starting from substitution of $W_4$ into the expression for $W_3$, one can step-by-step obtain
\begin{align}
&%\textstyle
W_3=2\sigma^2W_1+F W_2-\sigma^2\partial_\varphi W_2+\mathcal{O}(\mu)\,,
\nonumber\\
&%\textstyle
W_2=\sigma^2W_0+\mu\Big[-\frac{\sigma^2}{2}\partial_t W_0
 +F(F W_0-\sigma^2\partial_\varphi W_0)
\nonumber\\
&%\textstyle
\qquad\qquad
 -\frac{\sigma^2}{2}\partial_\varphi(F W_0)+\frac{\sigma^4}{2}\partial_\varphi^2W_0
\nonumber\\
&%\textstyle
\qquad\qquad
 -\sigma^2\partial_\varphi(F W_0-\sigma^2\partial_\varphi W_0)\Big]+\mathcal{O}(\mu^2)\,,
\nonumber\\[5pt]
&W_1=F W_0-\sigma^2\partial_\varphi W_0+\mu\Big[-(\partial_tF
 +F\partial_\varphi F)W_0
\nonumber\\
&%\textstyle
\qquad\qquad\qquad
 +\sigma^2(\partial_\varphi F)\partial_\varphi W_0\Big]+\mathcal{O}(\mu^2)\,.
\nonumber
\end{align}

Finally, in the $\mu^1$-order:
\begin{align}
\partial_tW_0+\partial_\varphi\big\{[F
 -\mu(\partial_tF+F\partial_\varphi F)]\,W_0\big\}
\quad
\nonumber\\
 {}
=\sigma^2\partial_\varphi\big[(1-\mu\partial_\varphi F)\,\partial_\varphi W_0\big]\,.
\label{eq024}
\end{align}
This is the {\em corrected Smoluchowski equation}~\cite{Gardiner-1983-1997,Wilemski-1976}.
An effective Langevin equation (in the Stratonovich interpretation) corresponding to the FPE~(\ref{eq024}) reads
\begin{equation}
%\textstyle
\dot\varphi=F-\mu\left(\partial_t+F\partial_\varphi +\frac{\sigma^2}{2}\partial_\varphi^2\right)F
 +\sigma\sqrt{\left|1-\mu\partial_\varphi F\right|}\,\xi(t)\,,
\label{eq:eff_Lang}
\end{equation}
where $\partial_tF(\varphi,t)$ is the partial derivative of $F$ with respect to $t$ under fixed $\varphi$.
Importantly, this equation accounts for nonstationarity of $F$ (for instance, Gardiner considers only the case of a stationary $F$~\cite{Gardiner-1983-1997}), which allows one to employ this equation for studies of self-organization in large ensembles where $F$ depends on integral order parameters evolving in time (see Sec.~\ref{ssec:CCs}).

\subsubsection{Higher order corrections}
The basic adiabatic elimination of a fast variable requires consideration of the first three moments $w_0$, $w_1$, $w_2$. The first correction for small $\mu$ requires $w_3$ and $w_4$. Numerical simulations of equation system~(\ref{eq003})--(\ref{eq005}) for $w_0$, $w_1$, ..., $w_{2m+2}$ with formal closure $w_{2m+3}=0$ delivers the accuracy order $\mu^m$. The truncated expansion with odd order of the last nonzero element, i.e., formal closure $w_{2m+2}=0$, still converges for $\mu\to 0$ or for very long series, $m\gg1$; however, the accuracy order in this case is significantly worsened.

\begin{figure*}[!t]
\centerline{
\sf
(a)\hspace{-12pt}
\includegraphics[width=0.4015\textwidth]%
 {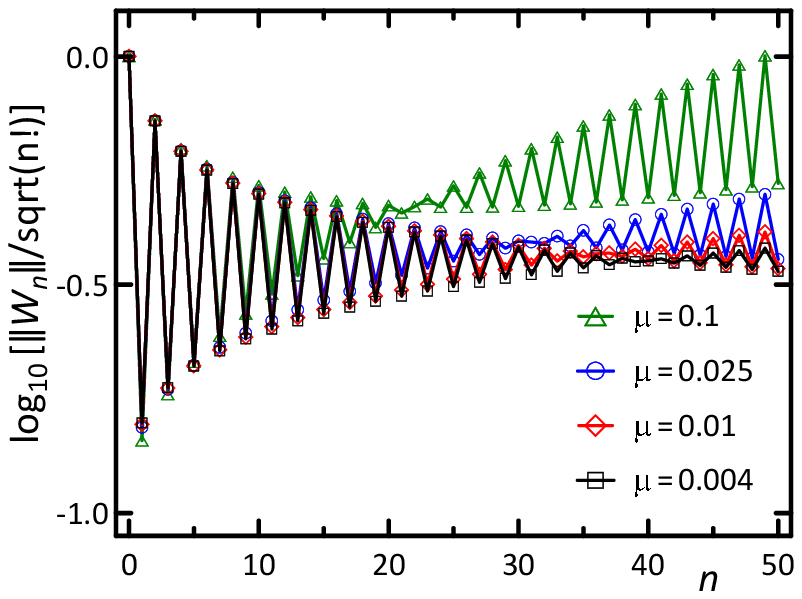 }
\qquad\qquad
(b)\hspace{-12pt}
\includegraphics[width=0.4015\textwidth]%
 {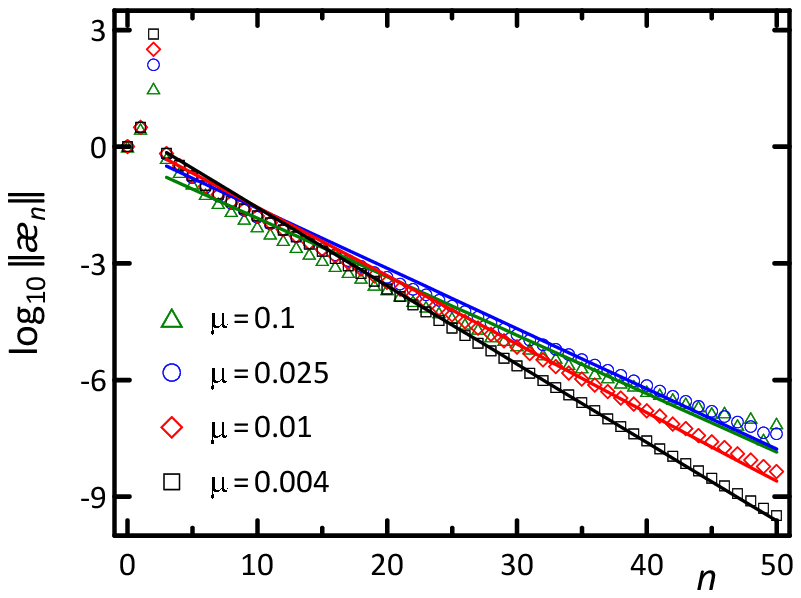}
}\vspace{9pt}

\centerline{
\sf
(c)\hspace{-12pt}
\includegraphics[width=0.4015\textwidth]%
 {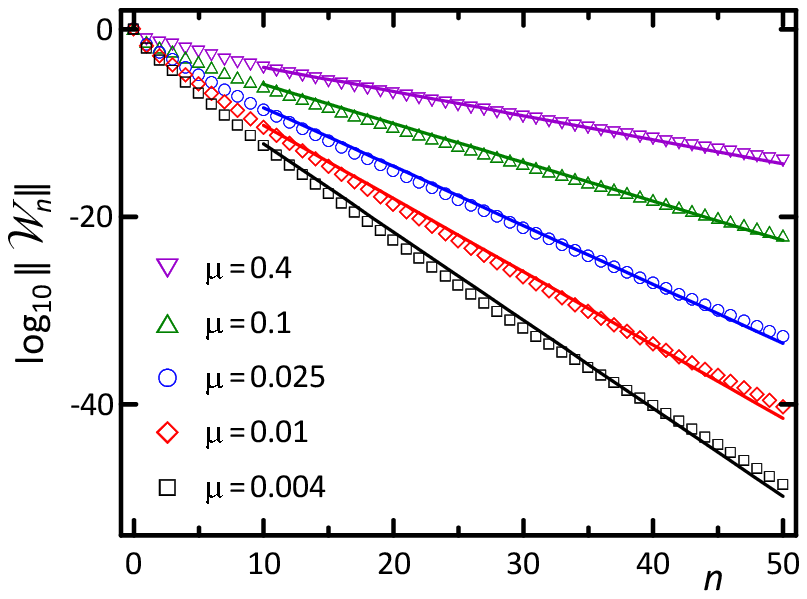}
\qquad\qquad
(d)\hspace{-12pt}
\includegraphics[width=0.4015\textwidth]%
 {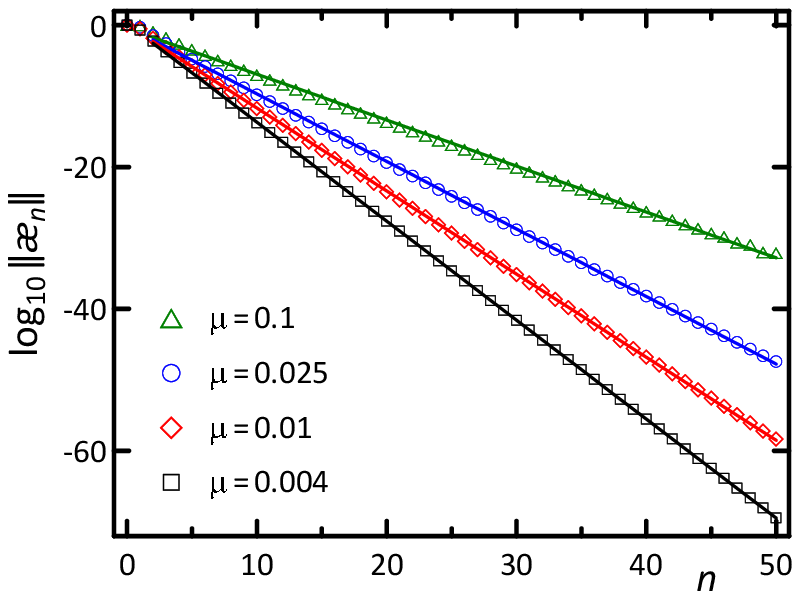}
}\vspace{7pt}

\caption{Hierarchy of smallness of high-order elements for different approaches; $L^1$-norm $||g(\varphi)||\equiv\int_0^{2\pi}|g(\varphi)|\,\mathrm{d}\varphi$.
The probability density functions $W_0(\varphi)$ for all approaches coincide with a relative accuracy on the level of the machine calculation accuracy.
(a):~moments, (b):~cumulants, (c):~Hermite basis, (d):~formal cumulants for the Hermite basis. (a,c):\ 100 elements are used for simulations, (b,d):\ 50 elements are used for simulations.
Equations are discretized in the $\varphi$-coordinate by means of the central difference schemes for derivatives and the number of nodes $N=100$.
The solid lines in panels~(b--d) serve as a guide to estimate how faithfully the high-order elements follow a geometric progression.
}
  \label{fig1}
\end{figure*}

In Figs.~\ref{fig1}(a) and \ref{fig2}(a), the formulated conclusions of the theoretical analysis are confirmed by the results of numerical simulation for the Kuramoto ensemble with small inertia and noise~\cite{Komarov-Gupta-Pikovsky-2014}. This ensemble corresponds to Eq.~(\ref{eq001}) with
$$
F=\omega+\mathrm{Im}(2he^{-i\varphi})
$$
and $h=\varepsilon a_1/2$, where $\varepsilon$ is the coupling coefficient.
The plotted data are calculated for $F=0.5+1.8\sin\varphi$, which self-organizes for the subpopulation of oscillators with natural frequency $\omega=0.5$ in a population with the bimodal distribution of natural frequencies with bandwidth $1$, noise amplitude $\sigma=1$, coupling $\varepsilon\approx3$; %~\cite{Komarov-Gupta-Pikovsky-2014};
for these parameter values the Kuramoto order parameter $\mathrm{Re}(a_1)\approx 0.6$. In Fig.~\ref{fig1}(a) we explicitly account for the scaling $W_n\propto\sqrt{n!}$\,: the $L^1$-norm $||W_n(\varphi)||\equiv\int_0^{2\pi}|W_n(\varphi)|\,\mathrm{d}\varphi$ is used and the quantity $\|W_n\|/\sqrt{n!}$ in the graph varies  in the range from $0.15$ to $1$, which is a small variation against the background of variation of $\sqrt{n!}$ for $n$ from $0$ to $50$.

%%%%%%%%%%%%%%%%%%%%%%%%%%%%%%%%%%%%%%%%%%%%%%%%%%%%%%%%%%%%
%%%%%%%%%%%%%%%%%%%%%%%%%%%%%%%%%%%%%%%%%%%%%%%%%%%%%%%%%%%%
%%%%%%%%%%%%%%%%%%%%%%%%%%%%%%%%%%%%%%%%%%%%%%%%%%%%%%%%%%%%
%%%%%%%%%%%%%%%%%%%%%%%%%%%%%%%%%%%%%%%%%%%%%%%%%%%%%%%%%%%%

\subsection{Cumulant representation}
\label{ssec32}
Equation system~(\ref{eq003})--(\ref{eq005}) for $w_n$, rewritten as
\begin{align}
nw_n+\mu\partial_t w_n=nF w_{n-1}-\mu\partial_\varphi w_{n+1}
%\qquad
%\nonumber\\
%\textstyle
%{}
+n(n-1)\frac{\sigma^2}{\mu}w_{n-2}\,,
\nonumber
\end{align}
gives for the generating function (characteristic function~\cite{Lukacs-1970})
\begin{equation}
f_w(s,\varphi,t)\equiv\sum_{n=0}^{+\infty}w_n(\varphi,t)\frac{s^n}{n!}
\label{eqCM0}
\end{equation}
the following evolution equation:
\begin{align}
\left(s\partial_s+\mu\partial_t\right)f_w=\left(sF-\mu\partial_s\partial_\varphi
 +s^2\frac{\sigma^2}{\mu}\right)f_w\,.
\nonumber
\end{align}
The procedure of derivation of the evolution equation for the generating function is described in~\cite{Goldobin-Dolmatova-2020} in detail and also implemented in \cite{Tyulkina-etal-2018} for ensembles of phase oscillators with additive noise and in \cite{Zheng-Kotani-Jimbo-2021,Goldobin-2021} for neural networks.

For the logarithm of generating function $\phi=\ln f_w$, $\partial f_w=f_w\partial\phi$, and
\begin{align}
%\textstyle
(s\partial_s+\mu\partial_t)\phi=sF+s^2\frac{\sigma^2}{\mu}
 -\mu\big[\partial_s\partial_\varphi\phi+(\partial_s\phi)(\partial_\varphi\phi)\big]\,.
%\nonumber
\label{eqCM1}
\end{align}
One can introduce cumulants of velocity $K_n(\varphi,t)$ via generating function
\begin{equation}
\phi(s,\varphi,t)\equiv\sum_{n=0}^{+\infty}K_n(\varphi,t)\frac{s^n}{n!}\,;
%\nonumber
\label{eqCM2}
\end{equation}
for such definition, the recursive formulas, allowing one to calculate moments and cumulants from each other, have the following form [at variance with formula~(\ref{eqCC05}); see Appendix~\ref{sec:app1}]:
\begin{equation}
\begin{array}{l}
\displaystyle
K_0=\ln{w_0}\,,
\\
\displaystyle
K_n=\frac{w_n}{w_0}-\sum_{l=1}^{n-1}\left({n-1 \atop l-1}\right)K_l \frac{w_{n-l}}{w_0}\,
\;\mbox{ for }n\ge1\,.
\end{array}
\label{eqCM3}
\end{equation}
where the binomial coefficients
 $\big({n \atop l}\big)=\frac{n!}{l!(n-l)!}$.
Substituting expansion~(\ref{eqCM2}) into Eq.~(\ref{eqCM1}) one finds
\begin{align}
%\textstyle
\mu\partial_tK_0&%\textstyle
=-\mu[\partial_\varphi K_1+K_1\partial_\varphi K_0]\,,
\label{eq025}\\[5pt]
%\textstyle
(n+\mu\partial_t)K_n&%\textstyle
=F\delta_{1n}+\frac{2\sigma^2}{\mu}\delta_{2n}
 -\mu\Big[\partial_\varphi K_{n+1}
\nonumber\\
&%\textstyle
\hspace{-1cm}
 +\sum\limits_{j=0}^{n}\left({n \atop j}\right)K_{j+1}\partial_\varphi K_{n-j}\Big]
\quad\mbox{ for }n\ge1\,.
\label{eq026}
\end{align}
For consistency with the representation of circular cumulants (Sec.~\ref{ssec:CCs} and Refs.~\cite{Tyulkina-etal-2018,Goldobin-etal-2018,Goldobin-Dolmatova-2019}) and ease of comparison, it can be convenient to introduce  $\varkappa_n=K_n/n!$ and rewrite the latter equation system in the following form:
\begin{align}
%\textstyle
\mu\partial_t\varkappa_0&%\textstyle
=-\mu[\partial_\varphi\varkappa_1 +\varkappa_1\partial_\varphi\varkappa_0]\,,
\qquad
\label{eq027}\\
%\textstyle
(n+\mu\partial_t)\varkappa_n&%\textstyle
=F\delta_{1n}+\frac{\sigma^2}{\mu}\delta_{2n}
 -\mu\Big[(n+1)\partial_\varphi\varkappa_{n+1}
 \nonumber\\
&%\textstyle
\quad
+\sum\limits_{j=1}^{n+1}j\varkappa_{j}\partial_\varphi\varkappa_{n+1-j}\Big]
\quad\mbox{ for }n\ge1\,.
\label{eq028}
\end{align}

Considering the first equations of chain~(\ref{eq025})--(\ref{eq026}),
\begin{equation}
%\textstyle
\begin{array}{l}
\qquad\quad\partial_tK_0=-\partial_\varphi K_1-K_1\partial_\varphi K_0\,,
\qquad
%\nonumber
\\[3pt]
%\textstyle
(1+\mu\partial_t)K_1
%\textstyle
=F -\mu\big[\partial_\varphi K_2
% \nonumber\\
%&\textstyle\quad
 +K_1\partial_\varphi K_1+K_2\partial_\varphi K_0\big]\,,
%\nonumber
\\[2pt]
%\textstyle
(2+\mu\partial_t)K_2
%\textstyle
=\frac{2\sigma^2}{\mu} -\mu\big[\partial_\varphi K_3 +K_1\partial_\varphi K_2
% \nonumber
\\
%&\textstyle
\hspace{2.5cm}
{}+2K_2\partial_\varphi K_1+K_3\partial_\varphi K_0\big]\,,
%\nonumber
\\[3pt]
%\textstyle
(3+\mu\partial_t)K_3
%\textstyle
=-\mu\big[\partial_\varphi K_4 +K_1\partial_\varphi K_3 +3K_2\partial_\varphi K_2
% \nonumber
\\[2pt]
%&\textstyle
\hspace{2.5cm}
{}+3K_3\partial_\varphi K_1+K_4\partial_\varphi K_0\big]\,,
%\nonumber
\\[3pt]
%\textstyle
(4+\mu\partial_t)K_4
%\textstyle
=-\mu\big[\partial_\varphi K_5 +K_1\partial_\varphi K_4 +4K_2\partial_\varphi K_3
% \nonumber
\\[2pt]
%&\textstyle
\hspace{1.6cm}
{}+6K_3\partial_\varphi K_2+4K_4\partial_\varphi K_1+K_5\partial_\varphi K_0\big]\,,
%\nonumber
%\\
%\qquad\qquad\dots\;,
\end{array}
\label{eq029}
\end{equation}
one can see that the elimination of a fast variable cannot be accomplished without analysis of at least the first three equations, since the noise intensity $\sigma^2$ appears only in the third equation. Below we will see that these equations are not only necessary but also sufficient for taking the limit $\mu\to0$. In the moment representation, the adiabatic elimination of a fast variable also required the first three equations: $w_0$, $w_1$, $w_2$ with Eqs.~(\ref{eq003})--(\ref{eq005}). However, in the moment representation, the $\mu^1$-correction for small $\mu$ requires $w_3$ and $w_4$, while in the cumulant representation, as we will see in the next section, this correction requires the same first three equations of the infinite chain~(\ref{eq025})--(\ref{eq026}) as the adiabatic elimination of a fast variable.

\subsubsection{Corrected Smoluchowski equation}
Let us compare the solutions of equation chain~(\ref{eq029}) with accuracy up to the $\mu^1$-contributions to equation system~(\ref{eq019})--(\ref{eq023}), with account for~(\ref{eq:wW}). First of all, the scaling of divergence of $K_n$ differs from the one of $w_n$\,: $K_0\sim K_1\sim\mu^0$, $K_2=\mu^{-1}\mathrm{const}+\mathcal{O}(1)$\,, $K_{n\ge3}\sim\mu^0$. This scaling suggests one to rewrite Eqs.~(\ref{eq029}) in a more informative form:
\begin{equation}
%\textstyle
\begin{array}{l}
\partial_tK_0%\textstyle
=-\partial_\varphi K_1-K_1\partial_\varphi K_0\,,
\qquad
%\nonumber
\\[4pt]
%\textstyle
K_1=F-(\mu K_2)\partial_\varphi K_0
 -\mu\big[\partial_tK_1
%% \partial_tF -(\mu K_2)\partial_t\partial_\varphi K_0
%\nonumber
%\\[2pt]
%&\textstyle\quad
%\hspace{2.5cm} {}
+\partial_\varphi K_2+K_1\partial_\varphi K_1\big]\,,
%\nonumber
\\[2pt]
%\textstyle
K_2%\textstyle
=\frac{\sigma^2}{\mu}
 -(\mu K_2)\partial_\varphi K_1
 -\frac{\mu}{2}\Big[\partial_t\left(K_2-\frac{\sigma^2}{\mu}\right) +\partial_\varphi K_3
%% -(\mu K_2)\partial_t\partial_\varphi K_1
% \nonumber
\\[2pt]
%&\textstyle
\hspace{2.5cm}
{} +K_1\partial_\varphi K_2 +K_3\partial_\varphi K_0\Big]\,,
%\nonumber
\\[4pt]
%\textstyle
K_3%\textstyle
=-(\mu K_2)\partial_\varphi K_2
 -\frac{\mu}{3}\big[
 \partial_tK_3
% -(\mu K_2)\partial_t\partial_\varphi K_2
 +\partial_\varphi K_4 +K_1\partial_\varphi K_3
% \nonumber
\\[2pt]
%&\textstyle
\hspace{2.5cm}
{} +3K_3\partial_\varphi K_1+K_4\partial_\varphi K_0\big]\,,
%\nonumber
\\[4pt]
%\textstyle
K_4%\textstyle
=-(\mu K_2)\partial_\varphi K_3
 -\frac{\mu}{4}\big[
 \partial_tK_4
% -(\mu K_2)\partial_t\partial_\varphi K_3
 +\partial_\varphi K_5 +K_1\partial_\varphi K_4
%\nonumber
\\[2pt]
%&\textstyle\quad
\hspace{1.5cm}
{} +6K_3\partial_\varphi K_2
%\nonumber\\
%&\textstyle\quad
 +4K_4\partial_\varphi K_1+K_5\partial_\varphi K_0\big]\,.
%\nonumber
%\\
%\qquad\dots\;.
\end{array}
\label{eq030}
\end{equation}

%%%%%%%%%%%%%%%%%%%%%%%%%%%%%%%%%%%%%%%%%%%%%%%%%%%%%%%%%%%%
%%%%%%%%%%%%%%%%%%%%%%%%%%%%%%%%%%%%%%%%%%%%%%%%%%%%%%%%%%%%
\begin{figure}[!thb]
\centerline{
\sf
(a)\hspace{-14pt}
\includegraphics[width=0.225\textwidth]%
 {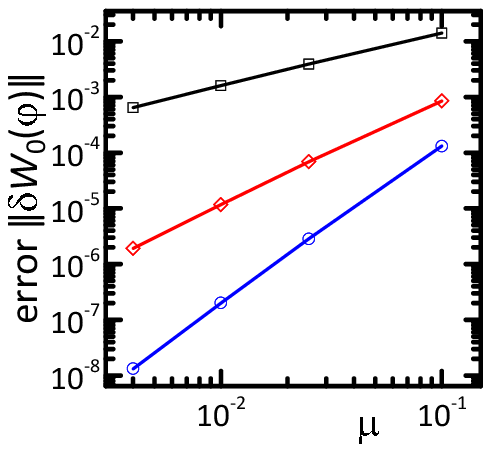}
\quad
(b)\hspace{-14pt}
\includegraphics[width=0.225\textwidth]%
 {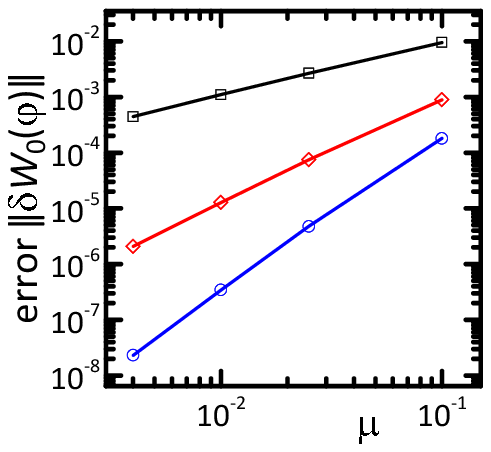}
}\vspace{9pt}

\centerline{
\sf
(c)\hspace{-14pt}
\includegraphics[width=0.225\textwidth]%
 {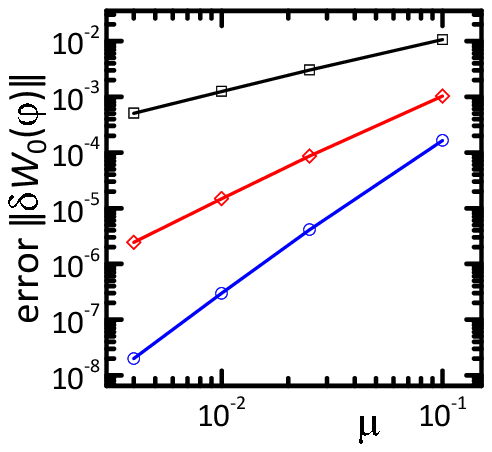}
\quad
(d)\hspace{-14pt}
\includegraphics[width=0.225\textwidth]%
 {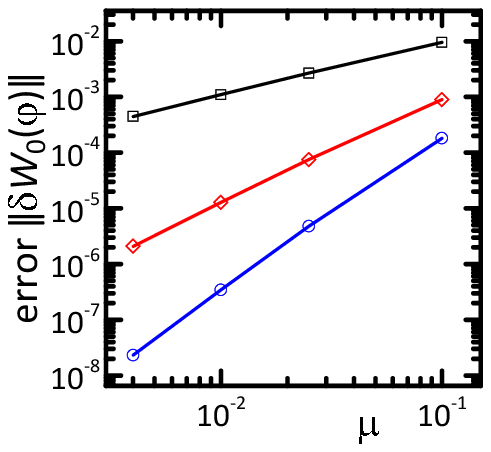}
}\vspace{7pt}

\caption{Error of calculation of the probability density $W_0(\varphi)$ is plotted vs $\mu$ for different approaches and orders of approximation.
%Here,
% $||\delta W_0||\equiv\int_0^{2\pi}|\delta W_0(\varphi)|\,\mathrm{d}\varphi$\,.
(a):~moments, (b):~cumulants, (c):~Hermite basis, (d):~formal cumulants for the Hermite basis. The order of approximation: $\mu^0$ (black squares), $\mu^1$ (red diamonds), $\mu^2$ (blue circles).
}
  \label{fig2}
\end{figure}
%%%%%%%%%%%%%%%%%%%%%%%%%%%%%%%%%%%%%%%%%%%%%%%%%%%%%%%%%%%%
%%%%%%%%%%%%%%%%%%%%%%%%%%%%%%%%%%%%%%%%%%%%%%%%%%%%%%%%%%%%

With the latter equation system one can see advantages of the cumulant representation: while $w_n\sim\mu^{-\mathrm{floor}[n/2]}$ [function $\mathrm{floor}(x)$ returns the largest integer $\le x$], for cumulants one finds $K_2\sim\mu^{-1}$, $K_{n\ne2}\sim\mu^0$. Furthermore, the $\mu^1$-correction requires $w_3$ and $w_4$ in the moment representation, while in the cumulant representation it is still enough to calculate $K_2$. Notice, the adiabatic elimination of velocity also requires $K_2$, i.e., the $\mu^0$- and $\mu^1$-approximations require the same number of cumulants: $K_0$, $K_1$, and $K_2$. Esq.~(\ref{eq030}) for the $\mu^1$-approximation takes a simplified form:
\begin{align}
%\textstyle
\partial_tK_0&%\textstyle
=-\partial_\varphi K_1-K_1\partial_\varphi K_0\,,
\nonumber\\[3pt]
%\textstyle
K_1&%\textstyle
=F-(\mu K_2)\partial_\varphi K_0
\nonumber\\
&%\textstyle
\quad
 -\mu\big(\partial_tK_1
 +\partial_\varphi K_2+K_1\partial_\varphi K_1\big)+\mathcal{O}(\mu^2)\,,
\nonumber\\
%\textstyle
K_2&%\textstyle
=\frac{\sigma^2}{\mu}
 -(\mu K_2)\partial_\varphi K_1
 +\mathcal{O}(\mu)\,,
% -\frac{\mu}{2}\big[-(\mu K_2)\partial_t\partial_\varphi K_1
% \nonumber\\
%&\textstyle
% +\partial_\varphi K_3+K_1\partial_\varphi K_2+K_3\partial_\varphi K_0\big] +\mathcal{O}(\mu^2)\,,
\nonumber\\
%\textstyle
K_3&%\textstyle
=-(\mu K_2)\partial_\varphi K_2+\mathcal{O}(\mu)
=\sigma^4\partial_\varphi^2K_1+\mathcal{O}(\mu)\,,
\nonumber\\[3pt]
%\textstyle
K_n&%\textstyle
=-(\mu K_2)\partial_\varphi K_{n-1}+\mathcal{O}(\mu)\qquad\mbox{ for } n\ge4\,.
\nonumber
\end{align}
Whence, step-by-step expressing $K_1$ and $K_2$ via $K_0$ with account for smallness of $\mu$, and then recursively expressing $K_n$ via $K_{n-1}$ for $n>2$, one can obtain
\begin{align}
%\textstyle
&\partial_tK_0%\textstyle
=-(\partial_\varphi+K_0^\prime)\big[F
 -\sigma^2K_0^\prime
\nonumber\\
&%\textstyle
\qquad\qquad
 +\mu(\partial_tF+F^\prime F +\sigma^2F^\prime K_0^\prime)\big]
 +\mathcal{O}(\mu^2)\,,
\label{eq031}\\
%\textstyle
&K_1%\textstyle
=F-  \sigma^2K_0^\prime
 -\mu\big(\partial_tF+F^\prime F
%\nonumber\\
%&\textstyle\qquad
 +\sigma^2F^\prime K_0^\prime\big)+\mathcal{O}(\mu^2)\,,
\nonumber\\
%\textstyle
&K_2%\textstyle
=\frac{\sigma^2}{\mu}
 -\sigma^2\partial_\varphi(F-\sigma^2K_0^\prime)
+\mathcal{O}(\mu)\,,
% -\frac{\mu\sigma^2}{2}\big[ -3\partial_t\partial_\varphi K_1
%\nonumber\\
%&\textstyle
% +3\sigma^2\partial_\varphi^3K_1
% -2(\partial_\varphi K_1)^2
% \nonumber\\
%&\textstyle
% -4K_1\partial_\varphi^2K_1+F\partial_\varphi^2K_1\big] +\mathcal{O}(\mu^2)\,,
\nonumber\\
%\textstyle
&K_n%\textstyle
=(-\sigma^2\partial_\varphi)^{n-1}(F
 -\sigma^2K_0^\prime)+\mathcal{O}(\mu)\quad\mbox{ for }n\ge3\,.
\nonumber
\end{align}
Here one can see that Eq.~(\ref{eq031}) is equivalent to corrected Smoluchowski equation~(\ref{eq024}) with
$K_0=\ln W_0$ [Eq.~(\ref{eqCM3})];
this equivalence is evident if one notice two identical equalities $\partial K_0=W_0^{-1}\partial W_0$\,,
$(\partial_\varphi+K_0^\prime)(\dots)=W_0^{-1}\partial_\varphi[(\dots)W_0]$\,.
The self-consistent evolution equation for $K_0$ turns out to be more lengthy than Eq.~(\ref{eq024}) for $w_0$ (recall, $w_0=W_0$).

It is instructive to extract the minimal approximate form of the first three equation of system~(\ref{eq030}) which is sufficient for the adiabatic elimination of velocity. According to scaling $K_0\sim K_1\sim\mu^0$, $K_2\sim\sigma^2/\mu$, we keep only the leading contributions:
\begin{align}
%\textstyle
\partial_tK_0&%\textstyle
=-\partial_\varphi K_1-K_1\partial_\varphi K_0\,,
\nonumber\\
%\textstyle
K_1&%\textstyle
=F-(\mu K_2)\partial_\varphi K_0
% -\mu\big(\partial_tK_1
%%\nonumber\\
%%&\textstyle\quad
% +\partial_\varphi K_2+K_1\partial_\varphi K_1\big)
 +\mathcal{O}(\mu^1)\,,
\nonumber\\
%\textstyle
K_2&%\textstyle
=\frac{\sigma^2}{\mu}
% -(\mu K_2)\partial_\varphi K_1
 +\mathcal{O}(\mu^0)\,.
% -\frac{\mu}{2}\big[-(\mu K_2)\partial_t\partial_\varphi K_1
% \nonumber\\
%&\textstyle
% +\partial_\varphi K_3+K_1\partial_\varphi K_2+K_3\partial_\varphi K_0\big] +\mathcal{O}(\mu^2)\,,
\nonumber
\end{align}
With this approximation accuracy the leading order of $K_3=(\mu K_2)\partial_\varphi K_2+\mathcal{O}(\mu^0)$ turns to $0$ and the higher cumulants $K_{n\ge3}\sim\mu^0$, but cannot be calculated. These three equations yield
\[
\partial_tK_0
=-(\partial_\varphi+K_0^\prime)\big[F
 -\sigma^2K_0^\prime\big]
 +\mathcal{O}(\mu^1)\,,
\]
which is identical to Eq.~(\ref{eq018}) [see explanations for the equivalence between Eqs.~(\ref{eq031}) and (\ref{eq024})].

Summarizing, cumulant equations~(\ref{eq025})--(\ref{eq026}) for finite small $\mu$ are more lengthy than the equations for moments $w_n$. However, the convergence properties of $K_n$ for $\mu\to0$ are better than that of $w_n$. The adiabatic elimination of velocity in terms of $K_n$ and $w_n$ requires the first three equations. Further, the $\mu^{1}$-correction to the Smoluchowski equation requires already 5 elements $w_n$ (see \cite{Wilemski-1976} for the multiple-dimension case), whereas in the cumulant representation, the same first three elements $K_0$, $K_1$, $K_2$ are found to be sufficient. Generally, the $\mu^m$-correction requires $K_{m+1}$ with accuracy up to the leading order, i.e., one has to consider the first $m+2$ cumulants. Meanwhile, in terms of $w_n$ (or $W_n$), one has to consider the first $2m+3$ moments.
In Figs.~\ref{fig1}(b) and \ref{fig2}(b) the formulated theoretical conclusions are illustrated and underpinned by the results of numerical simulations for the Kuramoto ensemble with small inertia and noise.

\subsection{Basis of Hermite functions}
\label{ssec33}
Conventional approach to the problem of elimination of a fast velocity from FPE is the usage of the basis of Hermite functions for  $v$~\cite{Gardiner-1983-1997,Komarov-Gupta-Pikovsky-2014}.
The procedure of the adiabatic elimination of velocity form FPE~(\ref{eq002}) for $\rho(v,\varphi)$,
\[
\partial_t\rho=-v\partial_\varphi\rho+\partial_v\left\{\frac{1}{\mu}\big[v-F(\varphi,t)\big]\rho\right\}
+\frac{\sigma^2}{\mu^2}\partial_v^2\rho\,,
\]
is linked to the operator
\begin{equation}
\hat{L}_1=\partial_u(u+\partial_u)\,.
\label{eqL1}
\end{equation}
One can see that $\hat{L}_1h_n(u)=-nh_n(u)$,
\[
h_n(u)=H_n(u)\frac{1}{\sqrt{2\pi}}e^{-u^2/2}\,,
\]
where $H_n(u)$ is the $n$th Hermite polynomial given by equation
\begin{equation}
H_n^{\prime\prime}-uH_n^\prime=-nH_n\,.
\label{eqH01}
\end{equation}

With the normalization condition
\[
\int\limits_{-\infty}^{+\infty}h_n(u)\,h_m(u)\,e^{u^2/2}\mathrm{d}u=\frac{n!\,\delta_{nm}}{\sqrt{2\pi}}\,,
\]
which gives $H_0=1$ and  $\int_{-\infty}^{+\infty}h_0(u)\,\mathrm{d}u=1$,
one has the recurrent formulas:
\begin{equation}
\textstyle
H_n^\prime=nH_{n-1}\,,
\label{eqH02}
\end{equation}
\begin{equation}
\textstyle
uH_n=nH_{n-1}+H_{n+1}\,.
\label{eqH03}
\end{equation}
%$H_n^\prime=nH_{n-1}$ и $uH_n=nH_{n-1}+H_{n+1}$.
With %Eqs.~(\ref{eqH02}) and (\ref{eqH03}),
these recurrent formulas,
FPE~(\ref{eq002}) (see also Eq.~(4) in~\cite{Komarov-Gupta-Pikovsky-2014}) for
\begin{equation}
\rho(v,\varphi,t)=\sum_{n=0}^\infty \frac{\sigma}{\sqrt{\mu}}\,h_n\!\!\left(\frac{\sqrt{\mu}}{\sigma}v\right)\mathcal{W}_n(\varphi,t)
\label{eq:rhoWH}
\end{equation}
yields
\begin{align}
%\textstyle
\sum\limits_nh_n\dot{\mathcal{W}}_n(\varphi,t) &=\sum\limits_n\Big[-\frac{\sigma}{\sqrt{\mu}}\left(nh_{n-1}
%\right.
%\nonumber\\
%&\textstyle\quad
%\left.{}
+h_{n+1}\right)\partial_\varphi \mathcal{W}_n(\varphi,t)
\nonumber\\
& -\frac{n}{\mu}h_n\mathcal{W}_n(\varphi,t)
%\nonumber\\
%&\textstyle\qquad
% {}
 +\frac{F}{\sigma\sqrt{\mu}}h_{n+1}
\mathcal{W}_n(\varphi,t)\Big]\,.
\nonumber
\end{align}
After projections onto modes $h_n(\sqrt{\mu}v/\sigma)$, one finds:
\begin{align}
\dot{\mathcal{W}}_0&=-\frac{\sigma}{\sqrt{\mu}}\partial_\varphi \mathcal{W}_1\,,
\label{eqH04}\\
\textstyle
\dot{\mathcal{W}}_n&=\frac{\sigma}{\sqrt{\mu}}\big[(\sigma^{-2}F-\partial_\varphi)\mathcal{W}_{n-1}
\nonumber\\
&\quad
 {}
 -(n+1)\partial_\varphi \mathcal{W}_{n+1}\big]
  -\frac{n}{\mu}\mathcal{W}_n\,
\quad
\;\mbox{ for } n\ge1\,.
\label{eqH05}
\end{align}
The zeroth mode of expansion in Hermite functions (\ref{eq:rhoWH}) gives the probability density of $\varphi$\,:
\[
W_0(\varphi,t)\equiv\int\limits_{-\infty}^{+\infty}\rho(v,\varphi,t)\,\mathrm{d}v=\mathcal{W}_0(\varphi,t)\,.
\]

\subsubsection{Elimination of a fast variable}
For small $\mu$ the infinite chain of equations~(\ref{eqH04})--(\ref{eqH05}) can be recast as
\begin{align}
\dot{\mathcal{W}}_0&=-\frac{\sigma}{\sqrt{\mu}}\partial_\varphi \mathcal{W}_1\,,
\label{eqH04-mu}\\
\textstyle
\mathcal{W}_n&=\frac{\sqrt{\mu}\,\sigma}{n}\big[(\sigma^{-2}F-\partial_\varphi)\mathcal{W}_{n-1}
\nonumber\\
&\quad
 {}
 -(n+1)\partial_\varphi \mathcal{W}_{n+1}\big]
  -\frac{\mu}{n}\partial_t\mathcal{W}_n
\quad
\;\mbox{ for } n\ge1\,.
\label{eqH05-mu}
\end{align}
From Eqs.~(\ref{eqH04-mu})--(\ref{eqH05-mu}) one can see that $\mathcal{W}_n\sim\mu^{n/2}$.

The obtained infinite chain of equations can be truncated, with  accounting in $\mathcal{W}_N$ (\ref{eqH05-mu}) only for the leading order contributions, $\mathcal{W}_{N}\approx(\sqrt{\mu}\,\sigma/N)(\sigma^{-2}F-\partial_\varphi)\mathcal{W}_{N-1}$.
Such approximation brings about
$\mathrm{error}(\mathcal{W}_N)\sim\mu^{N/2+1}$, $\mathrm{error}(\mathcal{W}_{N-1})\sim\mu^{N/2+1+1/2}$, \dots, $\mathrm{error}(\mathcal{W}_1)\sim\mu^{N/2+1+(N-1)/2}$, and $\mathrm{error}(\partial_t\mathcal{W}_0)\sim\mu^{N}$.
Thus, the formal truncation of chain~(\ref{eqH04-mu})--(\ref{eqH05-mu}) after $\mathcal{W}_N$ results in error $\sim\mu^{N}$ in the description of the evolution of the probability density $W_0(\varphi,t)=\mathcal{W}_0(\varphi,t)$. In particular, for $N=1$ we have the adiabatic elimination of velocity and the Smoluchowski equation~(\ref{eq018}) for the probability density $W_0(\varphi,t)$; for $N=2$, corrected Smoluchowski equation~(\ref{eq024}).
In Figs.~\ref{fig1}(c) and \ref{fig2}(c) the formulated theoretical conclusion are illustrated and underpinned with the results of numerical simulations for the Kuramoto ensemble with small inertia and noise.

\subsection{Analog of cumulant representation for the basis of Hermit functions}
\label{ssec34}
Let us construct an analog of cumulant representation for $v$ on the basis of the Hermit function representation. For the generating function
\begin{equation}
f_\mathcal{W}(s,\varphi,t)\equiv\sum_{n=0}^\infty \mathcal{W}_n(\varphi,t)s^n
\label{eqCH0}
\end{equation}
(for the sake of convenience, we use the series in $s^n$ instead of $s^n/n!$) one can obtain an evolution equation corresponding to Eqs.~(\ref{eqH04-mu})--(\ref{eqH05-mu}):
\[
%\textstyle
\partial_t{f}_\mathcal{W}=\frac{\sigma}{\sqrt{\mu}}\Big[s(\sigma^{-2}F-\partial_\varphi)f_\mathcal{W}
 -\partial_s\partial_\varphi f_\mathcal{W}\Big]-\frac{1}{\mu}s\partial_sf_\mathcal{W}\,.
\]

For the logarithm of generating function $\Phi=\ln f_\mathcal{W}$, $\partial\Phi=\partial f_\mathcal{W}/f_\mathcal{W}$, we obtain
\begin{align}
\partial_t{\Phi}&=\frac{\sigma}{\sqrt{\mu}}\Big[s(\sigma^{-2}F-\partial_\varphi\Phi)
%\nonumber\\
%&\textstyle\qquad\qquad
% {}
 -\partial_s\partial_\varphi\Phi-(\partial_s\Phi)(\partial_\varphi\Phi)\Big]
\nonumber\\
&\qquad\qquad
 -\frac{1}{\mu}s\partial_s\Phi\,.
%\nonumber
\label{eqCH1}
\end{align}
We introduce the coefficients of series
\begin{equation}
\Phi(s,\varphi,t)\equiv\sum_{n=0}^\infty\varkappa_n(\varphi,t)\,s^n\,;
\label{eqCH2}
\end{equation}
with such definition the recursive formulas for calculation of higher coefficients $\varkappa_n$ and $\mathcal{W}_n$ from each other [at variance with both Eqs.~(\ref{eqCC05}) and (\ref{eqCM3}); see Appendix~\ref{sec:app3}] take the form
\begin{equation}
\begin{array}{l}
\displaystyle
\varkappa_0=\ln{\mathcal{W}_0}\,,
\\
\displaystyle
\varkappa_n=\frac{\mathcal{W}_n}{\mathcal{W}_0} -\sum_{l=1}^{n-1}\frac{l}{n}\varkappa_l\frac{\mathcal{W}_{n-l}}{\mathcal{W}_0}\,
\quad\mbox{ for }n\ge1\,.
\end{array}
\label{eqCH3}
\end{equation}
Substitution of expansion~(\ref{eqCH2}) into Eq.~(\ref{eqCH1}) yields
\begin{align}
%\textstyle
\dot\varkappa_0
&%\textstyle
=-\frac{\sigma}{\sqrt{\mu}}(\partial_\varphi\varkappa_1 +\varkappa_1\partial_\varphi\varkappa_0)\,,
\label{eqH06}
\\
%\textstyle
\dot\varkappa_n
&%\textstyle
=\frac{\sigma}{\sqrt{\mu}}\bigg[\frac{F}{\sigma^{2}}\delta_{1n}-\partial_\varphi\varkappa_{n-1}
 -(n+1)\partial_\varphi\varkappa_{n+1}
\nonumber\\
&\quad
 {}
 -\sum\limits_{n_1+n_2 \atop =n+1}n_1\varkappa_{n_1}\partial_\varphi\varkappa_{n_2}\bigg]
 -\frac{n}{\mu}\varkappa_n
 \quad
 \mbox{ for }n\ge1\,.
\label{eqH07}
\end{align}
For small $\mu$, convenient is to rewrite the latter equation system as
\begin{align}
%\textstyle
\dot\varkappa_0
&%\textstyle
=-\frac{\sigma}{\sqrt{\mu}}(\partial_\varphi\varkappa_1 +\varkappa_1\partial_\varphi\varkappa_0)\,,
\label{eqH08}
\\
%\textstyle
\varkappa_n
&%\textstyle
=\frac{\sqrt{\mu}\,\sigma}{n}\bigg[\frac{F\delta_{1n}}{\sigma^{2}}-\partial_\varphi\varkappa_{n-1}
 -(n+1)\partial_\varphi\varkappa_{n+1}
\nonumber\\
&\quad
 {}
 -\sum\limits_{n_1+n_2 \atop =n+1}n_1\varkappa_{n_1}\partial_\varphi\varkappa_{n_2}\bigg]
 -\frac{\mu}{n}\partial_t\varkappa_n
 \quad
 \mbox{ for }n\ge1\,.
\label{eqH09}
\end{align}
Whence for the $\mu^1$-approximation one finds
\begin{align}
%\textstyle
\dot\varkappa_0
&%\textstyle
=-(\varkappa_0^\prime+\partial_\varphi)\big[F-\mu(\partial_t+F^\prime)F
\nonumber\\
&\qquad\qquad\qquad\quad
-\sigma^2(1-\mu F^\prime)\varkappa_0^\prime\big]+\mathcal{O}(\mu^2)\,,
\label{eqH10}
\\
%\textstyle
\varkappa_1
&%\textstyle
=\sqrt{\mu}\,\sigma\big\{\sigma^{-2}F-\partial_\varphi\varkappa_0
 -\mu\big[\sigma^{-2}(\partial_t+F^\prime)F
\nonumber\\
&\qquad\qquad\qquad\qquad
 {}
 -F^\prime\varkappa_0^\prime\big]\big\}+\mathcal{O}(\mu^{5/2})\,,
\label{eqH11}
\\
%\textstyle
\varkappa_2
&%\textstyle
=-\frac{\sqrt{\mu}\,\sigma}{2}\partial_\varphi\varkappa_1+\mathcal{O}(\mu^2)\,.
\label{eqH12}
\end{align}
Eq.~(\ref{eqH10}) is equivalent to Eq.~(\ref{eq024}) [see explanation after Eq.~(\ref{eq031})].

For system~(\ref{eqH08})--(\ref{eqH09}), $\varkappa_n\sim\mu^{n/2}$; the $\mu^N$-approxi\-ma\-tion requires truncation after $\varkappa_{N+1}$.
In this case there is no obvious decisive benefits of one of two representations: in terms of $\mathcal{W}_n$ or $\varkappa_n$. In terms of  $\varkappa_n$ the equations are somewhat more lengthy.
In this section the definition of generating function $f_\mathcal{W}(s,\varphi,t)$ via series of $\mathcal{W}_n(\varphi,t)\,s^n/n!$ is significantly inconvenient, since such definition results in the emergence of the term $\partial_s^{-1}f_\mathcal{W}$ in the evolution equation for $f_\mathcal{W}$. However, the term $\partial_s^{-1}f_\mathcal{W}$ cannot be represented by a simple and regular sum in terms of $\varkappa_n$.
In Figs.~\ref{fig1}(d) and \ref{fig2}(d) the formulated theoretical conclusion are illustrated and underpinned with the results of numerical simulations for the Kuramoto ensemble with small inertia and noise.

\section{Moment and cumulant representation for active Brownian particles}
\label{sec:ABP}
\subsection{The case of additive noise}\label{ssec:ABP1}
Consider the following Langevin equation:
\begin{equation}
\mu\ddot\varphi+\alpha\dot\varphi +\beta\dot\varphi^3=F(\varphi,t)+\sigma\xi(t)\,, %\quad\mu\ll1\,,
\label{eq-abp-01}
\end{equation}
where $\beta>0$. This equation with $\alpha<0$ is used for theoretical studies of dynamics of certain types of overactive Brownian particles~\cite{Erdmann-etal-2000,Erdmann-etal-2002,Erdmann-Ebeling-2005,Pikovsky-2023}.

For the Fokker--Planck equation
\begin{equation}
\partial_t\rho=-v\partial_\varphi\rho +\partial_v\left[\frac{\alpha v+\beta v^3 -F(\varphi,t)}{\mu}\rho\right]
+\frac{\sigma^2}{\mu^2}\partial_v^2\rho
\label{eq-abp-02}
\end{equation}
the moment representation gives an infinite equation chain
\begin{align}
%\textstyle
\alpha nw_n+\beta nw_{n+2}+\mu\partial_t w_n=nF w_{n-1}
%\qquad
\nonumber\\
%\textstyle
{}
-\mu\partial_\varphi w_{n+1} +n(n-1)\frac{\sigma^2}{\mu}w_{n-2}\,,
\label{eq-abp-03}
\end{align}
for which the evolution of the generating function $f_w(s,\varphi,t)=\sum_{n=0}^{+\infty}w_n(\varphi,t)\frac{s^n}{n!}$ (\ref{eqCM0}) obeys equation
\begin{align}
%\textstyle
(\alpha s\partial_s+\beta s\partial_s^3 +\mu\partial_t)f_w=\left(sF-\mu\partial_s\partial_\varphi
 +s^2\frac{\sigma^2}{\mu}\right)f_w\,.
\nonumber
\end{align}
For the logarithm of generating function $\phi=\ln f_w$, $\partial f_w=f_w\partial\phi$, one finds
\begin{align}
%\textstyle
(\alpha s\partial_s +\mu\partial_t)\phi
 +\beta s\left[\partial_s^3\phi+3\partial_s\phi\partial_s^2\phi +(\partial_s\phi)^3\right]
\qquad
\nonumber\\
%\textstyle
 =sF+s^2\frac{\sigma^2}{\mu}
 -\mu\left[\partial_s\partial_\varphi\phi+(\partial_s\phi)(\partial_\varphi\phi)\right]\,.
\nonumber
\end{align}
For $K_n$ defined by $\phi=\sum_{n=0}^{+\infty}K_n\frac{s^n}{n!}$ [Eq.~(\ref{eqCM2})],
\begin{widetext}
\begin{align}
%\textstyle
%\textstyle
\mu\partial_tK_0&=-\mu[\partial_\varphi K_1+K_1\partial_\varphi K_0]\,,
\label{eq-abp-04}\\[5pt]
%\textstyle
\left(\alpha +\frac{\mu\partial_t}{n}\right)K_n& +\beta \bigg[K_{n+2}
%\nonumber\\
%&
 +3\sum\limits_{j=1}^{n}\frac{(n-1)!}{(j-1)!(n-j)!}K_{j}K_{n+2-j}
%\nonumber\\
%&%\textstyle
%\quad
%\qquad\qquad\qquad\qquad\qquad
+\sum\limits_{j_1+j_2+j_3 \atop =n+2}\frac{(n-1)!}{(j_1-1)!(j_2-1)!(j_3-1)!}K_{j_1}K_{j_2}K_{j_3}\bigg]
\nonumber\\
&\qquad
%\textstyle
=F\delta_{1n}+\frac{\sigma^2}{\mu}\delta_{2n}
 -\frac{\mu}{n}\bigg[\partial_\varphi K_{n+1}
%\nonumber\\
%&\textstyle\quad
 +\sum\limits_{j=0}^{n}\frac{(n-1)!}{(j-1)!(n-j)!}K_{j+1}\partial_\varphi K_{n-j}\bigg]
\qquad\mbox{ for }n\ge1\,.
\label{eq-abp-05}
\end{align}
%{\color{red}
%Alternative representation in terms of $\varkappa_n=K_n/n!$\,:
%\begin{align}
%%\textstyle
%\mu\partial_t\varkappa_0&%\textstyle
%=-\mu[\partial_\varphi\varkappa_1 +\varkappa_1\partial_\varphi\varkappa_0]\,,
%\qquad
%\label{eq-abp-04kappa}\\
%%\textstyle
%(\alpha n+\mu\partial_t)\varkappa_n&%\textstyle
%+\beta\bigg[n(n+1)(n+2)\varkappa_{n+2}
%\nonumber\\
%&%\textstyle
%\quad
%\qquad
%+3\sum\limits_{j=1}^{n}j(n+1-j)(n+2-j)\varkappa_{j}\varkappa_{n+2-j}
%+\!\!\!\sum\limits_{j_1+j_2+j_3 \atop =n+2}\!\!j_1j_2j_3\varkappa_{j_1}\varkappa_{j_2}\varkappa_{j_3}\bigg]
%\nonumber\\
%&
%=F\delta_{1n}+\frac{\sigma^2}{\mu}\delta_{2n}
% -\mu\Big[(n+1)\partial_\varphi\varkappa_{n+1}
%% \nonumber\\
%%&\textstyle\quad
%+\sum\limits_{j=1}^{n+1}j\varkappa_{j}\partial_\varphi\varkappa_{n+1-j}\Big]
%\;\mbox{ for }n\ge1\,.
%\label{eq-abp-05kappa}
%\end{align}
%}

The first 5 equations of system~(\ref{eq-abp-04})--(\ref{eq-abp-05}):
\begin{equation}
\begin{array}{l}
%&\textstyle
\qquad
\partial_tK_0=-\partial_\varphi K_1-K_1\partial_\varphi K_0\,,
\qquad
%\nonumber
\\[3pt]
%&\textstyle
(\alpha+\mu\partial_t)K_1
+\beta[K_3+3K_1K_2+K_1^3]
% \nonumber\\[2pt]
%\qquad\qquad
 =F -\mu\big[\partial_\varphi K_2
 +K_1\partial_\varphi K_1+K_2\partial_\varphi K_0\big]\,,
%\nonumber
\\[3pt]
%&\textstyle
(\alpha+\frac{\mu\partial_t}{2})K_2 +\beta\big[K_4+3(K_2^2+K_1K_3+K_1^2K_2)\big]
%% \nonumber
% \\[2pt]
%%&\textstyle
%\qquad\qquad
 =\frac{\sigma^2}{\mu} -\frac{\mu}{2}\big[\partial_\varphi K_3
 +K_1\partial_\varphi K_2+2K_2\partial_\varphi K_1
%\\[2pt]
%\qquad\qquad\qquad\qquad\qquad
 +K_3\partial_\varphi K_0\big]\,,
%\nonumber
\\[3pt]
%&\textstyle
(\alpha+\frac{\mu\partial_t}{3})K_3
 +\beta\big[K_5+3(3K_3K_2+K_1K_4)
%\\[2pt]
%\qquad\qquad\qquad\qquad\qquad
 +3K_1^2K_3+6K_2^2K_1\big]
%\nonumber
\\[2pt]
%&\textstyle
\qquad\qquad \qquad
 =-\frac{\mu}{3}\big[\partial_\varphi K_4 +K_1\partial_\varphi K_3
% \nonumber\\
%&\textstyle\qquad\qquad
+3K_2\partial_\varphi K_2+3K_3\partial_\varphi K_1+K_4\partial_\varphi K_0\big]\,,
%\nonumber
\\[3pt]
%&\textstyle
(\alpha+\frac{\mu\partial_t}{4})K_4+\beta\big[K_6+3(4K_4K_2+3K_3^2+K_1K_5)
%\nonumber\\
%&\textstyle\qquad\qquad
 +6K_2^3+18K_1K_2K_3+3K_1^2K_4\big]
% \nonumber
\\[2pt]
%&\textstyle
\qquad\qquad \qquad
 =-\frac{\mu}{4}\big[\partial_\varphi K_5 +K_1\partial_\varphi K_4 +4K_2\partial_\varphi K_3
% \nonumber\\
%&\textstyle\qquad\qquad
+6K_3\partial_\varphi K_2+4K_4\partial_\varphi K_1+K_5\partial_\varphi K_0\big]\,.
%\nonumber
%\\
%%&
%\qquad\qquad\dots\;.
\end{array}
\label{eq-abp-06}
\end{equation}
\end{widetext}

A thorough consideration of equation system~(\ref{eq-abp-06}) suggests the scaling laws of $K_n$:
\begin{equation}
K_n\sim\left\{
\begin{array}{cr}
\displaystyle
  \mu^{-\frac{n}{4}} & \mbox{ for even }n\,, \\
\displaystyle
  \mu^{\frac{3}{4}-\frac{n}{4}} & \mbox{ for odd }n\,.
\end{array}\right.
\label{eq-abp-scaleK}
\end{equation}
With such scaling laws, the $\beta$- and $\sigma^2$-contributions for even $n$ in equation system~(\ref{eq-abp-06}) are dominating and the equation chain cannot be truncated without affecting the leading order in $\mu$. Similar issue takes place also for the elements with odd $n$, the leading order of which is defined by the force $F$. Thus, analytical calculations, even to the leading order, require accounting for the $\beta$-, $F$-, and $\sigma^2$-terms; and these calculations in terms of $K_n$ (or $w_n$) are extremely laborious.

It will be more productive to analyse the asymptotic behavior of the system within the framework of FPE~(\ref{eq-abp-02}), where we drop all the terms except the dominating ones --- with $\beta$, $F$, and $\sigma^2$. For a time-independent solution, this equation can be once integrated over $v$, whence the probability density flux $J=( %{\color{blue}-\alpha v}
-\beta v^3+F)\mu^{-1}\rho-(\sigma/\mu)^2\partial_v\rho$ must be uniform over $v$, but it also must be zero at infinity. Hence:
%With such scaling properties the $\beta$- and $\sigma^2$-terms for even $n$ in equation system~(\ref{eq-abp-06}) dominate and one cannot truncate the equation chain without affecting the leading order with respect to $\mu$. Moreover, one faces similar issue with odd $n$; which is coupled with the $F$-term in the leading order.
%Thus, the calculations in the leading order require $\beta$-, $F$- and $\sigma^2$-terms and these calculations in terms of $K_n$ (or $w_n$) are extremely challenging. This problem can be more efficiently solved with the Fokker--Planck equation~(\ref{eq-abp-02}) where all terms without $\beta$, $F$ or $\sigma^2$ are dropped. One finds
\begin{equation}
\rho=C(\varphi)e^{\frac{\mu}{\sigma^2}(-\frac{\beta v^4}{4}+F v)}+\cdots\,,
\label{eq-abp-rhoJ0}
\end{equation}
%\[
%\color{blue}
%\rho=C(\varphi)e^{\frac{\mu}{\sigma^2}(-\frac{\alpha v^2}{2}-\frac{\beta v^4}{4}+F v)}+\dots\,,
%\]
where dots stand for higher order corrections.
For
\begin{equation}
|F|\ll\beta^{1/4}\left(\frac{\sigma^2}{\mu}\right)^{3/4}
%\beta^{1/4}(\sigma^2/\mu)^{3/4}
\label{eq-abp-Fcond}
\end{equation}
%{\color{blue}
%и $|\alpha|\ll\sigma(\beta/\mu)^{1/2}$}
expression (\ref{eq-abp-rhoJ0}) can be simplified:
\[
\rho\approx C(\varphi)\left[1+\frac{\mu Fv}{\sigma^2}\right]e^{-\frac{\mu\beta v^4}{4\sigma^2}}.
\]
%\[
%\color{blue}
%\rho\approx C(\varphi)\left[1+\frac{\mu Fv}{\sigma^2}-\frac{\mu\alpha v^2}{2\sigma^2}\right]
%e^{-\frac{\mu\beta v^4}{4\sigma^2}}.
%\]
For this distribution, one can calculate moments $w_n=\int_{-\infty}^{+\infty}\rho v^n\mathrm{d}v$; with laborious but straightforward calculations yield
\begin{align}
w_0(\varphi)&\approx\frac{\pi}{\Gamma(\frac34)}\frac{\sqrt{\sigma}}{(\mu\beta)^{1/4}}C(\varphi)\,,
\label{eq-abp-w0C}\\
  w_{2m}(\varphi)&\approx\frac{\Gamma(\frac{m}{2}+\frac14)}{\Gamma(\frac14)} \left(\frac{2\sigma}{\sqrt{\beta\mu}}\right)^{m}w_0(\varphi)\,,
\label{eq-abp-07}\\
%\color{blue}
%  w_{2m}(\varphi)&
%\color{blue}
%\approx\frac{\Gamma(\frac{m}{2}+\frac14) -\Gamma(\frac{m}{2}+\frac34)\frac{\alpha}{\sigma}\big(\frac{\mu}{\beta}\big)^{1/2}}
%{\Gamma(\frac14)-\Gamma(\frac34)\frac{\alpha}{\sigma}\big(\frac{\mu}{\beta}\big)^{1/2}} \left(\frac{2\sigma}{\sqrt{\beta\mu}}\right)^{m}w_0(\varphi)\,,
%\nonumber\\
  w_{2m+1}(\varphi)&\approx\frac{4F}{\beta}\frac{\Gamma(\frac{m}{2}+\frac34)}{\Gamma(\frac14)} \left(\frac{2\sigma}{\sqrt{\beta\mu}}\right)^{m-1}w_0(\varphi)\,,
\label{eq-abp-08}
\end{align}
where $\Gamma(z)$ is the gamma function.
Corresponding cumulants (\ref{eqCM3}):
\begin{align}
&%\textstyle
K_0(\varphi)=\ln w_0(\varphi)\,,\qquad
K_2\approx\frac{\left[\Gamma(\frac34)\right]^2}{\pi\sqrt{2}}
  \frac{2\sigma}{\sqrt{\beta\mu}}\,,
\nonumber\\
&
%\qquad
K_4\approx-\left(\frac{3}{2\pi^2}\left[\textstyle\Gamma(\frac34)\right]^4-\frac14\right)
  \frac{4\sigma^2}{\beta\mu}\,,\quad
\nonumber\\
&%\textstyle
K_6\approx\frac{3\left[\Gamma(\frac34)\right]^2}{\pi\sqrt{2}} \left(\frac{5}{\pi^2}\left[\textstyle\Gamma(\frac34)\right]^4-1\right)
  \left(\frac{2\sigma}{\sqrt{\beta\mu}}\right)^3,\,\dots\,,
\nonumber\\
&%\textstyle
K_1(\varphi)\approx\frac{4F(\varphi)}{\beta}\frac{\left[\Gamma(\frac34)\right]^2}{\pi\sqrt{2}} \frac{\sqrt{\beta\mu}}{2\sigma}\,,
\nonumber\\
&
%\qquad
K_3(\varphi)\approx-\frac{4F(\varphi)}{\beta} \left(\frac{3}{2\pi^2}\left[\textstyle\Gamma(\frac34)\right]^4 -\frac14\right)\,,
\nonumber\\
&%\textstyle
K_5(\varphi)\approx3\frac{4F(\varphi)}{\beta}\frac{\left[\Gamma(\frac34)\right]^2}{\pi\sqrt{2}} \left(\frac{5}{\pi^2}\left[\textstyle\Gamma(\frac34)\right]^4 -1\right)\frac{2\sigma}{\sqrt{\beta\mu}}\,,
\nonumber\\
&\qquad\qquad\,\dots\,.
\nonumber
\end{align}
Here we used the identity $\Gamma(1/4)=\pi\sqrt{2}/\Gamma(3/4)$.
The calculated cumulants $K_n$ obey the scaling law~(\ref{eq-abp-scaleK}) for small $\mu$, deduced from the complete cumulant equations.

The flux of particles (probability density) $w_1(\varphi,t)$ is typically of primary practical interest; $w_1$ can be calculated from the expansion of distribution~(\ref{eq-abp-rhoJ0}) in a series of $F$. With nonlinear-in-$F$ corrections,
\begin{align}
&w_1=\frac{\int_{-\infty}^{+\infty}v\exp\Big[\frac{\mu}{\sigma^2}(-\frac{\beta v^4}{4}+Fv)\Big]\mathrm{d}v}
{\int_{-\infty}^{+\infty}\exp\Big[\frac{\mu}{\sigma^2}(-\frac{\beta v^4}{4}+Fv)\Big]\mathrm{d}v}w_0%\,.
\nonumber
\\
&%\color{red}
=\frac{\sqrt{2\sigma} w_0}{(\mu\beta)^{1/4}} \left[\frac{\left[\Gamma(\frac34)\right]^2}{\pi\sqrt{2}}F_{sc} +\left\{\frac{1}{24}-\frac{\left[\Gamma(\frac34)\right]^4}{4\pi^2}\right\}F_{sc}^3
\right.
% {\color{red} +}
%+\mathcal{O}(F_{sc}^5)
\nonumber\\
&%\color{red}
%\qquad\qquad
\left.+\left\{\frac{\left[\Gamma(\frac34)\right]^6}{8\sqrt{2}\pi^3} -\frac{\left[\Gamma(\frac34)\right]^2}{40\sqrt{2}\pi}\right\}F_{sc}^5+\mathcal{O}(F_{sc}^7)
\right]_{F_{sc}=\frac{\mu F}{\sigma^2}\frac{\sqrt{2\sigma}}{(\mu\beta)^{1/4}}}
\nonumber
\\
&=\left(\frac{\gamma_1\sqrt{\mu}}{\sigma\sqrt{\beta}}F +\frac{\gamma_3\mu^2 F^3}{\sigma^4\beta} +\frac{\gamma_5\mu^{7/2}F^5}{\sigma^7\beta^{3/2}}
+\cdots\right)w_0\,,
\label{eq-abp-w1}
\end{align}
where
\begin{align}
\gamma_1&=\frac{\sqrt{2}\left[\Gamma(\frac34)\right]^2}{\pi}=0.6759782400672847...
\,,
\label{eq-abp-gamma1}
\\
\gamma_3&=-\frac{\left[\Gamma(\frac34)\right]^4}{\pi^2}+\frac{1}{6}=-0.0618066238555651...
\,,
\label{eq-abp-gamma3}
\\
\gamma_5&=\frac{\left[\Gamma(\frac34)\right]^6}{\sqrt{2}\pi^3}-\frac{\left[\Gamma(\frac34)\right]^2}{5\sqrt{2}\pi}
=0.009623662408071...
\,.
\label{eq-abp-gamma5}
\end{align}
The smallness of dimensionless coefficient $\gamma_3$ and $\gamma_5$ is noticeable.

%%%%%%%%%%%%%%%%%%%%%%%%%%%%%%%%%%%%%%%%%%%%%%%%%%%%%%%%%%%%
%%%%%%%%%%%%%%%%%%%%%%%%%%%%%%%%%%%%%%%%%%%%%%%%%%%%%%%%%%%%
\begin{figure}[!t]
\centerline{
\sf
(a)\hspace{-12pt}
\includegraphics[width=0.4015\textwidth]%
 {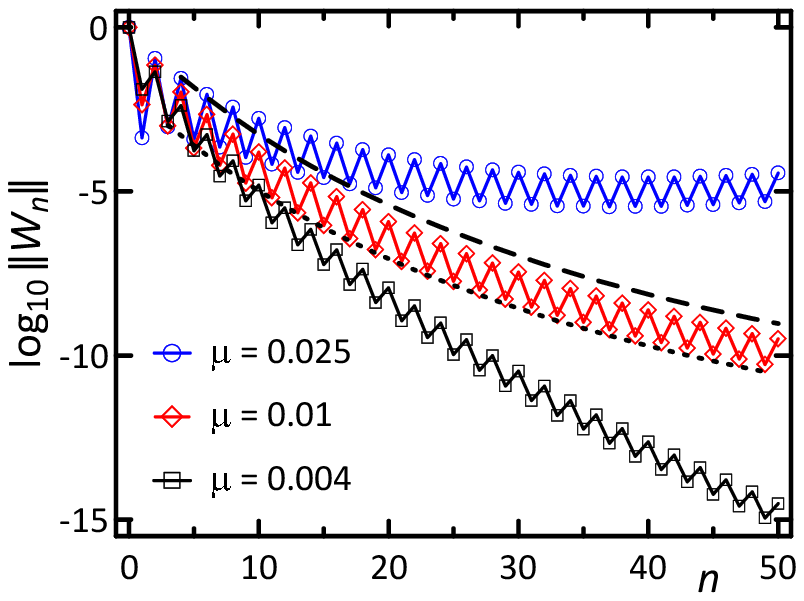}
}

\vspace{10pt}
%\qquad\qquad
\centerline{
\sf
(b)\hspace{-12pt}
\includegraphics[width=0.4015\textwidth]%
 {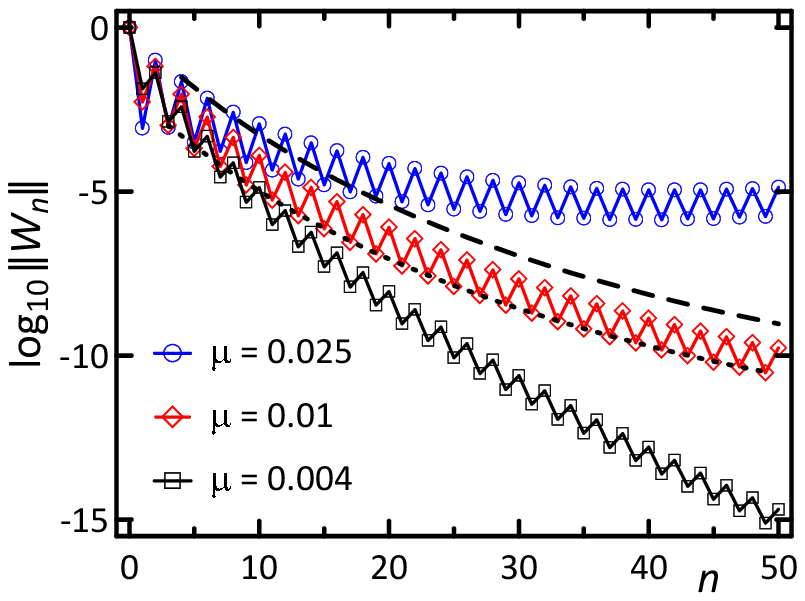}
}
\caption{Hierarchy of smallness of high-order elements for active Brownian particles with $\alpha=-1$, $\beta=1$ (\textit{a}) and passive particles with nonlinear friction $\alpha=+1$, $\beta=1$ (\textit{b}).
For convenience of presentation the same rescaling~(\ref{eq:wW}) is adopted as for the passive particles with linear friction.
Dashed line: asymptotic law~(\ref{eq-abp-07}), dotted line: law~(\ref{eq-abp-08}); $\mu=0.01$ for both curves.
Parameters of force $F(\varphi)$ and discretization in $\varphi$ are the same as in Fig.~\ref{fig1}.
A series of 50 terms is used.
%, here $||g(\varphi)||\equiv\int_0^{2\pi}|g(\varphi)|\mathrm{d}\varphi$.
}
  \label{fig3}
\end{figure}
%%%%%%%%%%%%%%%%%%%%%%%%%%%%%%%%%%%%%%%%%%%%%%%%%%%%%%%%%%%%
%%%%%%%%%%%%%%%%%%%%%%%%%%%%%%%%%%%%%%%%%%%%%%%%%%%%%%%%%%%%

Generally, numerical simulations of system~(\ref{eq-abp-03}) for active Brownian particles require lengthy expansion series and can suffer from numerical instabilities. To deal with these challenges in this work we used modification~\cite{Permyakova-Goldobin-2025} of the exponential time differencing method~\cite{Cox-Matthews-2002}, which allows for high accuracy and performance of numerical simulations of ``stiff'' systems \cite{Matthews-Cox-2000a,Matthews-Cox-2000b}. In Fig.~\ref{fig3}, the results of numerical simulations are presented for the same $F$, as in Figs.~\ref{fig1} and \ref{fig2}, but nonlinear dissipation law~(\ref{eq-abp-01}). The numerical simulation of truncated chain of the moment equations with sufficient number of elements can be seen to give a regular behavior which is in agreement with theoretical asymptotic laws~(\ref{eq-abp-07})--(\ref{eq-abp-08}), in spite of a fast growth of elements $w_n$ for $\mu\to0$.

Presumably, the employment of the cumulant representation should be fruitful mainly for the systems, where the distribution of a fast variable is similar to the Gaussian one. The case of passive Brownian particles is an example of such systems, because the Fluctuation--dissipation theorem~\cite{Callen-Welton-1951,Kubo-1966,Hanggi-Thomas-1982}, which is valid for passive Brownian particles, requires the Gaussian distribution in a statistically stationary state. A reasonable proximity to the Gaussian distribution can be also expected for those active Brownian particles whose leading part of the dissipation term is in agreement with the fluctuation term.

\subsubsection{Basis of eigenfunctions of $\hat{L}_1$}
For the case of active Brownian particles~\cite{Milster-etal-2017,Lighthill-1952,Blake-1971,Ebbens-Howse-2010}, the moment and cumulant representations can be implemented straightforwardly, whereas the basis of the eigenfunctions of operator $\hat{L}_1$ requires significant adaptation. While for passive particles the eigenfunctions of $\hat{L}_1$ (\ref{eqL1}) are the Hermite functions, for system~(\ref{eq-abp-01}) in FPE~(\ref{eq-abp-02}), the term $\partial_v[\mu^{-1}(\alpha v+\beta v^3)\rho+(\sigma/\mu)^2\partial_v\rho]$ corresponds to $\hat{L}_1=-Au+u^3+\partial_u$ for $\alpha<0$, $\beta>0$, where $A=(-\alpha/\sigma)\sqrt{\mu/\beta}$ and $u=(\beta\mu)^{1/4}\sigma^{-1/2}v$\,, and the basis functions are different. One either has to employ the basis depending on parameter $A$ or use the basis with $A=0$ but deal with equations which are nondiagonal even in the leading order. In both cases, new basis functions need to be found. Thus, for the usage of the representation of the basis functions of operator $\hat{L}_1$, individual mathematical preparation is needed for each new variant of the problem setup, which can be problematic.

\subsection{Adiabatic elimination of velocity for active Brownian particle with additive noise}
To explicitly take into account the scaling law~(\ref{eq-abp-scaleK}), also observed in (\ref{eq-abp-07})--(\ref{eq-abp-08}) for $w_n$, we substitute
\begin{equation}
w_n
=\left\{
\begin{array}{cr}
\displaystyle
\mu^{-\frac{n}{4}}U_n\,
& \mbox{ for even }n\,,\\[3pt]
\displaystyle
\mu^{\frac{3}{4}-\frac{n}{4}}U_n\,
& \mbox{ for odd }n\,,
\end{array}
\right.
\label{eq:wU}
\end{equation}
into equation system~(\ref{eq-abp-03}). Hence,
\begin{widetext}
\begin{align}
\partial_tU_0&=-\sqrt{\mu}\partial_\varphi U_1\,,
\label{eq-abp-fe01}
\\
\alpha\sqrt{\mu}U_1+\beta U_3+\mu^\frac32\partial_t U_1&
=FU_0-\sqrt{\mu}\partial_\varphi U_2\,,
\label{eq-abp-fe02}
\\
\alpha U_n+\frac{\beta}{\sqrt{\mu}}U_{n+2}+\frac{\mu}{n}\partial_t U_n&
=\mu FU_{n-1}-\frac{\mu^\frac32}{n}\partial_\varphi U_{n+1}+(n-1)\frac{\sigma^2}{\sqrt{\mu}}U_{n-2}\,%,
\quad\mbox{ for } n=2m\,,
\label{eq-abp-fe03}
\\
\alpha\sqrt{\mu} U_n+\beta U_{n+2}+\frac{\mu^\frac32}{n}\partial_t U_n&
=FU_{n-1}-\frac{\sqrt{\mu}}{n}\partial_\varphi U_{n+1}+(n-1)\sigma^2U_{n-2}\,%,
\qquad\mbox{for } n=2m+1\,,
\label{eq-abp-fe04}
\end{align}
where $m=1,2,3,...$\,. Collecting terms with the identical exponent of $\mu$, taking the smallness of $\mu$ into account and introducing ``slow'' time $\tau=\sqrt{\mu}t$, one can recast equation system~(\ref{eq-abp-fe01})--(\ref{eq-abp-fe04}) as
\begin{align}
\partial_\tau U_0&=-\partial_\varphi U_1\,,
\label{eq-abp-fe05}
\\
\beta U_3-FU_0&
=-\sqrt{\mu}(\alpha U_1+\partial_\varphi U_2)+\mathcal{O}(\mu^2)\,,
\label{eq-abp-fe06}
\\
\beta U_{n+2}-(n-1)\sigma^2U_{n-2}&
=-\sqrt{\mu}\alpha U_n +\mu^\frac32 FU_{n-1}+\mathcal{O}(\mu^2)\,%,
\qquad\mbox{ for } n=2m\,,
\label{eq-abp-fe07}
\\
\beta U_{n+2}-FU_{n-1}-(n-1)\sigma^2U_{n-2}&
=-\sqrt{\mu}\left(\alpha U_n+\frac{\partial_\varphi U_{n+1}}{n}\right)+\mathcal{O}(\mu^2)\,%,
\quad\mbox{ for } n=2m+1\,.
\label{eq-abp-fe08}
\end{align}
\end{widetext}

Considering the limit $\mu\to0$ for system (\ref{eq-abp-fe05})--(\ref{eq-abp-fe08}), we find that, to the leading order, Eqs.~(\ref{eq-abp-fe06})--(\ref{eq-abp-fe08}) are equivalent to the problem
\[
0=J=\frac{-\beta v^3+F}{\mu}\rho-\frac{\sigma^2}{\mu^2}\partial_v\rho\,.
\]
[To see this we multiply the latter equation by $v^n$ for $n=0,1,2,...$, integrate over $v$ and obtain an equation system the leading order of which is identical to that of (\ref{eq-abp-fe06})--(\ref{eq-abp-fe08}) with scaling~(\ref{eq:wU}) taken into account.]
In turn, this equation also corresponds to the leading order of FPE~(\ref{eq-abp-02}). In Sec.~\ref{ssec:ABP1}, solution~(\ref{eq-abp-rhoJ0}) [or (\ref{eq-abp-07})--(\ref{eq-abp-08})] was obtained for the latter problem. The first equation~(\ref{eq-abp-fe05}) of the system is the integral of FPE~(\ref{eq-abp-02}) over $v$, where $U_1$ (or $w_1$) are given by solution~(\ref{eq-abp-08}). In original variables, one finds
\begin{equation}
\partial_t w_0(\varphi,t)=
%-\partial_\varphi w_1(\varphi,t)\approx
-\partial_\varphi\left(\frac{\gamma_1\sqrt{\mu}}{\sigma\sqrt{\beta}} F(\varphi,t)\,w_0(\varphi,t)\right)\,.
\label{eq-abp-SmoluchEq0}
\end{equation}
This continuity equation is equivalent to the deterministic dynamics with velocity
\begin{equation}
\dot\varphi=\frac{\gamma_1\sqrt{\mu}}{\sigma\sqrt{\beta}}F(\varphi,t)\,.
\label{eq-abp-Vmu0}
\end{equation}

Even though the final equation~(\ref{eq-abp-Vmu0}) effectively describes deterministic dynamics, this result is essentially linked to fluctuations. In the absence of fluctuations $\sigma\xi(t)$, the dynamics of system~(\ref{eq-abp-01}) with $\mu\to0$ is a ballistic motion with velocities $v_{\pm}(F)$, which are the most right and most left solutions of the cubic equation $\beta v^3+\alpha v-F=0$\,. Switching between the regimes of ballistic motion requires large values of force $F$, for which the cubic equation has single solution: $|F|>F_\mathrm{\ast}=2(-\alpha/3)^{3/2}/\sqrt{\beta}$\,. Moreover, the noise cannot be too weak, since the employed expression for $w_1$ was derived under condition~(\ref{eq-abp-Fcond}). For weak noise ($\sigma^2\sim\mu$) the dependence of $w_1$ on $F$ becomes nonlinear and is approximately given by formula~(\ref{eq-abp-w1}).

At this level of accuracy with respect to $\mu$ the {\it effective} dynamics of active Brownian particle~(\ref{eq-abp-01}) turned out to be deterministic, in contrast to the case of passive particles, where the leading order of accuracy $\mu^0$ (\ref{eq018}) gives the diffusion of distribution $W_0$.
The description of fluctuations and diffusion in the effective dynamics of particles requires one to account for the next order correction with respect to $\mu$. To do so, within the framework of equation system~(\ref{eq-abp-fe05})--(\ref{eq-abp-fe08}), one has to keep the terms $\propto\sqrt{\mu}$.

\subsection{Corrected Smoluchowski equation for active Brownian particles with additive noise}
Inspection of equation system~(\ref{eq-abp-fe05})--(\ref{eq-abp-fe08}) suggests the following expansion with respect to small parameter $\mu$: $U_n=U_n^{(0)}(\varphi,\tau,\tau_2,...)+\sqrt{\mu}U_n^{(1)}(\varphi,\tau,\tau_2,...)+\mathcal{O}(\mu)$\,, where ``slow'' times $\tau_m\equiv\mu^{m/2}t$ and partial derivative $\partial_t=\sqrt{\mu}\partial_\tau+\mu\partial_{\tau_2}+\mu^{3/2}\partial_{\tau_3}+\cdots$ (customarily for the standard multiple scale method~\cite{Nayfeh-1981-1984}). Since $U_0(\varphi,\sqrt{\mu}t)=w_0(\varphi,t)=\int_{-\infty}^{+\infty}\rho(v,\varphi,t)\,\mathrm{d}v$ is the current particle density distribution, for which the evolution in time is to be calculated, natural is to adopt the normalization condition
\begin{equation}
U_0=U_0^{(0)}(\varphi,\tau,\tau_2,...)\,,\qquad U_0^{(m\ge1)}=0\,.
\label{eq-abp-U0m}
\end{equation}
In the $\mu^0$-order, system~(\ref{eq-abp-fe05})--(\ref{eq-abp-fe08}) yields
\begin{align}
\partial_\tau U_0^{(0)}&=-\partial_\varphi U_1^{(0)},
\label{eq-abp-fe09}
\\
\beta U_3^{(0)}-FU_0^{(0)}&
=0\,,
\label{eq-abp-fe10}
\\
\beta U_{n+2}^{(0)}-(n-1)\sigma^2U_{n-2}^{(0)}&
=0\,%,
\;\mbox{ for } n=2m\,,
\label{eq-abp-fe11}
\\
\beta U_{n+2}^{(0)}-FU_{n-1}^{(0)}-(n-1)\sigma^2U_{n-2}^{(0)}&
=0\,%,
\;\mbox{ for } n=2m+1\,.
\label{eq-abp-fe12}
\end{align}
The solution to this problem is given by Eqs.~(\ref{eq-abp-07}), (\ref{eq-abp-08}), (\ref{eq:wU}) and was obtained in the $v$-space~(\ref{eq-abp-rhoJ0}); it leads to the continuity equation~(\ref{eq-abp-SmoluchEq0}).
\\
In the $\mu^{1/2}$-order:
\begin{widetext}
\begin{align}
\partial_{\tau_2} U_0^{(0)}&=-\partial_\varphi U_1^{(1)},
\label{eq-abp-fe13}
\\
\beta U_3^{(1)} %{\color{red} -FU_0^{(1)}}
&
=-\alpha U_1^{(0)}-\partial_\varphi U_2^{(0)},
\label{eq-abp-fe14}
\\
\beta U_{n+2}^{(1)}-(n-1)\sigma^2U_{n-2}^{(1)}&
=-\alpha U_n^{(0)}\,%,
\qquad\qquad\qquad\mbox{ for } n=2m\,,
\label{eq-abp-fe15}
\\
\beta U_{n+2}^{(1)}-FU_{n-1}^{(1)}-(n-1)\sigma^2U_{n-2}^{(1)}&
=-\alpha U_n^{(0)}-\frac{1}{n}\partial_\varphi U_{n+1}^{(0)}\,%,
\quad\mbox{for } n=2m+1\,.
\label{eq-abp-fe16}
\end{align}
\end{widetext}

With given $\{U_n^{(0)}\}$ the problem for $\{U_n^{(1)}\}$ can be solved in a matrix form approximately by truncating $\{U_{n>M}^{(1)}\}=0$ with sufficiently large $M$. As one can see from Fig.~\ref{fig3}, with such truncation one not only can obtain algebraic results of high accuracy but also conduct a direct numerical simulation for very small values of $\mu$, where the dynamical system~(\ref{eq-abp-03}) is a ``stiff'' one.

\subsubsection{Solution of the problem~(\ref{eq-abp-fe13})--(\ref{eq-abp-fe16}) in the $v$-space}
The problem for $\{U_n^{(1)}\}$ can be solved analytically in the $v$-space. Let us rewrite Eq.~(\ref{eq-abp-02}), integrating over $v$ from $-\infty$ to $+\infty$ (case i) and to a finite value (case ii). In case i:
\begin{align}
\partial_t\overline{\rho}&=-\partial_\varphi\overline{v\rho}\,,
\label{eq-abp-fe17}
\end{align}
where $\overline{(\cdots)}=\int_{-\infty}^{+\infty}\cdots\mathrm{d}v$.
Note identities $\overline{\rho}=w_0$ and $\overline{v\rho}=w_1$.
In case ii:
\begin{align}
\partial_t\int\limits_{-\infty}^v\mathrm{d}v_1\rho(v_1,\varphi,t)+\partial_\varphi\int\limits_{-\infty}^v\mathrm{d}v_1v_1\rho(v_1,\varphi,t)
\nonumber\\
=\left[\frac{\alpha v+\beta v^3 -F(\varphi,t)}{\mu}
+\frac{\sigma^2}{\mu^2}\partial_v\right]\rho(v,\varphi,t)\,.
\label{eq-abp-fe18}
\end{align}
Eq.~(\ref{eq-abp-fe17}) secures that the left hand side (l.h.s.) of Eq.~(\ref{eq-abp-fe18}) tends to zero for $v\to+\infty$, which allows one to take off one differentiation  $\partial_v$ from Eq.~(\ref{eq-abp-02}).

Within the moment representation~(\ref{eq-abp-03}), Eq.~(\ref{eq-abp-fe17}) corresponds to the first equation of the infinite chain ($n=0$), and Eq.~(\ref{eq-abp-fe18}) corresponds to all other equations of the chain ($n=1,2,...$). The first correspondence is obvious. To proof the second one, we multiply (\ref{eq-abp-fe18}) by $v^{n-1}$ and integrate over all $v$.
%{\color{red}
Further, in the l.h.s.\ part of equation one can use the integration by parts to obtain
\begin{align}
\int\limits_{-\infty}^{+\infty}\mathrm{d}v\,v^{n-1}\int\limits_{-\infty}^{v}\mathrm{d}v_1 \left\{\partial_t\rho(v_1,\varphi,t) +\partial_\varphi\big[v_1\rho(v_1,\varphi,t)\big]\right\}
\nonumber\\
=\left.\frac{v^n}{n}\int\limits_{-\infty}^{v}\mathrm{d}v_1 \left\{\partial_t\rho(v_1,\varphi,t) +\partial_\varphi\big[v_1\rho(v_1,\varphi,t)\big]\right\}\right|_{-\infty}^{+\infty}
\nonumber\\
-\int\limits_{-\infty}^{+\infty}\mathrm{d}v \frac{v^{n}}{n}\left\{\partial_t\rho+\partial_\varphi\big[v\rho\big]\right\}.
\nonumber
\end{align}
In the second line for $v\to\pm\infty$ we see uncertainty of type infinity (factor $v^n$) multiplied by zero (the integral). To resolve this uncertainty we replace the limits $|_{-\infty}^{+\infty}$ with $|_{-B}^{+B}$ and consider $B\to\infty$. For $B=+\infty$ the integral tends to zero by virtue of (\ref{eq-abp-fe17}) and for large finite $B$ its deviation from zero is determined by the tails of the probability density distribution along $v$\,: quantities $(\cdots)|_{-\infty}^{-B}$ and $(\cdots)|_{-\infty}^{B}=(\cdots)|_{-\infty}^{+\infty} -(\cdots)|_{B}^{+\infty}$ are of the order of magnitude of $\sim\int_B^{+\infty}\rho(v,\phi,t)\,\mathrm{d}v$. If the asymptotic decay of $\rho$ is a power-law one, $\rho\propto1/|v|^{m+1}$, then the uncertainty $\lim_{B\to+\infty}(v^n\int_\infty^v\mathrm{d}v_1\left\{\cdots\right\})\big|_{-B}^{+B}\propto \lim_{B\to+\infty}B^{n-m}$ is zero for $n<m$. For a typical exponentially fast decay of $\rho$ for large $v$ this uncertainty is always resolved as 0. Thus, for physically realistic $\rho(v)$ %{\color{green}
we obtain
\begin{align}
&\int\limits_{-\infty}^{+\infty}\mathrm{d}v\,v^{n-1}\int\limits_{-\infty}^{v}\mathrm{d}v_1 \left\{\partial_t\rho(v_1,\varphi,t) +\partial_\varphi\big[v_1\rho(v_1,\varphi,t)\big]\right\}=
\nonumber\\
&
-\int\limits_{-\infty}^{+\infty}\mathrm{d}v \frac{v^{n}}{n}\left\{\partial_t\rho+\partial_\varphi\big[v\rho\big]\right\}=
\nonumber\\
&
\int\limits_{-\infty}^{+\infty}\mathrm{d}v\left[\frac{\alpha v^n+\beta v^{n+2} -v^{n-1}F(\varphi,t)}{\mu}
-\frac{\sigma^2}{\mu^2}(n-1)v^{n-2}\right]\rho
\nonumber
\end{align}
and can see that
%}
Eq.~(\ref{eq-abp-fe18}) corresponds to the equations of chain~(\ref{eq-abp-03}) with $n=1,2,...$\,, i.e.\ all but the first one ($n=0$), which corresponds to (\ref{eq-abp-fe17}).
%}

Collecting the terms contributing to the leading order of the problem [Eqs.~(\ref{eq-abp-fe10})--(\ref{eq-abp-fe12})] in the l.h.s.\ part of equation, and all other terms in the r.h.s.\ part, we write:
\begin{align}
&\left[\beta v^3 -F(\varphi,t)
+\frac{\sigma^2}{\mu}\partial_v\right]\rho(v,\varphi,t)
\nonumber\\
&\qquad
=-\alpha v\rho(v,\varphi,t)
+\mu\int\limits_{-\infty}^v\mathrm{d}v_1
%%\left\{
\partial_t\rho(v_1,\varphi,t)
%%-\frac{\partial_t\overline{\rho}}{\overline{\rho}}\rho(v_1,\varphi,t)
%%\right\}
\nonumber\\
&\qquad\qquad\qquad
+\mu\int\limits_{-\infty}^v\mathrm{d}v_1
%%\left\{
\partial_\varphi\left[v_1\rho(v_1,\varphi,t)\right]
%%-\frac{\partial_\varphi\overline{v\rho}}{\overline{\rho}}\rho(v_1,\varphi,t)
%%\right\}
\,.
\label{eq-abp-fe20}
\end{align}

For comparison to the expansion $U_n=U_n^{(0)}+\sqrt{\mu}U_n^{(1)}+\mu U_n^{(2)}+\cdots$, we make expansion  $\rho=\rho^{(0)}+\sqrt{\mu}\rho^{(1)}+\mu\rho^{(2)}+\cdots$ with the normalization condition $\overline{\rho^{(0)}}=\overline{\rho}$, $\overline{\rho^{(n\ge1)}}=0$
and hierarchy of timescales $\partial_t=\sqrt{\mu}\partial_\tau+\mu\partial_{\tau_2}+\mu^{3/2}\partial_{\tau_3}+\cdots$\,.
Then Eq.~(\ref{eq-abp-fe17}) takes the form of
\begin{align}
(\sqrt{\mu}\partial_\tau+\mu\partial_{\tau_2}+\mu^{3/2}\partial_{\tau_3}+\cdots)\overline{\rho^{(0)}}=
\qquad\qquad
\nonumber\\
-\partial_\varphi\left(
\overline{v\rho^{(0)}} +\sqrt{\mu}\overline{v\rho^{(1)}} +\mu\overline{v\rho^{(2)}} +\cdots\right)\,.
\label{eq-abp-fe21}
\end{align}
In the leading order of this equation
\begin{equation}
\sqrt{\mu}\partial_\tau\overline{\rho^{(0)}}
=-\partial_\varphi\overline{v\rho^{(0)}},
\label{eq-abp-fe22}
\end{equation}
and the next-order approximation is
\begin{equation}
\sqrt{\mu}\partial_{\tau_2}\overline{\rho^{(0)}}
=-\partial_\varphi\overline{v\rho^{(1)}} .
\label{eq-abp-fe23}
\end{equation}
Further, for Eq.~(\ref{eq-abp-fe20}) we construct consecutive approximations, which give the equation chain:
\begin{align}
&\left[\beta v^3 -F(\varphi,t)
+\frac{\sigma^2}{\mu}\partial_v\right]\rho^{(0)}=0\,,
\label{eq-abp-fe24}
\\
&\left[\beta v^3 -F(\varphi,t)
+\frac{\sigma^2}{\mu}\partial_v\right]\rho^{(1)}
%\nonumber\\
%&\qquad
 =-\frac{\alpha v}{\sqrt{\mu}}\rho^{(0)}
\nonumber\\
&\qquad\qquad
+\mu\int\limits_{-\infty}^v\mathrm{d}v_1
%%\left\{
\partial_\tau\rho^{(0)}(v_1,\varphi,t)
%%-\frac{\partial_\tau\overline{\rho^{(0)}}}{\overline{\rho^{(0)}}}\rho^{(0)}(v_1,\varphi,t)
%%\right\}
\nonumber\\
&\qquad\qquad
+\sqrt{\mu}\int\limits_{-\infty}^v\mathrm{d}v_1
%%\left\{
\partial_\varphi\left[v_1\rho^{(0)}(v_1,\varphi,t)\right]
%%-\frac{\partial_\varphi\overline{v\rho^{(0)}}}{\overline{\rho^{(0)}}}\rho^{(0)}(v_1,\varphi,t)
%%\right\}
\,,
\label{eq-abp-fe25}
%\\
%\hat{L}\rho^{(2)}
% &=-\frac{\alpha v}{\sqrt{\mu}}\rho^{(1)}
%+\sqrt{\mu}\int\limits_{-\infty}^v\mathrm{d}v_1\left\{\partial_t\rho^{(1)}(v_1,\varphi,t)
%-\frac{\partial_t\overline{\rho^{(0)}}}{\overline{\rho^{(0)}}}\rho^{(1)}(v_1,\varphi,t)
%\right\}
%\nonumber\\
%&
%+\sqrt{\mu}\int\limits_{-\infty}^v\mathrm{d}v_1\left\{
%\partial_\varphi\left[v_1\rho^{(1)}(v_1,\varphi,t)\right]
%-\frac{\rho^{(1)}(v_1,\varphi,t)\partial_\varphi\overline{v\rho^{(0)}} +\rho^{(0)}(v_1,\varphi,t)\partial_\varphi\overline{v\rho^{(1)}}}{\overline{\rho^{(0)}}}
%\right\}\,,
%\label{eq-abp-fe26}
\\
&\qquad\qquad\qquad\qquad
\dots\;.
\nonumber
\end{align}
By construction, such iterative procedure of consecutive approximations yields a converging expansion for small $\mu$. We restrict ourselves to the first two orders of expansion: Eq.~(\ref{eq-abp-fe24}), for which solution~(\ref{eq-abp-rhoJ0}) was obtained earlier in the text, and Eq.~(\ref{eq-abp-fe25}) for calculation of $\rho^{(1)}$. Taking the scaling law (\ref{eq:wU}) for $\mu\to0$ into account, we can see that the moment representation of the mathematical problem~(\ref{eq-abp-fe22}) and (\ref{eq-abp-fe24}) is equivalent to the equation system~(\ref{eq-abp-fe09})--(\ref{eq-abp-fe12}), and that of the problem~(\ref{eq-abp-fe23}) and (\ref{eq-abp-fe25}) is equivalent to the equation system~(\ref{eq-abp-fe13})--(\ref{eq-abp-fe16}). Moreover, for $\mu\to0$, in the problem for $\rho^{(1)}$ the contributions with $\partial_\tau\rho^{(0)}$ drop out [see Eqs.~(\ref{eq-abp-fe14})--(\ref{eq-abp-fe16}), where no time-derivatives are present]; therefore, to this order of accuracy, Eq.~(\ref{eq-abp-fe25}) can be reduced to
\begin{align}
\left[\beta v^3 -F(\varphi,t)
+\frac{\sigma^2}{\mu}\partial_v\right]\rho^{(1)}
%\nonumber\\
%&\qquad
 =-\frac{\alpha v}{\sqrt{\mu}}\rho^{(0)}
\qquad
\nonumber\\
+\sqrt{\mu}\int\limits_{-\infty}^v\mathrm{d}v_1
%%\left\{
\partial_\varphi\left[v_1\rho^{(0)}(v_1,\varphi,t)\right]
%%-\frac{\partial_\varphi w_1^{(0)}}{w_0^{(0)}}\rho^{(0)}(v_1,\varphi,t)
%%\right\}
\,.
\label{eq-abp-fe27}
\end{align}

Given the condition~(\ref{eq-abp-Fcond}) is met, which is realistic for small $\mu$ and finite $F$ and $\sigma$, for the calculation of the leading order of the term $\overline{v\rho^{(1)}}$, present in Eqs.~(\ref{eq-abp-fe21}) and (\ref{eq-abp-fe23}), we can drop the $F$-term in Eq.~(\ref{eq-abp-fe27}). Further, we explicitly decompose $\rho^{(1)}$ into the symmetric- and asymmetric-in-$v$ parts, $\rho^{(1)}=\rho_{1s}+\rho_{1a}$, $\rho_{1s}(v)=\rho_{1s}(-v)$, $\rho_{1a}(v)=-\rho_{1a}(-v)$. Eq.~(\ref{eq-abp-fe27}) with the $F$-term dropped yields for these parts:
\begin{align}
\left[\beta v^3+\frac{\sigma^2}{\mu}\partial_v\right]\rho_{1s}
%\nonumber\\
%&\qquad
 &=-\frac{\alpha v}{\sqrt{\mu}}\rho^{(0)}|_{F=0}
%% -\sqrt{\mu}\frac{\partial_\varphi w_1^{(0)}}{w_0^{(0)}}\int\limits_{-\infty}^v\mathrm{d}v_1 \rho^{(0)}(v_1,\varphi,t|F=0)
\,,
\label{eq-abp-fe28}
\\
\left[\beta v^3+\frac{\sigma^2}{\mu}\partial_v\right]\rho_{1a}
&=\sqrt{\mu}\int\limits_{-\infty}^v\mathrm{d}v_1
v_1\partial_\varphi\rho^{(0)}(v_1,\varphi,t|F=0)\,.
\label{eq-abp-fe29}
\end{align}
Here we used that $\rho^{(0)}(v,\varphi,t)$ is a symmetric function of $v$ for $F=0$.
Since $\overline{v\rho_{1s}}=0$, for the calculation of $\overline{v\rho^{(1)}}=\overline{v\rho_{1a}}$ it is enough to solve Eq.~(\ref{eq-abp-fe29}). Substituting $\rho^{(0)}(F=0)$ from (\ref{eq-abp-rhoJ0}), we find
\begin{align}
\left[\beta v^3+\frac{\sigma^2}{\mu}\partial_v\right]\rho_{1a}
=\sqrt{\mu}\int\limits_{-\infty}^v\mathrm{d}v_1
v_1e^{-\frac{\mu\beta v_1^4}{4\sigma^2}}\partial_\varphi C(\varphi)
\nonumber\\
%&\color{red}
%=\sqrt{\mu}\partial_\varphi C\frac{\sigma}{\sqrt{\mu\beta}} \int\limits_{-\infty}^V\mathrm{d}V_1^2\;e^{-V_1^4}
%\nonumber\\
=-\frac{\sqrt{\pi}\sigma\partial_\varphi C}{2\sqrt{\beta}}\left[1-\mathrm{erf}(V^2)\right],
\quad v\equiv\frac{\sqrt{2\sigma}\,V}{(\mu\beta)^{1/4}}\,,
\label{eq-abp-fe30}
\end{align}
where the error function $\mathrm{erf}(x)\equiv(2/\sqrt{\pi})\int_0^xe^{-x_1^2}\mathrm{d}x_1$\,.
In terms of $V$ Eq.~(\ref{eq-abp-fe30}) reads
\begin{align}
\left(\partial_V+4V^3\right)\rho_{1a}
&=-\sqrt{\frac{\pi}{2\sigma}}\left(\frac{\mu}{\beta}\right)^\frac34
\left[1-\mathrm{erf}(V^2)\right]\partial_\varphi C
\,.
\label{eq-abp-fe31}
\end{align}
Solving the latter equation by the method of variation of a constant under the asymmetry condition for $\rho_{1a}$, we obtain
\begin{align}
&\rho_{1a}=-\sqrt{\frac{\pi}{2\sigma}}\left(\frac{\mu}{\beta}\right)^\frac34
 \partial_\varphi C \int\limits_0^V\mathrm{d}V_1\left[1-\mathrm{erf}(V_1^2)\right]e^{V_1^4-V^4},
\nonumber\\
&\qquad
\overline{v\rho^{(1)}}=-G_1\frac{\sqrt{2\pi\sigma}\mu^{1/4}}{\beta^{5/4}}\partial_\varphi C
\,,
\label{eq-abp-fe32}
\\
&\qquad
G_1\equiv\int\limits_{-\infty}^{+\infty}\mathrm{d}V\,V\int\limits_0^V \mathrm{d}V_1\left[1-\mathrm{erf}(V_1^2)\right]e^{V_1^4-V^4}.
\nonumber
\end{align}
The analytical expression (\ref{eq:app401}) for constant $G_1$ is derived in Appendix~\ref{sec:app4}: $G_1=0.49859365698...$\,.

Thus, the problem~(\ref{eq-abp-fe23}) and (\ref{eq-abp-fe25}) in the $v$-space, with account for (\ref{eq-abp-w0C}), yields
\begin{align}
&\partial_{\tau_2}w_0^{(0)}
%{\color{red}=
%\frac{G_1}{\sqrt{\mu}}\frac{\sqrt{2\pi\sigma}\mu^{1/4}}{\beta^{5/4}} \frac{\Gamma(\frac34)\,(\mu\beta)^{1/4}}{\pi\sqrt{\sigma}}\partial_\varphi^2w_0^{(0)}}
=\frac{G_2}{\beta}\partial_\varphi^2w_0^{(0)},
\label{eq-abp-fe33}
\\
G_2&=\frac{\sqrt{2}\Gamma(\frac34)}{\sqrt{\pi}}
%\int\limits_{-\infty}^{+\infty}\mathrm{d}V\,V\int\limits_0^V \mathrm{d}V_1\left[1-\mathrm{erf}(V_1^2)\right]e^{V_1^4-V^4}
G_1
=\frac{\pi}{\sqrt{2}} -{}_3F_2\left(\frac{1}{4},\frac{1}{2},1;\frac{3}{4},\frac{5}{4};1\right)
\nonumber\\
&=0.48749549439936...\,,
\label{eq-abp-fe34}
\end{align}
where the generalized hypergeometric function ${}_3F_2$ is given by Eq.~(\ref{eq:app402}).
Eq.~(\ref{eq-abp-fe33}) is a sought solution of the problem~(\ref{eq-abp-fe13})--(\ref{eq-abp-fe16}). Essentially, we calculated $U_1^{(1)}$ given by the infinite equation chain~(\ref{eq-abp-fe14})--(\ref{eq-abp-fe16}) and the zeroth-order solution $\{U_n^{(0)}\}$; up to a constant coefficient, it is the derivative $\partial_\varphi U_0^{(0)}$. Substitution of $U_1^{(1)}$ into Eq.~(\ref{eq-abp-fe13}) gives an effective diffusion of the probability density $U_0^{(0)}$ with ``slow'' time $\tau_2$.

\subsubsection{Corrected Smoluchowski equation describing effective diffusion}
We can again consider Eq.~(\ref{eq-abp-fe21}) restricting ourselves to the first two orders of expansion,
$(\sqrt{\mu}\partial_\tau+\mu\partial_{\tau_2}+\cdots)\overline{\rho^{(0)}}
=-\partial_\varphi\big(
\overline{v\rho^{(0)}} +\sqrt{\mu}\overline{v\rho^{(1)}} +\cdots\big)$. We restore the derivative $\partial_t$ in its l.h.s.\ part and substitute above-calculated $\overline{v\rho^{(0)}}=w_1^{(0)}$ [see Eq.~(\ref{eq-abp-08}) with $m=0$] and (\ref{eq-abp-fe33}) into the r.h.s.\ part. Hence, we obtain a corrected Smoluchowski equation accounting for fluctuations and the diffusive component in the motion of active Brownian particle~(\ref{eq-abp-01}):
\begin{equation}
\partial_tw_0+\partial_\varphi\left[\frac{\gamma_1\sqrt{\mu}}{\sigma\sqrt{\beta}}F(\varphi,t)\,w_0\right] =\frac{G_2\mu}{\beta}\partial_\varphi^2w_0\,.
\label{eq-abp-enhSmol}
\end{equation}
Here $\gamma_1$ and $G_2$ are given by formulas~(\ref{eq-abp-gamma1}) and (\ref{eq-abp-fe34}), respectively, and the normalization condition $w_0=w_0^{(0)}$ (since $w_0^{(m\ge1)}=0$) was taken into account.

Calculating $\overline{v\rho^{(1)}}$ in the previous section we dropped the corrections related to $F$. From Eq.~(\ref{eq-abp-w1}) one can see that the next-order correction for the deterministic part of the flux [the second term in Eq.~(\ref{eq-abp-enhSmol})], associated with $F$, is $\sim\mu^2F^3/\sigma^4$ and small as compared to the derived diffusion terms. Thus, the employed approximation did not affect the strong accuracy order of corrected Smoluchowski equation~(\ref{eq-abp-enhSmol}).

Even though for $\mu\to0$ the last term of Eq.~(\ref{eq-abp-enhSmol}) is small against the background of the second term, it is essentially important, since the second term gives an effective deterministic dynamics, for which the the distribution heterogeneities do not dissipate, while the last term describes diffusion and makes the equation robust (structurally stable). One more peculiarity of the derived equation distinguishing it from the corrected Smoluchowski equation~(\ref{eq024}) for passive particles is the absence of the terms linked to the time-derivative $\partial_tF$. For a passive particle such derivative was absent only for a stationary $F(\varphi)$ and emerged in the corrected Smoluchowski equation as a result of a rigorous derivation. For active particle~(\ref{eq-abp-01}), analogous term does not emerge in the course of a rigorous derivation, since it is of higher order of smallness.

At first glance, the derivation of Eq.~(\ref{eq-abp-enhSmol}) is based solely on solving the problem in the $v$-space. However, this derivation heavily relies on the results of analysis of the moment equations. It was within the framework of the moment equations that the expansions were analyzed and the terms negligible in the considered expansion orders were identified.
The moment equations also allowed us to find the minimal form of the auxiliary problems in the $v$-space, on the basis of which Eq.~(\ref{eq-abp-enhSmol}) was derived. Furthermore, the moment equations were found to be utilitary for numerical simulations of the system for finite $\mu$ (Fig.~\ref{fig3}). At small values of $\mu$ for long but finite chains of moment equations, the truncation does not induce numerical instabilities in spite of a fast growth of $w_n$ with $n$. Numerical simulations exhibit a reasonably fast convergence of series: for several tens of moments $w_n$ the dynamics of the macroscopically observable $w_0(\varphi,t)$ becomes insensitive to the truncation order.

\section{Comparison to results presented in the literature}
\label{sec:litr}
This section does not present anything close to a comprehensive literature review: here we put our analysis and the derived results into the context of some relevant reference papers and books in the field.

%$\bullet$\;[S.\ Milster, J.\ N\"otel, I.M.\ Sokolov, and L.\ Schimansky-Geier,
%{\em Eliminating inertia in a stochastic model of a micro-swimmer with constant speed},
%Eur.\ Phys.\ J.\ ST {\bf 226}, 2039--2055 (2017)]
In paper~\cite{Milster-etal-2017}, the problem of adiabatic elimination of velocity (or inertia term) is analysed for the cases of both a passive Brownian particle and an active particle with a nearly constant propulsion speed on the plane. The first case analysis is provided for a didactic purpose; in Sec.~2.2 of~\cite{Milster-etal-2017} there is no $\varphi$-dependent force (in terms of~\cite{Milster-etal-2017}, ``$x$-dependent''), calculations are conducted for the first three moments of velocity and the linear-in-$\mu$ correction is neglected.
For the second case, the particle diffusion is associated with stochastic variation of the orientation of the velocity: in the limit of vanishing inertia the impact of the fluctuations of the particle speed vanishes against the background of the dynamics of the velocity angle. Besides the fact that in~\cite{Milster-etal-2017} the nonlinearity type resulting in a nearly constant value of speed differs from that in Eq.~(\ref{eq-abp-01})~\cite{Pikovsky-2023,Erdmann-etal-2000,Erdmann-etal-2002,Erdmann-Ebeling-2005},
more importantly Eq.~(\ref{eq-abp-enhSmol}) derived in this paper describes the diffusion related to stochastic switchings between two propulsion directions in a one-dimensional setup. This mechanism requires higher orders of expansion for $\mu\ll1$, than the diffusion mechanism related to a continuous random walk of the velocity angle.

%In Sec.\ 2.2., there is no $\varphi$-dependent force (in their terms, $x$-dependent), the calculation is performed for the first three moments of the velocity: a first-order in $\mu$ correction is omitted.

%$\bullet$\;[R.\ Becker, {\em Theorie der W\"arme} (Springer, Berlin, 1985), Chap.\ VI B]

In~\cite{Becker-1985} (Chapter VI B) and \cite{Haken-1977} (Chapter 7), the $\mu^1$-correction in the equation for a passive Brownian particle is omitted. The scaling law of the velocity moments for $\mu\to0$ are not considered.

%$\bullet$\;[H.\ Haken, {\em Synergetics---An Introduction}, 2nd edition (Springer, Berlin, 1977), Chap.\ 7]
%
%A first-order correction is not present.

%$\bullet$\;[C.W.\ Gardiner, {\em Handbook of Stochastic Methods} (Springer, 1983)]

In book~\cite{Gardiner-1983-1997} the derivations in Sects.~6.4 (Adiabatic Elimination of Fast Variables) and 6.4.1 (Abstract Formulation in Terms of Operators and Projectors) correspond to calculations of $w_0$ and $w_1$ with the $\mu^1$-correction for $w_2$ omitted.
In Sec.~6.4.2 of \cite{Gardiner-1983-1997}, Gardiner derives the evolution equation for $w_0$. In Sec.~6.4.3 it is also noted that the derived equation is valid for $t\gg\mu$; the same statement can be made for Eqs.~(\ref{eq020})--(\ref{eq021}), where we ignore the boundary layer $t\sim\mu$ in time. In Sec.~6.4.5 Gardiner constructs a regular expansion in $\mu$ and provides the equation for the particular case of Brownian motion. In terms of our paper the equation reads
\begin{align}
\partial_tw_0+\partial_\varphi\big[(F
 -\mu F\partial_\varphi F)\,w_0\big]
% \quad
%\nonumber\\
% {}
  =\sigma^2\partial_\varphi\big[(1-\mu\partial_\varphi F)\,\partial_\varphi w_0\big]\,,
\label{eq-Smol}
\end{align}
where in comparison with Eq.~(\ref{eq024}) the only missing term is the $\partial_tF$-contribution, which is absent since Gardiner considers only static potentials as a source of force $F$. Eq.~(\ref{eq-Smol}) is a \textit{corrected Smoluchowski equation}.

%Secs.\ 6.4.\ Adiabatic Elimination of Fast Variables and 6.4.1.\ Abstract Formulation in Terms of Operators and Projectors: This corresponds to calculation of $w_0$ and $w_1$, where for $w_2$ the $\mu^1$-corrections are omitted. In 6.4.2, he derives the equation for $w_0$; in 6.4.3, he notes that the derived equation is valid for $t\gg\mu$ (we see the same with Eqs.~(\ref{eq020})--(\ref{eq021}), where the time boundary layer $t\sim\mu$ is ignored). In Sec.~6.4.5, he constructs the regular expansion in $\mu$, and provides the equation for a particular case of the Brownian motion; in our terms it is
%\begin{align}
%\partial_tw_0+\partial_\varphi\big[(F
% -\mu F\partial_\varphi F)\,w_0\big]
%% \quad
%%\nonumber\\
%% {}
%  -\sigma^2\partial_\varphi\big[(1-\mu\partial_\varphi F)\,\partial_\varphi w_0\big]=0\,,
%\label{eq-Smol}
%\end{align}
%compared to Eq.~(\ref{eq024}), the only lacking term here is due to $\partial_tF$ which is absent in the Gardiner analysis, since he considers time-independent potentials. This is the {\em corrected Smoluchowski equation}.

%$\bullet$\;[C.W.\ Gardiner, {\em Adiabatic elimination in stochastic systems. I.~Formulation of methods and application to few-variable systems}, Phys.\ Rev.\ A {\bf 29}, 2814 (1984)]

In Ref.~\cite{Gardiner-1984}, the original stochastic equations have a more general and sophisticated form than in our paper on the one hand; on the other hand, they obey the Fluctuation--dissipation theorem for a nonlinear dissipation law and other generalizations (which excludes the case of active particles from the theory scope). In Sec.~III.B of \cite{Gardiner-1984} the case of Eq.~(\ref{eq001}) is considered but without $\mu^1$-corrections.

%The starting equations are more complicated that our ones but they obey the fluctuation-dissipation theorem for the nonlinear friction law and other generalizations. Sec.\ III.B deals with the our case; however, no account for $\mu^1$-corrections is made.

%$\bullet$\;[G.\ Wilemski, {\em On the derivation of Smoluchowski equations with corrections in the classical theory of Brownian motion}, J.\ Stat.\ Phys. {\bf 14}(2), 153--169 (1976)]

In Ref.~\cite{Wilemski-1976}, an equation of type~(\ref{eq024}) is derived with the $\partial_tF$-term; moreover, the $\partial_t^2F$-term is obtained for arbitrary dimensionality of space (Eq.~(26) on page~160~\cite{Wilemski-1976}). The derivation procedure is equivalent to calculation of $w_3$ and $w_4$.

%Eq.~(\ref{eq024}) is reported with $\partial_tF$-term (and $\partial_t^2F$-term) for arbitrary space dimension (Eq.~(26) on p.~160). The derivation procedure is equivalent to the calculation of $w_3$ and $w_4$.

\section{Application to collective dynamics of populations of noisy oscillators with small inertia}
\label{sec:applic}
For an important class of systems with $F(\varphi,t)=\omega(t)+\mathrm{Im}[2h(t)e^{-i\varphi}]$, for small inertia and weak noise corrected Smoluchowski equation~(\ref{eq024}) can be written in the Fourier space:
\begin{align}
\dot{a}_n & =n\big[i\omega_1(t) a_n +h_1a_{n-1} -h_1^\ast a_{n+1}
\nonumber\\
 &\qquad\qquad
  +h_2a_{n-2} -h_2^\ast a_{n+2}\big]-n^2\sigma^2a_n\,,
\label{eq51:05}
\end{align}
where $a_{-n}=a_n^\ast$, $a_0=1$, $\omega_1=\omega-\mu\dot\omega$, $h_1=h-\mu(\dot{h}-i\omega h)$, $h_2=\mu h^2$.
Infinite equation chain~(\ref{eq51:05}) gives for the first two circular cumulants ($\kappa_1=a_1$ and $\kappa_2=a_2-a_1^2$):
\begin{align}
&\dot{\kappa}_1 =i\omega_1\kappa_1 +h_1 -h_1^\ast(\kappa_1^2+\kappa_2)+h_2\kappa_1^\ast
\nonumber\\
 &\qquad\qquad\quad
  -h_2^\ast(2\kappa_3+3\kappa_2\kappa_1+\kappa_1^3)-\sigma^2\kappa_1\,,
\label{eq52:01}\\
&\dot{\kappa}_2 =(2i\omega_1-4\sigma^2-4h_1^\ast\kappa_1)\kappa_2
-4h_1^\ast\kappa_3 +2h_2(1-|\kappa_1|^2)
\nonumber\\
 &\qquad
-6h_2^\ast(2\kappa_4+2\kappa_3\kappa_1 +\kappa_2^2+\kappa_2\kappa_1^2) -2\sigma^2\kappa_1^2\,.
\label{eq52:02}
\end{align}
Assuming deviations from the OA manifold (which is given by $a_n=(a_1)^n$) to be small, one can approximately close this equation system by setting $\kappa_3=\kappa_4=0$~\cite{Goldobin-etal-2018,Goldobin-2019,Goldobin-Dolmatova-2019} and obtain
\begin{align}
&\dot{\kappa}_1 =(i\omega_1-\sigma^2)\kappa_1 +h_1 -h_1^\ast(\kappa_1^2+\kappa_2)+h_2\kappa_1^\ast
% {\color{red} -}
\nonumber\\
 &\qquad\qquad\qquad\quad
  -h_2^\ast(3\kappa_2\kappa_1+\kappa_1^3)\,,
\label{eq5200:01}\\
&\dot{\kappa}_2 =(2i\omega_1-4\sigma^2-4h_1^\ast\kappa_1)\kappa_2
-2\sigma^2\kappa_1^2 +2h_2(1-|\kappa_1|^2)
\nonumber\\
 &\qquad\qquad\qquad\quad
-6h_2^\ast(\kappa_2^2+\kappa_2\kappa_1^2)\,.
\label{eq5200:02}
\end{align}
Low-dimensional equation system~(\ref{eq5200:01})--(\ref{eq5200:02}) is the two circular cumulant (2CC) model reduction; it is the main result of this section. In this system, we account for  possible nonstationarity of $\omega$ and $h$, which can be explicit or caused by the dependence of these quantities on the Kuramoto--Daido order parameters $a_n$ (within the framework of a two cumulant reduction one expresses $a_n=\kappa_1^n+\frac{n(n-1)}{2}\kappa_2\kappa_1^{n-2}$~\cite{Tyulkina-etal-2018,Goldobin-etal-2018,Goldobin-Dolmatova-2019} and any dependence is reduced to the dependence on $\kappa_1$ and $\kappa_2$).

\subsection{Time scales and conditions on smallness of parameters}
\label{ssec61}
For passive Brownian particles and phase oscillators with effective inertia, the reference dynamics rate (time) scales are determined by three numbers: $1/\mu$, $|F|$, $\sigma^2$.
In the inequalities expressing the applicability conditions for any model reductions one must have the combinations of these numbers of the same dimension on both sides.
The same can be formulated as the rescaling invariance of the original Eq.~(\ref{eq001}), which is invariant with respect to the rescaling
\begin{equation}
t\mapsto\eta{t}, \quad \sigma\mapsto \left.\sigma\middle/\sqrt{\eta}\right., \quad
\mu\mapsto\eta\mu, \quad F\mapsto \left.F\middle/\eta\right.
\end{equation}
for any positive $\eta$.
Therefore, all equalities and conditions expressed by inequalities must be also invariant under this rescaling transformation.

In particular, the condition of inertia smallness for the corrected Smoluchowski equation (with $\mu^1$-correction) is
\begin{equation}
\mu\abs{F} \ll 1\,;
\label{eq:muFll1}
\end{equation}
for an oscillator population in low synchrony regimes, this condition simplifies to  $\mu|\omega|\ll 1$. The later restriction impedes the applicability of the corrected Smoluchowski equation and the approaches based on it for a rigorous analysis for broadband and heavy-tailed frequency distributions.

For the CC approach and few-CC truncations of an infinite equation chain, the noise intensity is formally required to be small, i.e.\ $\sigma^2\ll X$, where $X$ is some reference value of the dimension of an inverse reference time. In the zero-inertia case, the only other time scale is  $1/|F|$; therefore, the scale invariant condition must read
\begin{equation}
\sigma^{2} \ll |F|\,.
\label{eq:sigma2llF}
\end{equation}
In the case of nonzero small inertia, the fundamental condition is \eqref{eq:muFll1}; combining this condition with the one of noise weakness~\eqref{eq:sigma2llF} yields the hierarchy of inequalities
\begin{equation}
\mu\sigma^2 \ll \mu|F| \ll 1\,.
\label{eq:chainll}
\end{equation}

\subsection{Comparison to exact analytical solutions in the weak synchrony limit}
In~\cite{Munyayev-etal-2020}, time-independent solutions of the corrected Smoluchowski equation were derived analytically for infinite equation chains for the circular moments (CM). The CM solutions do not allow for the stability analysis and are blind to the collective oscillation regimes. The 2CC model~(\ref{eq5200:01})--(\ref{eq5200:02}) can be employed for the study of both. One can examine its accuracy %{\color{red} not only with numerical simulations but also}
by comparison to the analytical solutions.
Noteworthily, for the CM approach without truncation of infinite equation chains, the restricting condition is $\mu|F|\ll 1$, and the additional condition $\mu\sigma^{2} \ll 1$ is excessive.
However, the applicability of truncated CC expansions requires~\eqref{eq:chainll}.

In this section we deal with the regimes, where $h \to 0$ and the Kuramoto order parameter is small (can be finite). In this case, condition \eqref{eq:chainll} simplifies to
$\mu\sigma^2 \ll \mu|\omega| \ll 1$.
The range of admissible values of $\omega$ is bounded, which influences the result accuracy depending on the frequency distribution width. For narrow distributions, the 2CC results deviate from the CM solution for regimes with nonlarge values of the Kuramoto order parameter near the excitation threshold of the collective mode. For wider frequency distributions, the results of two approaches become more similar, but the further increase of the distribution width results in a growing deviation of both approaches from the accurate solution of the original Fokker--Planck equation with inertia.
On the other hand, in the case of $\omega=0$, the 2CC reduction is reliably accurate only for large enough $|h|$ (but still lesser than $1/\mu$), that is for a moderate degree of synchrony, whereas a systematic error appears near the phase transition threshold even though the magnitude of this error may be small.

Let us see this explicitly with a specific example; we compare the analytical solution for a time-independent regime derived in~\cite{Munyayev-etal-2020} to the asymptotic (for $h\to0$) time-independent solution of Eqs.~\eqref{eq5200:01}--\eqref{eq5200:02}.
The two leading terms of the expansion of the analytical solution $a_1=\kappa_1$~\cite{Munyayev-etal-2020} for $h\to0$ read
\begin{widetext}
\begin{align}
 & a_1(\omega) = \frac{\mathrm{I}_{1-i\omega/\sigma^2}\big(\frac{2h}{\sigma^2}\big)} {\mathrm{I}_{-i\omega/\sigma^2}\big(\frac{2h}{\sigma^2}\big)}
 \left[1+\frac{i\mu\sigma^2 \sinh\frac{\pi\omega}{\sigma^2}}{\pi \mathrm{I}_{-i\omega/\sigma^2}\big(\frac{2h}{\sigma^2}\big)\, \mathrm{I}_{i\omega/\sigma^2}\big(\frac{2h}{\sigma^2}\big)}\right]
%\nonumber
%\\
% &
  = \frac{1+i\mu\omega}{\sigma^2 - i\omega} h - \frac{\sigma^2+i\omega + i\mu\omega\left(5\sigma^2 - i\omega\right)}{\left(\sigma^4+\omega^2\right)\left(2\sigma^2 -i\omega\right)} h^3 + \mathcal{O}(h^5)\,,
  \label{eq:momenth0}
\end{align}
where $\mathrm{I}_\nu(z)$ is the modified Bessel function.
The two leading terms of the expansion of solution $a_1=\kappa_1$ of the 2CC model reduction for $h \to 0$ are
\begin{align}
  a_1(\omega) = \frac{1+i\mu\omega}{\sigma^2 - i\omega} h
  -\frac{\left(1-i\mu\omega\right)\left[\left(1-\mu\omega^2\right)\left(\sigma^2+i\omega\right)
  -\mu\sigma^2\left(\sigma^2-5i\omega\right)\right]}{\left(\sigma^4+\omega^2\right)
  \left(2\sigma^2 -i\omega\right)} h^3 + \mathcal{O}(h^5)\,.
\label{eq:cumulh0}
\end{align}
\end{widetext}
Comparing~\eqref{eq:momenth0} with \eqref{eq:cumulh0}, one can see that the linear-in-$h$ terms are identical, but the $h^3$-terms mismatch by the contributions $\propto\mu\sigma^2$ and $\propto\mu^2\omega^2$, that is in the higher orders of smallness of hierarchy~\eqref{eq:chainll}.

\begin{figure}[!b]
\centerline{
\includegraphics[width=0.465\textwidth]{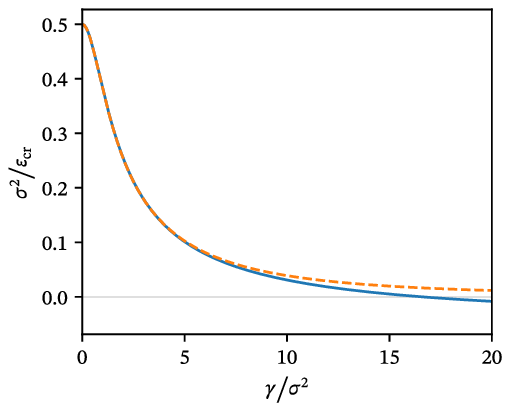}}
\caption{The dependence of the inverse critical coupling $\varepsilon_\mathrm{cr}$ versus the half-width $\gamma$ of a uniform distribution of natural frequencies $\omega$ is plotted for the corrected Smoluchowski equation (solid line) and for the original Fokker--Planck equation with inertia (dashed line). Parameters: $\mu\sigma^2=0.1$.}
\label{fig4}
\end{figure}

The linear- and cubic-in-$h$ terms provide important information about synchronization transitions. In particular, the critical coupling value of $\varepsilon_{\mathrm{cr}}$ of the Kuramoto-transition is given by the integral of the coefficient of the linear-in-$h$ term over $\omega$. For example, for the Kuramoto ensemble with natural frequency distribution $g(\omega)$, one has $h=\varepsilon R/2$ and $R=\int g(\omega)a_1(\omega)\mathrm{d}\omega$. But the type of the transition (sub- or supercritical) is determined by the sign of the integral of the coefficient of the $h^3$-term over $\omega$. Hence, because of the error $\propto\mu\sigma^2$, the 2CC model reduction gives a biased value of the critical inertia $\mu_\ast$, where the type of transition changes. %(calculated in~\cite{Munyayev-etal-2020}).
For instance, for a Lorentzian distribution $g(\omega)= \gamma/[\pi(\gamma^2 + \omega^2)]$, the critical value calculated with the CM solution is $\mu_\ast=\sigma^2/(\gamma^2+3\sigma^2\gamma)$, whereas the 2CC model reduction yields $\mu_\ast=\sigma^2/(\gamma^2+3\sigma^2\gamma+\sigma^4)$. For a bimodal distribution $g(\omega)=[\delta(\omega-\gamma)+\delta(\omega+\gamma)]/2$, the critical inertia given by the CM solution is $\mu_\ast=2\sigma^2(\sigma^4-2\gamma^2)/\gamma^2(\gamma^2+13\sigma^4)$, and the 2CC model gives $\mu_\ast=2\sigma^2(\sigma^4-2\gamma^2)/(\gamma^4+9\sigma^4\gamma^2+\sigma^8)$. The results of the 2CC model are identical to the CM analytical solution for $\gamma\gg\sigma^2$. Thus, the CC approach can be used for a rigorous analysis for $\sigma^2 \ll \gamma \ll 1/\mu$ and treated only as an approximation otherwise.

High degrees of synchrony require $|a_1|=|\kappa_1|\approx1$ and hence small $|\kappa_2|$; one typically observes a fast decay of higher CCs~\cite{Goldobin-Dolmatova-2019} and few-CC reductions become accurate. In this section, this is the case of higher $|h|$, where the error of the 2CC solutions becomes small again. Notice, however, that further increase of $|h|$ results in $|h|\sim1/\mu$ and the corrected Smoluchowski equation becomes an inaccurate approximation of the original Fokker--Plank equation with inertia.

Finally, in order to see the importance of the condition $\mu|\omega|\ll 1$ (or $\mu\gamma\ll1$) we compare the coefficient of the linear term,
\begin{equation}
c_1=\frac{1+i\mu\omega}{\sigma^2-i\omega},
\end{equation}
to the known exact solution~\cite{Acebron-Bonilla-Spigler-2000}
\begin{equation}
c_1 = \frac{e^{\mu\sigma^2}}{\sigma^2}\sum_{n = 0}^{\infty} \frac{\mu\sigma^2+n}{\mu\sigma^2+n-i\mu\omega}\frac{\left(-\mu\sigma^2\right)^n}{n!}.
\label{eq:sum_acebron}
\end{equation}
In Fig.~\ref{fig4}, the inverse critical coupling
\begin{equation}
\frac 1{\varepsilon_{\mathrm{cr}}} = \frac12\int_{-\infty}^{\infty} \fun{g}{\omega} c_1 \D{\omega}
\end{equation}
is plotted versus the distribution half-width $\gamma$ for the uniform distribution $g(\omega)$ within the interval $[-\gamma,\gamma]$. %и распределения Коши $\gamma/\pi(\gamma^2 + \omega^2)$.
For a small inertia, with the corrected Smoluchowski equation, the critical coupling becomes infinite ($\sigma^2/\varepsilon_\mathrm{cr}=0$) for a finite distribution width, which does not occur in reality. The deviation from the exact solution becomes noticeable at $\gamma\gtrsim 1/\mu$.

\begin{figure*}
\centerline{
\includegraphics[width=0.975\textwidth]{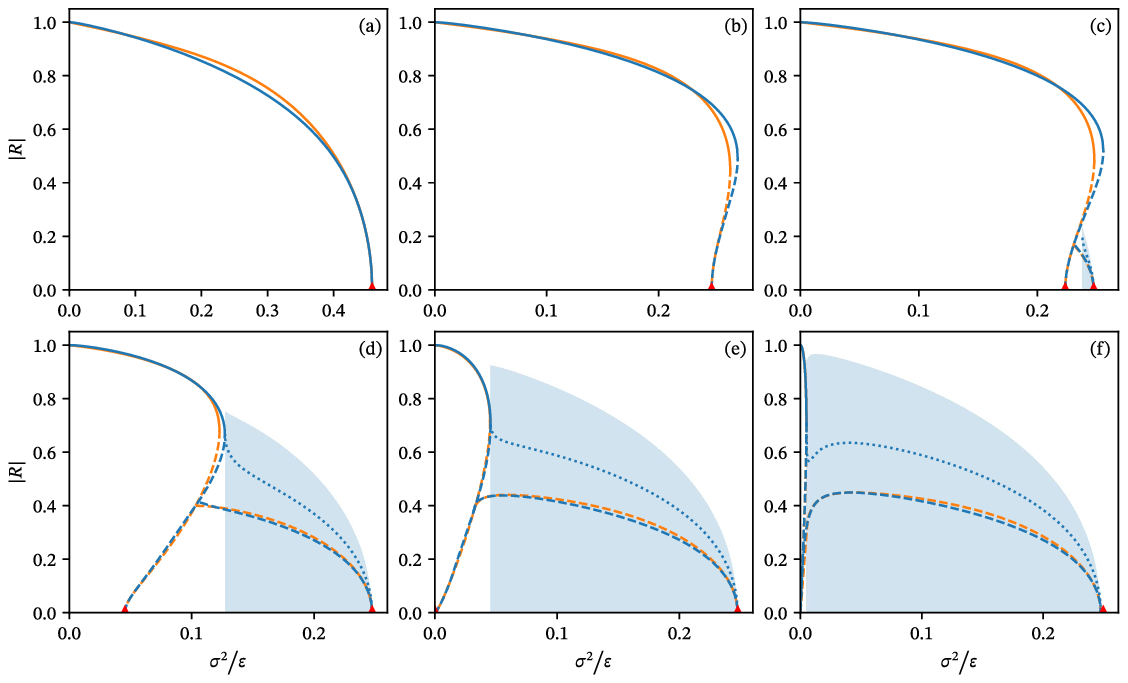}}
\caption{The dependencies of the global Kuramoto order parameter $|R|$ versus $\sigma^2/\varepsilon$ are plotted for a population of phase rotators with a bimodal frequency distribution in the thermodynamic limit. Blue lines: 2CC model (\ref{eq5200:01})--(\ref{eq5200:02}), orange: exact time-independent solutions of the corrected Smoluchowski equation. Lines are solid (dashed) for stable (unstable) solutions (stability was analyzed only for the blue lines, and for the orange lines it was inferred by analogy with the blue ones). The rms value of $|R|$ for oscillatory regimes is plotted with dotted lines; the shading shows the range of variation of $|R|$ for oscillatory solutions. The red triangles mark the critical values of coupling $\varepsilon_\mathrm{cr}$ calculated with Eq.~\eqref{eq:sum_acebron}.
Parameters: $\mu\sigma^2=0.01$ and $\gamma/\sigma^2=0.3$ (a), $1$ (b), $1.1$ (c), $3$ (d), $10$ (e), and $100$ (f).}
\label{fig5}
\end{figure*}

\subsection{Bimodal distribution}
In this section we employ the 2CC model (\ref{eq5200:01})--(\ref{eq5200:02}) for studying phase transitions in the population with the bimodal frequency distribution $g(\omega)=[\delta(\omega-\gamma)+\delta(\omega+\gamma)]/2$.
Namely, Eqs.\ (\ref{eq5200:01}) and (\ref{eq5200:02}) were written for each subpopulation (with $\omega=\pm\gamma$), coupled through $h=\varepsilon[\kappa_1(\omega=+\gamma)+\kappa_1(\omega=-\gamma)]/4$. The resulting 8-variable system (two pairs of coupled complex equations) was solved numerically.
For the bimodal frequency distribution, the picture of the phase transitions between regimes with different level of global synchrony quantified by the Kuramoto order parameter $R$ is quite rich and well studied in the no-inertia case~\cite{Bonilla-etal-1998,Martens-etal-2009,Campa-2020,Kostin-etal-2023}. Some time-independent states are oscillatory unstable and one observes stable collective oscillations. Both oscillatory instability and collective oscillations can be studied within the framework of low-dimensional 2CC model, but cannot be studied with the method of analytical CM solutions developed in~\cite{Munyayev-etal-2020} for time-independent macroscopic states. Moreover, the 2CC model with $h_1$ allows us to handle the regimes with time-dependent $h$ (and hence with time-dependent $a_1(\omega)$ and $R$).
In Fig.~\ref{fig5}, we report the phase diagrams of macroscopic regimes; the dependence of the global Kuramoto order parameter $|R|$ versus $\sigma^2/\varepsilon$ is plotted.

One can see that for small values of the order parameter time-independent solutions are accurately described by the 2CC model.
For small $\gamma/\sigma^2$ one observes mismatch for moderate synchronization levels ($0.5\lesssim|R|\lesssim0.8$); the 2CC approach misestimates the inertia correction for these states.
For larger values of $\gamma/\sigma^2$ (Fig.~\ref{fig5}e), the 2CC model accurately reproduces the stable time-independent solutions of the corrected Smoluchowski equation.
Finally, for large values of $\gamma/\sigma^2$ (Fig.~\ref{fig5}f), the solutions of the 2CC model and the corrected Smoluchowski equation are practically identical, but both models become inaccurate reduction of the original Fokker--Planck equation with inertia. In particular, in the limit $\gamma/\sigma^2\to0$ the inertia-induced shift of the Kuramoto-transition point vanishes (i.e., $\varepsilon_\mathrm{cr}=4\sigma^2$), while the corrected Smoluchowski equation (and the 2CC model) suggests $\varepsilon_\mathrm{cr}=4/(1-\mu\sigma^2)$. The absolute value of the inaccuracy turns out to be small for the considered bimodal distribution if $\mu\sigma^2\ll1$.

Summarizing, the numerical comparison for a bimodal distribution is found to be in a decent agreement with the results of Buckingham's method of dimensional analysis (Sec.~\ref{ssec61}) and confirms that the applicability of few-CC models with inertial corrections is given by the inequality chain~\eqref{eq:chainll}. Noticeably, the solution with only the two first CCs captures the effects of noise and inertia on time-independent states reasonably well. Moreover, it adequately reproduces the bifurcation scenario for bimodal distributions reported earlier in the literature for the no-inertia case~\cite{Bonilla-etal-1998,Martens-etal-2009,Campa-2020,Kostin-etal-2023}.
The circular cumulant approach appears a promising tool for such studies in the case with inertia.

\section{Conclusion}
\label{sec:concl}
For the Langevin equation with small inertia or large dissipation the problem of elimination of velocity (a fast variable) and reduction of the description to an effective dynamics of a single variable $\varphi$ has been addressed. Four approaches to this problem have been considered in detail:

\paragraph{Moment formalism:}
\label{par:M}
representation in terms of $w_n(\varphi)=\int_{-\infty}^{+\infty}v^n\rho(v,\varphi)\,\mathrm{d}v$\,;
%\\
calculations with Eqs.~(\ref{eq006})--(\ref{eq009}), see Figs.~\ref{fig1}(a) and \ref{fig2}(a) [or Eq.~(\ref{eq-abp-03}) for active Brownian particles, Fig.~\ref{fig3}].
\\
Adiabatic elimination requires the elements $w_n$ (or $W_n$) of the order $n$ from $0$ to $2$; the $\mu^1$-correction requires the elements with $n=0-4$; the $\mu^m$-correction: 0--$(2m+2)$.
The infinite chain of equations for $w_n$ is optimally truncated after an even-order element, $n=2m$, since keeping an odd-order element as a last nonzero one induces large truncation error and decreases the order of solution accuracy.

\paragraph{Cumulant formalism:}
\label{par:C}
representation in terms of $K_n(\varphi)$ (or $\varkappa_n=K_n/n!$) defined by recursive formulas~(\ref{eqCM3});
%\\
calculations with Eqs.~(\ref{eq027})--(\ref{eq028}), see Figs.~\ref{fig1}(b) and \ref{fig2}(b).
\\
Adiabatic elimination requires the elements $K_n$ with $n=0-2$; the $\mu^1$-correction: 0--2 (for adiabatic elimination the same three equations are used, but the higher-order contributions are dropped); the $\mu^m$-correction: 0--$(m+1)$.

\paragraph{The basis of Hermite functions}
\label{par:H}
$h_n(u)$ which are the eigenfunctions of operator $\hat{L}_1=\partial_u(u+\partial_u)$:
\[
%\textstyle
\rho(v,\varphi,t)=\sum_{n=0}^\infty \frac{\sigma}{\sqrt{\mu}}\,h_n\!\!\left(\frac{\sqrt{\mu}}{\sigma}v\right)\mathcal{W}_n(\varphi,t)\,;
\]
representation in terms of $\mathcal{W}_n$;
%\\
calculations with Eqs.~(\ref{eqH04})--(\ref{eqH05}), see Figs.~\ref{fig1}(c) and \ref{fig2}(c).
\\
Adiabatic elimination requires the elements with $n=0-1$; the $\mu^1$-correction: 0--2; the $\mu^m$-correction: 0--$(m+1)$.

\paragraph{Analog of the cumulant formalism for the representation of the Hermite function basis:}
\label{par:CH}
representation in terms of $\varkappa_n$ defined by recursive formulas~(\ref{eqCH3});
%\\
calculations with Eqs.~(\ref{eqH06})--(\ref{eqH07}), see Figs.~\ref{fig1}(d) and \ref{fig2}(d).
\\
Adiabatic elimination requires the elements with $n=0-1$;  the $\mu^1$-correction: 0--2; the $\mu^m$-correction: 0--$(m+1)$.

%{\color{blue}
The moment (a) and cumulant (b) representations can be immediately employed for numerical simulation of macroscopic dynamics of populations of active Brownian particles~\cite{Milster-etal-2017,Lighthill-1952,Blake-1971,Ebbens-Howse-2010} (Fig.~\ref{fig3}). Generally, calculations with system~(\ref{eq-abp-03}) for active Brownian particles with small but finite inertia require lengthy series and can suffer from numerical instabilities. To overcome these difficulties we employed modification~\cite{Permyakova-Goldobin-2025} of the exponential time differencing method~\cite{Cox-Matthews-2002}.

These representations are also suitable for theoretical studies. Within the framework of the fast variable elimination procedure for active particles, we have derived an effective stochastic dynamics description for one-dimensional overactive particles: see Fokker--Planck-type equation~(\ref{eq-abp-enhSmol}). In two and three dimensions, the diffusion/deterministic dynamics of a particle with small inertia is related to random walk/dynamics of the velocity angle~\cite{Erdmann-etal-2000,Erdmann-etal-2002,Erdmann-Ebeling-2005,Milster-etal-2017,Aranson-Pikovsky-2022,Pikovsky-2023}. In one dimension, this degree of freedom is absent and diffusion is contributed exclusively by the sporadic velocity reversals (through zero, at variance with rotational revolutions). For small inertia, this mechanism is negligible in higher dimensions and its mathematical theory is laborious (Sec.~\ref{sec:ABP}). The diffusion and forced drift terms in FPE~(\ref{eq-abp-enhSmol}), with constants $G_2$ and $\gamma_1$ given by Eqs.~(\ref{eq-abp-fe34}) and (\ref{eq-abp-gamma1}), are one of the main results of this paper.

Approaches~(c) and (d) using the Hermite function basis are most efficient~\cite{Komarov-Gupta-Pikovsky-2014} for systems with a linear dissipation law. However, their generalization to nonlinear laws, including active Brownian particles, requires individual mathematical preparation for each new law, which can be problematic.

The second main utilitarian result of this paper is derived for a linear dissipation law.
We have employed the corrected Smoluchowski equation~(\ref{eq024}) with time-dependent force $F(\varphi,t)$ to construct the generalization of the Ott--Antonsen Ansatz for oscillators with small effective inertia: see Sec.~\ref{sec:applic} and Eqs.~(\ref{eq5200:01}) and (\ref{eq5200:02}). These equations constitute a closed 4-dimensional (two complex variables) equation system governing macroscopic dynamics of the Kuramoto order parameter $\kappa_1=a_1$ and the deviation from the Ott--Antonsen Ansatz $\kappa_2=a_2-a_1^2$.

\acknowledgments{The authors are thankful to Prof.\ Arkady Pikovsky for fruitful discussions and, in particular, for clarifying the applicability conditions of small-inertia asymptotics,
and acknowledge financial support from RSF Grant No.\ 23-12-00180 (Secs.\ 3, 4 and Appendixes) and RSF Grant No.\ 22-12-00348-P (Secs.\ 2, 5, 6).}

\section*{DATA AVAILABILITY}
The data that support the findings of this article are available
on request.

%}

\appendix
\section{Recursive formulas for $K_n$ and $w_n$}
\label{sec:app1}
The standard relation between cumulants and moments of a single variable needs to be modified, since it relies on the properties $w_0=1$ and $K_0=0$, which are broken in our case. For $f_w(s,\varphi,t)=\exp[\phi(s,\varphi,t)]$ we can write $\partial_sf_w=f_w\partial_s\phi$ and substitute series (\ref{eqCM0}) and (\ref{eqCM2}):
\begin{equation}
\sum_{n=1}^{+\infty}w_{n}\frac{s^{n-1}}{(n-1)!} =\sum_{m=0}^{+\infty}w_{m}\frac{s^m}{m!} \sum_{l=1}^{+\infty}K_{l}\frac{s^{l-1}}{(l-1)!}\;.
\label{eq:app101}
\end{equation}
In the r.h.s.\ part of this equation we separate the $w_0$-terms and write
\begin{align}
&w_0\sum_{l=1}^{+\infty}K_{l}\frac{s^{l-1}}{(l-1)!} +\sum_{m=1}^{+\infty}w_{m}\frac{s^m}{m!} \sum_{l=1}^{+\infty}K_{l}\frac{s^{l-1}}{(l-1)!}
\nonumber\\
&\qquad
=w_0\sum_{l=1}^{+\infty}K_{l}\frac{s^{l-1}}{(l-1)!} +\sum_{n=2}^{+\infty}\sum_{l=1}^{n-1}\frac{w_{n-l}K_{l}s^{n-1}}{(n-l)!(l-1)!} \;,
\nonumber
\end{align}
where in the double sum we introduced $n=m+l$ and got rid of $m$. With the latter expression in the r.h.s.\ part of Eq.~(\ref{eq:app101}) we collect the coefficients of terms $s^{n-1}/(n-1)!$ and obtain
\begin{equation}
w_{n}=w_0K_n +\sum_{l=1}^{n-1}\frac{(n-1)!}{(n-l)!(l-1)!}w_{n-l}K_{l}
\quad\mbox{ for }n\ge1\,.
\label{eq:app102}
\end{equation}
For the $0$th order elements we set $s=0$ in definition $\phi(0,\varphi,t)=\ln{f_w(0,\varphi,t)}$ and find
\begin{equation}
K_0=\ln{w_0}\,.
\label{eq:app103}
\end{equation}
Eqs.~(\ref{eq:app103}) and (\ref{eq:app102}) are identical to the recursive formulas~(\ref{eqCM3}).

\section{Recursive formula for circular cumulants $\kappa_n$ and moments $a_n$}
\label{sec:app2}
For the distribution of a single cyclic variable considered in Sec.~\ref{ssec:CCs} we can use the result of Appendix~\ref{sec:app1} but without dependence of $f_w$ and $\phi$ on the second variable. Technically, we substitute $w_0\to a_0=1$, $w_{n\ge1}\to a_n$, $K_n\to(n-1)!\kappa_n$. Hence, Eq.~(\ref{eq:app103}) yields a trivial result $\kappa_0=0$ (as it should be for a single variable distribution) and Eq.~(\ref{eq:app102}) takes the form of
\begin{equation}
\frac{a_n}{(n-1)!}=\kappa_n +\sum_{l=1}^{n-1}\frac{a_{n-l}\kappa_{l}}{(n-l)!}\,,
\label{eq:app201}
\end{equation}
which is identical to Eq.~(\ref{eqCC05}).

\section{Recursive formulas for $\varkappa_n$ and $\mathcal{W}_n$}
\label{sec:app3}
The case of generating functions $f_\mathcal{W}(s,\varphi,t)$ (\ref{eqCH0}) and $\Phi(s,\varphi,t)$ (\ref{eqCH2}) can be obtained from the case of $f_w$ (\ref{eqCM0}) and $\phi$ (\ref{eqCM2}) of Appendix~\ref{sec:app1} by means of the substitution $(w_n,K_n)\to(n!\mathcal{W}_n,n!\varkappa_n)$. Hence, in place of Eqs.~(\ref{eq:app102}) and (\ref{eq:app103}), one finds
\begin{align}
&\qquad\qquad\qquad
 \varkappa_0=\ln{\mathcal{W}_0}\,,
\label{eq:app301}
\\
&\mathcal{W}_n=\mathcal{W}_0\varkappa_n +\sum_{l=1}^{n-1}\frac{l}{n}\mathcal{W}_{n-l}\varkappa_{l}
\quad\mbox{ for }n\ge1\,,
\label{eq:app302}
\end{align}
which is identical to the recursive formulas~(\ref{eqCH3}).

\section{Analytical calculation of constants $G_1$ and $G_2$}
\label{sec:app4}
We make use of the symmetry of the integrand of the integral with respect to $V$ in the definition of $G_1$, change the order of integration operations over the area $V_1\ge0,V\ge V_1$, evaluate the inner integral over $V$, and introduce $z=V_1^2$:
\begin{align}
G_1%&=\int\limits_{-\infty}^{+\infty}\mathrm{d}V\,Ve^{-V^4}\int\limits_0^V \mathrm{d}V_1\left[1-\mathrm{erf}(V_1^2)\right]e^{V_1^4}
%\nonumber\\
&=2\int\limits_{0}^{+\infty}\mathrm{d}V\,Ve^{-V^4}\int\limits_0^V \mathrm{d}V_1\left[1-\mathrm{erf}(V_1^2)\right]e^{V_1^4}
\nonumber\\
&=2\int\limits_{0}^{+\infty}\mathrm{d}V_1\int\limits_{V_1}^{+\infty} \mathrm{d}V\,Ve^{-V^4}\left[1-\mathrm{erf}(V_1^2)\right]e^{V_1^4}
\nonumber\\
%&\color{red}
%=\frac{\sqrt{\pi}}{2}\int\limits_{0}^{+\infty}\mathrm{d}V_1\left[1-\mathrm{erf}(V_1^2)\right]^2e^{V_1^4}
%\nonumber\\
&
=\frac{\sqrt{\pi}}{4}\int\limits_{0}^{+\infty}\mathrm{d}z\frac{\left[1-\mathrm{erf}(z)\right]^2e^{z^2}}{\sqrt{z}}.
\nonumber
\end{align}
This is the table integral, Eq.~(2.8.20.12) in~\cite{Prudnikov-Brychkov-Marichev-1992}:
\begin{align}
G_1&=\frac{\Gamma(\frac{1}{4})}{2}\left[\sqrt{\frac{\pi}{2}} -\frac{{}_3F_2(\frac{1}{4},\frac{1}{2},1;\frac{3}{4},\frac{5}{4};1)}{\sqrt{\pi}}\right]
\nonumber\\
&=0.49859365698...\,,
\label{eq:app401}
\end{align}
where the generalized hypergeometric function
\begin{equation}
\begin{array}{l}
{}_3F_2\left(\frac{1}{4},\frac{1}{2},1;\frac{3}{4},\frac{5}{4};1\right)
=\sum_{l=0}^{+\infty}\frac{2^l(2l-1)!!}{(4l-1)!!!!(4l+1)}
\\
=1+\frac{2}{3\times5} +\frac{2^2\times3}{3\times7\times9} +\frac{2^3\times3\times5}{3\times7\times11\times13}
%+\frac{2^4\times3\times5\times7}{3\times7\times11\times15\times17}
+\cdots\,.
\end{array}
\label{eq:app402}
\end{equation}
Constant
\begin{align}
G_2&=\frac{\sqrt{2}\Gamma(\frac34)}{\sqrt{\pi}}G_1=\frac{\pi}{\sqrt{2}} -{}_3F_2\left(\frac{1}{4},\frac{1}{2},1;\frac{3}{4},\frac{5}{4};1\right)
\nonumber\\
&=0.48749549439936...\,.
\nonumber
%\label{eq:app403}
\end{align}


\begin{thebibliography}{60}

\bibitem{Haken-1977}
 H.\ Haken,
 {\em Self-Organization}, in
 {\em Synergetics--An Introduction} %2nd ed.\
 (Springer, Berlin, 1977),
 pp.\ 191--223.
 \\
 \url{https://doi.org/10.1007/978-3-642-96363-6_7}

\bibitem{Gardiner-1983-1997}
 C.~W.\ Gardiner,
 {\em Handbook of Stochastic Methods},
 2nd ed.\
 (Springer, Berlin, 1997).

\bibitem{Becker-1985}
 R.\ Becker,
 {\em Schwankungen und Brownsche Bewegung}, in
 {\em Theorie der Warme}
 edited by W.~Ludwig
 (Springer, Berlin, 1985),
 pp.\ 277--314.
\\
 \url{http://doi.org/10.1007/978-3-662-10440-8_6}
\\
 {[{\it English translation}:
 R.\ Becker {\em Fluctuations and Brownian Motion}, in
 {\em Theory of Heat} edited by G.~Leibfried, 2nd ed.
 (Springer, Berlin, Heidelberg, 1967),
 pp.\ 302--343.
 \url{https://doi.org/10.1007/978-3-642-49255-6_6}]}


\bibitem{Winfree-1967}
 A.~T.\ Winfree,
 {\em Biological rhythms and the behavior of populations of coupled oscillators},
 J.\ Theor.\ Biol. {\bf 16}, 15 %--42
 (1967).
 \\
 \url{https://doi.org/10.1016/0022-5193(67)90051-3}

\bibitem{Kuramoto-1975}
 Y.\ Kuramoto,
 {\em Self-entrainment of a population of coupled non-linear oscillators},
 in
 {\em International Symposium on Mathematical Problems in Theoretical Physics},
 Springer Lecture Notes in Physics No.\ 39, edited by H.~Araki
 (Springer, New York, 1975), pp.\ 420--422.


\bibitem{Wilemski-1976}
 G.\ Wilemski,
 {\em On the derivation of Smoluchowski equations with corrections in the classical theory of Brownian motion},
 J.\ Stat.\ Phys. {\bf 14}, %(2),
 153 %--169
 (1976).
 \\
 \url{https://doi.org/10.1007/BF01011764}


\bibitem{Gardiner-1984}
 C.~W.\ Gardiner,
 {\em Adiabatic elimination in stochastic systems. I.\ Formulation of methods and application to few-variable systems},
 Phys.\ Rev.\ A {\bf 29}, 2814 %--2822
 (1984).
 \\
 \url{https://doi.org/10.1103/PhysRevA.29.2814}

\bibitem{Goldobin-Klimenko-2020}
 D.~S.\ Goldobin and L.~S.\ Klimenko,
 {\em Small and finite inertia in stochastic systems: Moment and cumulant formalisms},
 AIP Conf.\ Proc. {\bf 2216}, 070001 (2020).
 \\
 \url{https://doi.org/10.1063/5.0003459}


\bibitem{Yoshimura-Arai-2008}
 K.\ Yoshimura and K.\ Arai,
 {\em Phase Reduction of Stochastic Limit Cycle Oscillators},
 Phys.\ Rev.\ Lett. {\bf 101}, 154101 (2008).
 \\
 \url{https://doi.org/10.1103/PhysRevLett.101.154101}

\bibitem{Teramae-etal-2009}
  J.~N.\ Teramae, H.\ Nakao, and G.~B.\ Ermentrout,
 {\em Stochastic Phase Reduction for a General Class of Noisy Limit Cycle Oscillators},
 Phys.\ Rev.\ Lett. {\bf 102}, 194102 (2009).
 \url{https://doi.org/10.1103/PhysRevLett.102.194102}

\bibitem{Goldobin-etal-2010}
 D.~S.\ Goldobin, J.~N.\ Teramae, H.\ Nakao, and G.~B.\ Ermentrout,
 {\em Dynamics of Limit-Cycle Oscillators Subject to General Noise},
 Phys.\ Rev.\ Lett. {\bf 105}, 154101 (2010).
 \\
 \url{https://doi.org/10.1103/PhysRevLett.105.154101}



\bibitem{Acebron-Bonilla-Spigler-2000}
 J.~A.\ Acebr\'on, L.~L.\ Bonilla, and R.\ Spigler,
 {\em Synchronization in populations of globally coupled oscillators with inertial effects},
 Phys.\ Rev.\ E {\bf 62}, %(3),
 3437, %--3454
 (2000).
 \\
 \url{https://doi.org/10.1103/PhysRevE.62.3437}


\bibitem{Komarov-Gupta-Pikovsky-2014}
 M.\ Komarov, S.\ Gupta, and A.\ Pikovsky,
 {\em Synchronization transitions in globally coupled rotors in the presence of noise and inertia: Exact results},
 Europhys.\ Lett. {\bf 106}(4), 40003 (2014).
 \\
 \url{ https://doi.org/10.1209/0295-5075/106/40003}

\bibitem{Olmi-etal-2014}
 S.\ Olmi, A.\ Navas, S.\ Boccaletti, and A.\ Torcini,
 {\em Hysteretic transitions in the Kuramoto model with inertia},
 Phys.\ Rev.\ E {\bf 90}, %(4),
 042905 (2014).
 \\
 \url{https://doi.org/10.1103/PhysRevE.90.042905}

\bibitem{Olmi-2015}
 S.\ Olmi,
 {\em Chimera states in coupled Kuramoto oscillators with inertia},
 Chaos {\bf 25}, %(12),
 123125 (2015).
 \\
 \url{https://doi.org/10.1063/1.4938734}

\bibitem{Laing-2019}
 C.~R.\ Laing,
 {\em Dynamics and stability of chimera states in two coupled populations of oscillators},
 Phys.\ Rev.\ E {\bf 100}, %(4),
 042211 (2019).
 \\
\url{ https://doi.org/10.1103/PhysRevE.100.042211}



\bibitem{Bountis-etal-2014}
 T.\ Bountis, V.~G.\ Kanas, J.\ Hizanidis, and A. Bezerianos,
 {\em Chimera states in a two-population network of coupled pendulum-like elements},
 Eur.\ Phys.\ J.\ ST {\bf 223}, %(4),
 721 %--728
 (2014).
 \\
 \url{https://doi.org/10.1140/epjst/e2014-02137-7}

\bibitem{Jaros-Maistrenko-Kapitaniak-2015}
 P.\ Jaros, Yu.\ Maistrenko, and T.\ Kapitaniak,
 {\em Chimera states on the route from coherence to rotating waves},
 Phys.\ Rev.\ E {\bf 91}, %(2),
 022907 (2015).
 \\
 \url{https://doi.org/10.1103/PhysRevE.91.022907}


\bibitem{Munyayev-etal-2020}
 V.~O.\ Munyayev, L.~A.\ Smirnov, V.~A.\ Kostin, G.~V.\ Osipov, and A. Pikovsky,
 {\em Analytical approach to synchronous states of globally coupled noisy rotators},
 New J.\ Phys. {\bf 22}, 023036 (2020).
 \\
 \url{https://doi.org/10.1088/1367-2630/ab6f93}

\bibitem{Munyayev-etal-2022}
 V.~O.\ Munyaev, M.~I.\ Bolotov, L.~A.\ Smirnov, G.~V.\ Osipov, and I.\ Belykh,
 {\em Stability of rotatory solitary states in Kuramoto networks with inertia},
 Phys.\ Rev.\ E {\bf 105}, 024203 (2022).
 \\
 \url{https://doi.org/10.1103/PhysRevE.105.024203}

\bibitem{Munyayev-etal-2023}
 V.~O.\ Munyaev, M.~I.\ Bolotov, L.~A.\ Smirnov, G.~V.\ Osipov, and I.\ Belykh,
 {\em Cyclops States in Repulsive Kuramoto Networks: The Role of Higher-Order Coupling},
 Phys.\ Rev.\ Lett. {\bf 130}, 107201 (2023).
 \\
 \url{https://doi.org/10.1103/PhysRevLett.130.107201}



\bibitem{Alexandrov-Gorsky-2024}
 A.\ Alexandrov and A.\ Gorsky,
 {\em Penrose method for Kuramoto model with inertia and noise},
 Chaos Soliton.\ Fract. {\bf 183}, 114938 (2024).
 \\
 \url{https://doi.org/10.1016/j.chaos.2024.114938}


\bibitem{Zharkov-Altudov-1978}
 G.~F.\ Zharkov and Yu.~K.\ Al'tudov,
 {\em Alternating-current Josephson effect},
 Zh.\ Eksp.\ Teor.\ Fiz. {\bf 74}, 1727 (1978)
 [Sov.\ Phys.\ JETP {\bf 47}, 901 (1978)].
 \url{http://www.jetp.ras.ru/cgi-bin/dn/e_047_05_0901.pdf}

\bibitem{Schoner-Haken-1987}
 G.\ Sch\"oner and H.\ Haken,
 {\em A Systematic Elimination Procedure for Ito Stochastic Differential Equations and the Adiabatic Approximation},
 Z.\ Physik B -- Condensed Matter {\bf 68}, 89 %--103
 (1987).
 \\
 \url{https://doi.org/10.1007/BF01307868}


\bibitem{Milster-etal-2017}
 S.\ Milster, J.\ N\"otel, I.~M.\ Sokolov, and L.\ Schimansky-Geier,
 {\em Eliminating inertia in a stochastic model of a micro-swimmer with constant speed},
 Eur.\ Phys.\ J.\ ST {\bf 226}, %(9),
 2039 %--2055
 (2017).
 \\
 \url{https://doi.org/10.1140/epjst/e2017-70052-8}


\bibitem{Aranson-Pikovsky-2022}
 I.~S.~Aranson and A.~Pikovsky,
 {\em Confinement and Collective Escape of Active Particles},
 Phys.\ Rev.\ Lett. {\bf 128}, 108001 (2022).
 \\
 \url{https://doi.org/10.1103/PhysRevLett.128.108001}

\bibitem{Pikovsky-2023}
 A.~Pikovsky,
 {\em Deterministic active particles in the overactive limit},
 Chaos {\bf 33}, 113114 (2023).
 \\
 \url{https://doi.org/10.1063/5.0172125}


\bibitem{Tyulkina-etal-2018}
 I.~V.\ Tyulkina, D.~S.\ Goldobin, L.~S.\ Klimenko, and A.\ Pikovsky,
 {\em Dynamics of Noisy Oscillator Populations beyond the Ott-Antonsen Ansatz},
 Phys.\ Rev.\ Lett. {\bf 120}, 264101 (2018).
 \\
 \url{https://doi.org/10.1103/PhysRevLett.120.264101}


\bibitem{Goldobin-etal-2018}
 D.~S.\ Goldobin, I.~V.\ Tyulkina, L.~S.\ Klimenko, and A.\ Pikovsky,
 {\em Collective mode reductions for populations of coupled noisy oscillators},
 Chaos {\bf 28}, 101101 (2018).
 \\
 \url{https://doi.org/10.1063/1.5053576}

\bibitem{Goldobin-Dolmatova-2019}
 D.~S.\ Goldobin and A.~V.\ Dolmatova,
 {\em Ott-Antonsen ansatz truncation of a circular cumulant series},
 Phys.\ Rev.\ Research {\bf 1}, %(3),
 033139 (2019).
 \\
 \url{https://doi.org/10.1103/PhysRevResearch.1.033139}


\bibitem{Ott-Antonsen-2008}
 E.\ Ott and T.~M.\ Antonsen,
 {\em Low dimensional behavior of large systems of globally coupled oscillators},
 Chaos {\bf 18}, 037113 (2008).
 \url{https://doi.org/10.1063/1.2930766}

\bibitem{Ott-Antonsen-2009}
 E.\ Ott and T.~M.\ Antonsen,
 {\em Long time evolution of phase oscillator systems},
 Chaos {\bf 19}, 023117  (2009).
 \\
 \url{https://doi.org/10.1063/1.3136851}

\bibitem{Watanabe-Strogatz-1993}
 S.\ Watanabe and S.~H.\ Strogatz,
 {\em Integrability of a globally coupled oscillator array},
 Phys.\ Rev.\ Lett. {\bf 70}, 2391 %--2394
 (1993).
 \\
 \url{https://doi.org/10.1103/PhysRevLett.70.2391}

\bibitem{Watanabe-Strogatz-1994}
 S.\ Watanabe and S.~H.\ Strogatz,
 {\em Constant of motion for superconducting josephson arrays},
 Phys.\ D {\bf 74}(3-4), 197 %--253
 (1994).
 \\
 \url{https://doi.org/10.1016/0167-2789(94)90196-1}

\bibitem{Pikovsky-Rosenblum-2008}
 A.\ Pikovsky and M.\ Rosenblum,
 {\em Partially Integrable Dynamics of Hierarchical Populations of Coupled Oscillators},
 Phys.\ Rev.\ Lett. {\bf 101}, 264103 (2008).
 \\
 \url{https://doi.org/10.1103/PhysRevLett.101.264103}


\bibitem{Marvel-Mirollo-Strogatz-2009}
 S.~A.\ Marvel, R.~E.\ Mirollo, and S.~H.\ Strogatz,
 {\em Identical phase oscillators with global sinusoidal coupling evolve by M\"obius group action},
 Chaos {\bf 19}, %(4),
 043104 (2009).
 \\
\url{ https://doi.org/10.1063/1.3247089}


\bibitem{Tanaka-2007}
 D.\ Tanaka,
 {\em General Chemotactic Model of Oscillators},
 Phys.\ Rev.\ Lett. {\bf 99}, 134103 (2007).
 \\
 \url{https://doi.org/10.1103/PhysRevLett.99.134103}


\bibitem{OKeeffe-etal-2017}
 K.~P.\ O'Keeffe, H.\ Hong, and S.~H.\ Strogatz,
 {\em Oscillators that sync and swarm},
 Nat.\ Commun. {\bf 8}, 1504 (2017).
 \\
 \url{https://doi.org/10.1038/s41467-017-01190-3}



\bibitem{Juniper-etal-2015}
 M.~P.~N.\ Juniper, A.~V.\ Straube, R.\ Besseling, {\em et al.}
 {\em Microscopic dynamics of synchronization in driven colloids},
 Nat.\ Commun. {\bf 6}, 7187 (2015).
 \\
 \url{https://doi.org/10.1038/ncomms8187}

\bibitem{Tierno-Johansen-Straube-2021}
 P.\ Tierno, T.~H.\ Johansen, and A.~V.\ Straube,
 {\em Thermally active nanoparticle clusters enslaved by engineered domain wall traps},
 Nat.\ Commun. {\bf 12}, 5813 (2021).
 \\
 \url{https://doi.org/10.1038/s41467-021-25931-7}

\bibitem{Kourov-Samoilova-Straube-2025}
 M.\ Kourov, A.\ Samoilova, and A.\ Straube,
 {\em Dynamics of a chain of magnetically interacting particles in a one-dimensional periodic (energy) landscape},
 Bull.\ Russ.\ Acad.\ Sci.\ Phys. {\bf 89}, 1086 %--1092
 (2025).
 \\
 \url{https://doi.org/10.1134/S1062873825711742}



\bibitem{Morren-etal-2006}
 J.\ Morren, S.~W.~H.\ de Haan, W.~L.\ Kling, and J.~A.\ Ferreira,
 {\em Wind turbines emulating inertia and supporting primary frequency control},
 IEEE T.\ Power Syst. {\bf 21}(1), 433 %--434
 (2006).
 \\
 \url{https://doi.org/10.1109/TPWRS.2005.861956}


\bibitem{Short-Infield-Freris-2007}
 J.~A.\ Short, D.~G.\ Infield, and L.~L.\ Freris,
 {\em Stabilization of Grid Frequency Through Dynamic Demand Control},
 IEEE T.\ Power Syst. {\bf 22}(3), 1284 %--1293
 (2007).
 \\
 \url{https://doi.org/10.1109/TPWRS.2007.901489}

%\bibitem{Goldobin-etal-PND-2017}
% Голдобин Д.С., Долматова А.В., Розенблюм М., Пиковский А.
% Синхронизация в ансамблях Курамото–Сакагучи при конкурирующем влияния общего шума и глобальной связи // Известия вузов. Прикладная нелинейная динамика. 2017. Т. 25, № 6. C. 5–37.
% DOI: 10.18500/0869-6632-2017-25-6-5-37


\bibitem{Klinshov-Franovic-2019}
 V.\ Klinshov and I.\ Franovic,
 {\em Two scenarios for the onset and suppression of collective oscillations in heterogeneous populations of active rotators},
 Phys.\ Rev.\ E {\bf 100}, 62211 (2019).
 \\
 \url{https://doi.org/10.1103/PhysRevE.100.062211}

\bibitem{Pazo-Montbrio-2014}
 D.\ Paz\'o and E.\ Montbri\'o,
 {\em Low-dimensional dynamics of populations of pulse-coupled oscillators},
 Phys.\ Rev.\ X {\bf 4}, 011009 (2014).
 \\
 \url{https://doi.org/10.1103/PhysRevX.4.011009}

\bibitem{Laing-2014}
 C.~R.\ Laing,
 {\em Derivation of a neural field model from a network of theta neurons},
 Phys.\ Rev.\ E {\bf 90}, %(1),
 010901(R) (2014).
 \\
 \url{https://doi.org/10.1103/PhysRevE.90.010901}


\bibitem{Daido-1996}
 H.\ Daido,
 {\em Onset of cooperative entrainment in limit-cycle oscillators with uniform all-to-all interactions: bifurcation of the order function},
 Phys.\ D {\bf 91}(1--2), 24 %--66
 (1996).
 \\
 \url{https://doi.org/10.1016/0167-2789(95)00260-X}


\bibitem{Ley-Verdebout-2017}
 C.\ Ley and T.\ Verdebout,
 {\em Modern Directional Statistics}
 (Chapman and Hall/CRC, Boca Raton, 2017).
 \\
 \url{https://doi.org/10.1201/9781315119472}


\bibitem{Goldobin-Dolmatova-2020}
 D.~S.\ Goldobin and A.~V.\ Dolmatova,
 {\em Circular cumulant reductions for macroscopic dynamics of Kuramoto ensemble with multiplicative intrinsic noise},
 J.\ Phys.\ A: Math.\ Theor. {\bf 53}, %(8),
 08LT01 (2020).
 \\
 \url{https://doi.org/10.1088/1751-8121/ab6b90}

\bibitem{diVolo-etal-2022}
 M.\ di Volo, M.\ Segneri, D.~S.\ Goldobin, A.\ Politi, and A.\ Torcini,
 {\em Coherent oscillations in balanced neural networks driven by  endogenous fluctuations},
 Chaos {\bf 32}, %(2),
 023120 (2022).
 \url{https://doi.org/10.1063/5.0075751}


\bibitem{Pikovsky-Rosenblum-Kurths-2003}
 A.\ Pikovsky, M.\ Rosenblum, and J.\ Kurths,
 {\em Synchronization. A Universal Concept in Nonlinear Sciences}
 (Cambridge University Press, Cambridge, 2003). 432~p.


\bibitem{Lukacs-1970}
 E.\ Lukacs,
 {\em Characteristic Functions} 2nd ed.\
 (Griffin, London, 1970).


\bibitem{Zheng-Kotani-Jimbo-2021}
 T.\ Zheng, K.\ Kotani, and Y.\ Jimbo,
 {\em Distinct effects of heterogeneity and noise on gamma oscillation in a model of neuronal network with different reversal potential},
 Sci.\ Rep. {\bf 11}, 12960 (2021).
 \\
 \url{https://doi.org/10.1038/s41598-021-91389-8}


\bibitem{Goldobin-2021}
 D.~S. Goldobin,
 {\em Mean-field models of populations of quadratic integrate-and-fire neurons with noise on the basis of the circular cumulant approach},
 Chaos {\bf 31}, %(8),
 083112 (2021).
 \url{https://doi.org/10.1063/5.0061575}


\bibitem{Erdmann-etal-2000}
 U.\ Erdmann, W.\ Ebeling, L.\ Schimansky-Geier, and F.\ Schweitzer,
 {\em Brownian particles far from equilibrium},
 The European Physical Journal B-Condensed Matter and Complex Systems
 {\bf 15}, 105 %--113
 (2000).
 \\
 \url{https://doi.org/10.1007/s100510051104}


\bibitem{Erdmann-etal-2002}
 U.\ Erdmann, W.\ Ebeling, and V.~S.\ Anishchenko,
 {\em Excitation of rotational modes in two-dimensional systems of driven Brownian particles},
 Phys.\ Rev.\ E {\bf 65}, 061106 (2002).
 \\
 \url{https://doi.org/10.1103/PhysRevE.65.061106}


\bibitem{Erdmann-Ebeling-2005}
 U.\ Erdmann and W.\ Ebeling,
 {\em On the attractors of twodimensional Rayleigh oscillators including noise},
 International Journal of Bifurcation and Chaos {\bf 15}(11), 3623 %--3633
 (2005).
 \url{https://doi.org/10.1142/S0218127405014271}


\bibitem{Permyakova-Goldobin-2025}
 E.~V.\ Permyakova and D.~S.\ Goldobin,
 {\em High-order schemes of exponential time differencing for stiff systems with nondiagonal linear part},
 J.\ Comput.\ Phys. {\bf 520}, 113493 (2025).
 \\
 \url{https://doi.org/10.1016/j.jcp.2024.113493}


\bibitem{Cox-Matthews-2002}
 S.~M.\ Cox and P.~C.\ Matthews,
 {\em Exponential time differencing for stiff systems},
 J.\ Comput.\ Phys. {\bf 176}(2), 430 %--455
 (2002).
 \url{https://doi.org/10.1006/jcph.2002.6995}



\bibitem{Matthews-Cox-2000a}
 P.~C.\ Matthews and S.~M.\ Cox,
 {\em One-dimensional pattern formation with Galilean invariance near a stationary bifurcation},
 Phys.\ Rev. E {\bf 62}, %(2),
 R1473(R) (2000).
 \\
 \url{https://doi.org/10.1103/PhysRevE.62.R1473}


\bibitem{Matthews-Cox-2000b}
 P.~C.\ Matthews and S.~M.\ Cox,
 {\em Pattern formation with a conservation law},
 Nonlinearity {\bf 13}, %(4),
 1293 %--1320
 (2000).
 \\
 \url{https://doi.org/10.1088/0951-7715/13/4/317}


\bibitem{Callen-Welton-1951}
 H.~B.\ Callen and T.~A.\ Welton,
 {\em Irreversibility and Generalized Noise},
 Phys.\ Rev. {\bf 83}, %(1),
 34 %--40
 (1951).
 \\
 \url{https://doi.org/10.1103/PhysRev.83.34}


\bibitem{Kubo-1966}
 R.\ Kubo,
 {\em Fluctuation-dissipation theorem},
 Rep.\ Prog.\ Phys. {\bf 29}, %(1),
 255 %--284
 (1966).
 \\
 \url{https://doi.org/10.1088/0034-4885/29/1/306}


\bibitem{Hanggi-Thomas-1982}
 P.~H\"anggi and H.\ Thomas,
 {\em Stochastic processes: time evolution, symmetries and linear response},
 Phys.\ Rep. {\bf 88}(4), 207 %--319
 (1982).
 \\
 \url{https://doi.org/10.1016/0370-1573(82)90045-X}


\bibitem{Lighthill-1952}
 M.~J.\ Lighthill,
 {\em On the squirming motion of nearly spherical deformable bodies through liquids at very small reynolds numbers},
 Commun.\ Pure Appl.\ Math. {\bf 5}, 109 %--118
 (1952).
 \url{https://doi.org/10.1002/cpa.3160050201}


\bibitem{Blake-1971}
 J.~R.\ Blake,
 {\em A spherical envelope approach to ciliary propulsion},
 J.\ Fluid Mech. {\bf 46}, 199 %--208
 (1971).
 \\
 \url{https://doi.org/10.1017/S002211207100048X}


\bibitem{Ebbens-Howse-2010}
 S.~J.\ Ebbens and J.~R.\ Howse,
 {\em In pursuit of propulsion at the nanoscale},
 Soft Matter {\bf 6}, %(4),
 726 %--738
 (2010).
 \\
 \url{https://doi.org/10.1039/B918598D}


%\bibitem{Shklyaev-2015}
% Shklyaev S 2015
% {\em Europhys.\ Lett.} {\bf 110} %(5)
% 54002


\bibitem{Nayfeh-1981-1984}
 A.~H.\ Nayfeh,
 {\em Introduction to Perturbation Techniques}
 (Wiley-VCH, Weinheim, 2024).


%\bibitem{Marvel-Strogatz-2009}
% Marvel S.A., Strogatz S.H.
% Invariant submanifold for series arrays of Josephson junctions // Chaos. --- 2009. --- Vol.\ 19. --- 013132.
% https://doi.org/10.1063/1.3087132
%
%\bibitem{Cestnik-Pikovsky-2022a}
% R.\ Cestnik and A.\ Pikovsky,
% ``Hierarchy of Exact Low-Dimensional Reductions for Populations of Coupled Oscillators,''
% Phys.\ Rev.\ Lett. {\bf 128}(5), 054101 (2022).
% https://doi.org/10.1103/PhysRevLett.128.054101
%
%\bibitem{Cestnik-Pikovsky-2022b}
% R.\ Cestnik and A.\ Pikovsky,
% ``Exact finite-dimensional reduction for a population of noisy oscillators and its link to Ott-Antonsen and Watanabe-Strogatz theories,''	Chaos {\bf 32}(11), 113126 (2022).
% https://doi.org/10.1063/5.0106171
%
%\bibitem{Tyulkina-etal-2018-2019}
% И.В.\ Тюлькина, Д.С.\ Голдобин, Л.С.\ Клименко и А.С.\ Пиковский,
% ``Двухгрупповые решения для динамики ансамблей фазовых систем типа Отта-Антонсена,''
% Известия Вузов. Радиофизика {\bf 61}(8--9), 718--728 (2018).
%
%\bibitem{Mirollo-2012}
% R.~E.\ Mirollo,
% ``The asymptotic behavior of the order parameter for the infinite-N Kuramoto model,''
% Chaos \textbf{22}, 043118 (2012).
% https://doi.org/10.1063/1.4766596
%
%\bibitem{Pietras-Daffertshofer-2016}
% B.\ Pietras and A.\ Daffertshofer,
% ``Ott-Antonsen attractiveness for parameter-dependent oscillatory systems,''
% Chaos \textbf{26}, 103101 (2016).
% https://doi.org/10.1063/1.4963371


\bibitem{Goldobin-2019}
 D.~S.\ Goldobin,
 {\em Relationships between the Distribution of Watanabe-Strogatz Variables and Circular Cumulants for Ensembles of Phase Elements},
 Fluct.\ Noise Lett. {\bf 18}(2), 1940002 (2019).
 \\
 \url{https://doi.org/10.1142/S0219477519400029}


\bibitem{Bonilla-etal-1998}
 L.~L.\ Bonilla, C.~J.\ P\'erez Vicente, and R.\ Spigler,
 {\em Time-periodic phases in populations of nonlinearly coupled oscillators with bimodal frequency distributions},
 Phys.\ D \textbf{113}(1), 79 %--97
 (1998).
 \\
 \url{https://doi.org/10.1016/S0167-2789(97)00187-5}

\bibitem{Martens-etal-2009}
 E.~A.\ Martens, E.\ Barreto, S.~H.\ Strogatz, E.\ Ott, P.\ So, and T.~M.\ Antonsen,
 {\em Exact results for the Kuramoto model with a bimodal frequency distribution},
 Phys.\ Rev.\ E \textbf{79}, 026204 (2009).
 \\
 \url{https://doi.org/10.1103/PhysRevE.79.026204}

\bibitem{Campa-2020}
 A.\ Campa,
 {\em Phase diagram of noisy systems of coupled oscillators with a bimodal frequency distribution},
 J.\ Phys.\ A: Math.\ Theor. \textbf{53}, 154001 (2020).
 \\
 \url{https://doi.org/10.1088/1751-8121/ab79f2}

\bibitem{Kostin-etal-2023}
 V.~A.\ Kostin, V.~O.\ Munyaev, G.~V.\ Osipov, and L.~A.\ Smirnov,
 {\em Synchronization transitions and sensitivity to asymmetry in the bimodal Kuramoto systems with Cauchy noise},
 Chaos \textbf{33}, 083155 (2023).
 \\
 \url{https://doi.org/10.1063/5.0160006}


\bibitem{Prudnikov-Brychkov-Marichev-1992}
 A.~P.\ Prudnikov, Yu.~A.\ Brychkov, and O.~I.\ Marichev,
 {\em Integrals and Series. Vol.\ 2: Special Functions}
 (Gordon and Breach Science Publishers, New York, 1992).


\end{thebibliography}
\end{document}